\documentclass[twocolumn]{aastex62}

\def\chandra{{\itshape Chandra\/}}

\def\hst{{\itshape HST\/}}

\def\spitzer{{\itshape Spitzer\/}}
\def\swift{{\itshape Swift\/}}
\def\herschel{{\itshape Herschel\/}}
\def\wise{{\itshape WISE\/}}
\def\galex{{\itshape GALEX\/}}
\def\uhuru{{\itshape Uhuru\/}}

\def\xray{\hbox{X-ray}}

\def\etal{{et\,al.}}

\def\ltsima{$\; \buildrel < \over \sim \;$}
\def\simlt{\lower.5ex\hbox{\ltsima}}
\def\gtsima{$\; \buildrel > \over \sim \;$}
\def\simgt{\lower.5ex\hbox{\gtsima}}
\def\kms{\ifmmode{~{\rm km~s^{-1}}}\else{~km s$^{-1}$}\fi}
\def\lsim{\lower0.3em\hbox{$\,\buildrel <\over\sim\,$}}
\def\gsim{\lower0.3em\hbox{$\,\buildrel >\over\sim\,$}}

\def\msol{$M_\odot$}
\def\h2{H$_2$}
\def\flux{erg~cm$^{-2}$~s$^{-1}$}
\def\lum{erg~s$^{-1}$}

\def\arcsec{\mbox{$^{\prime\prime}$}}

\def\aap{A\&A}
\def\apj{ApJ}
\def\apjl{ApJL}
\def\apjs{ApJS}
\def\aj{AJ}
\def\mnras{MNRAS}
\def\araa{ARA\&A}

\def\nat{Nature}

\def\ngal{24}  
\def\ngc{595}  
\def\nbkg{104}  
\def\nbkgf{47}  
\def\nfield{1238}  
\def\nxsrc{3923}
\def\nx{1937}

\usepackage{underscore}
\usepackage{longtable}
\usepackage{amsmath}
\shorttitle{Field LMXBs in Elliptical Galaxies}
\shortauthors{Lehmer et al.}

\begin{document}

%
\title{X-ray Binary Luminosity Function Scaling Relations in Elliptical Galaxies: Evidence for Globular Cluster Seeding of Low-Mass X-ray Binaries in Galactic Fields}

\correspondingauthor{Bret Lehmer}
\email{lehmer@uark.edu}

\author{Bret~D.~Lehmer}
\affil{Department of Physics, University of Arkansas, 226 Physics Building, 825 West Dickson Street, Fayetteville, AR 72701, USA}

\author{Andrew~P.~Ferrell}
\affil{Department of Physics, University of Arkansas, 226 Physics Building, 825 West Dickson Street, Fayetteville, AR 72701, USA}

\author{Keith~Doore}
\affil{Department of Physics, University of Arkansas, 226 Physics Building, 825 West Dickson Street, Fayetteville, AR 72701, USA}

\author{Rafael~T.~Eufrasio}
\affil{Department of Physics, University of Arkansas, 226 Physics Building, 825 West Dickson Street, Fayetteville, AR 72701, USA}

\author{Erik~B.~Monson}
\affil{Department of Physics, University of Arkansas, 226 Physics Building, 825 West Dickson Street, Fayetteville, AR 72701, USA}

\author{David~M.~Alexander}
\affil{Department of Physics, University of Durham, South Road,
Durham DH1 3LE, UK}

\author{Antara Basu-Zych}
\affiliation{NASA Goddard Space Flight Center, Code 662, Greenbelt, MD 20771, USA}
\affiliation{Center for Space Science and Technology, University of
Maryland Baltimore County, 1000 Hilltop Circle, Baltimore, MD 21250, USA}

\author{William~N.~Brandt}
\affil{Department of Astronomy \& Astrophysics, Pennsylvania State
University, University Park, PA 16802, USA}
\affil{Institute for Gravitation and the Cosmos, The Pennsylvania State University, 525 Davey Lab, University Park, PA 16802, USA}
\affil{Department of Physics, The Pennsylvania State University, University Park, PA 16802, USA}

\author{Greg~Sivakoff}
\affil{Department of Physics, University of Alberta, CCIS 4-183
Edmonton, AB T6G 2E1, Canada}

\author{Panayiotis Tzanavaris}
\affiliation{NASA Goddard Space Flight Center, Code 662, Greenbelt, MD 20771, USA}
\affiliation{Center for Space Science and Technology, University of
Maryland Baltimore County, 1000 Hilltop Circle, Baltimore, MD 21250, USA}

\author{Mihoko Yukita}
\affiliation{The Johns Hopkins University, Homewood Campus, Baltimore, MD
21218, USA}

\author{Tassos Fragos}
\affiliation{Geneva Observatory, Geneva University, Chemin des Maillettes
51, 1290 Sauverny, Switzerland}

\author{Andrew Ptak}
\affiliation{NASA Goddard Space Flight Center, Code 662, Greenbelt, MD 20771, USA}
\affiliation{The Johns Hopkins University, Homewood Campus, Baltimore, MD
21218, USA}

%
\begin{abstract}
%
We investigate X-ray binary (XRB) luminosity function (XLF) scaling relations
for \chandra\ detected populations of low-mass XRBs (LMXBs) within the
footprints of \ngal\ early-type galaxies.  Our sample includes \chandra\ and
\hst\ observed galaxies at $D \simlt 25$~Mpc that have estimates of the
globular cluster (GC) specific frequency ($S_N$) reported in the literature.
As such, we are able to directly classify \xray-detected sources as being
either coincident with unrelated background/foreground objects, GCs, or sources
that are within the fields of the galaxy targets.  We model the GC and field
LMXB population XLFs for all galaxies separately, and then construct global
models characterizing how the LMXB XLFs vary with galaxy stellar mass and
$S_N$.  We find that our {\it field} LMXB XLF models require a component that
scales with $S_N$, and has a shape consistent with that found for the GC LMXB
XLF.  We take this to indicate that GCs are ``seeding'' the galactic field LMXB
population, through the ejection of GC-LMXBs and/or the diffusion of the GCs in
the galactic fields themselves.  However, we also find that an important LMXB
XLF component is required for all galaxies that scales with stellar mass,
implying that a substantial population of LMXBs are formed ``in situ,'' which
dominates the LMXB population emission for galaxies with $S_N \simlt 2$.  For
the first time, we provide a framework quantifying how directly-associated GC
LMXBs, GC-seeded LMXBs, and in-situ LMXBs contribute to LMXB XLFs in the
broader early-type galaxy population.

%
\end{abstract}
%

\section{Introduction}

Due to its subarcsecond imaging resolution, \chandra\ has revolutionized our
understanding of \xray\ binary (XRB) formation and evolution by dramatically
improving our ability to study XRBs in extragalactic environments (see, e.g.,
Fabbiano~2006 for a review).  Extragalactic XRBs probe the compact object
populations and accretion processes within parent stellar populations that can
vary considerably from those represented in the Milky Way (MW; e.g., starbursts
and massive elliptical galaxies).  Low-mass XRBs (LMXBs) are of broad importance in
efforts to understand XRBs, as they are the most numerous XRB populations in
the MW (Grimm \etal\ 2002; Liu \etal\ 2007) and likely dominate the XRB
emissivity of the Universe from $z \approx$~0--2 (Fragos \etal\ 2013).  With
\chandra, these populations are readily resolved into discrete point sources in
relatively nearby ($D \simlt 30$~Mpc) elliptical galaxies; however, there is still debate about
their formation pathways.

LMXB populations are thought to form through two basic channels: (1) Roche-lobe
overflow of normal stars onto compact-object companions in isolated binary
systems that form in situ within galactic fields; and (2) dynamical
interactions (e.g., tidal capture  and multibody exchange with
constituent stars in primordial binaries) in high stellar density environments
like globular clusters (GCs; Clark~1975; Fabian \etal\ 1975; Hills~1976), and
possibly some high-density galactic regions (e.g., Voss \& Gilfanov 2007; Zhang
\etal\ 2011).  The ``in-situ LMXBs'' form on stellar evolutionary
timescales (typically $\simgt$1~Gyr) following past star-formation events.  In
contrast, the  ``GC LMXBs'' form continuously over time as stochastic
interactions between stars tighten binary orbits and induce mass transfer.  

Since the early results from \uhuru, it has been known that the number of LMXBs
per unit stellar mass coincident with GCs is a factor of $\sim$50--100 times larger
than that observed for the Galactic field (Clark~1975; Katz~1975), clearly
indicating the importance of the GC LMXB formation channel.  
GC LMXBs have been
studied extensively in the literature, showing that stellar interaction rates
and metallicity are the primary factors that influence the formation of these
systems
(see, e.g., Pooley \etal\ 2003; Heinke \etal\ 2003; 
Jord{\'a}n \etal\ 2007; Sivakoff \etal\ 2007; Maxwell \etal\ 2012; Kim \etal\ 2013; Cheng \etal\ 2018).

Fewer studies have been able to explore the notable population of 
LMXBs that have been observed within galactic {\it
fields}, which apparently trace the distributions of the old stellar
populations (e.g., in late-type galaxy bulges and early-type galaxies).  Given
the very high formation efficiencies of GC LMXBs, and similarities in the
\xray\ properties of field versus GC LMXBs, it has been speculated that the
field LMXB population may have also formed dynamically within GC environments,
and then subsequently been planted within galactic fields, potentially through
the ejection of LMXBs from GCs (Grindlay \& Hertz~1985; Hut \etal\ 1992; Kremer
\etal\ 2018) or the dissolution of GCs (e.g., Grindlay~1984).  Several studies
have confirmed strong correlations between the LMXB population emission per
optical luminosity, $L_{\rm X}/L_{\rm opt}$, and the GC specific frequency:
$S_N \equiv N_{\rm GC} 10^{0.4(M_{V}^T + 15)}$, which is the number of GCs per
$V$-band luminosity (e.g., Irwin~2005; Juett~2005; Humphrey \& Buote 2008;
Boroson \etal\ 2011; Zhang \etal\ 2012).  However, a non-zero intercept of the
$L_{\rm X}/L_{\rm opt}$--$S_N$ correlation implied that a non-negligible
population of LMXBs that are unassociated with GCs must be present and dominant
at low-$S_N$, suggesting the in-situ formation channel is likely very important
(e.g., Irwin~2005).  

The majority of early \chandra\ studies of LMXB populations within elliptical
galaxies investigated correlations between the total LMXB \xray\ luminosity
function (XLF) and host-galaxy stellar mass ($M_\star$) and were unable to
segregate field versus GC sources directly (e.g., Kim \& Fabbiano~2004;
Gilfanov~2004; Humphrey \& Buote~2008; Zhang \etal\ 2012).  These
investigations identified breaks in the LMXB XLF around 0.5--8~keV luminosities
$L \approx 5 \times 10^{37}$~\lum\ and $\approx 3 \times 10^{38}$~\lum, and
showed that the XLF normalization increases with stellar mass and $S_N$.  In
the case of Zhang \etal\ (2012), a positive correlation was also observed
between stellar age and $S_N$, indicating that stellar age may also
be a driving physical factor.

Over the last decade, \chandra\ studies have directly isolated field LMXBs by
removing \xray\ sources with direct \hst\ counterparts that are associated with
either GCs or unrelated foreground stars, background galaxies, and active
galactic nuclei (AGN; e.g., Kim \etal\ 2009; Voss \etal\ 2009; Paolillo \etal\
2011; Luo \etal\ 2013; Lehmer \etal\ 2014; Mineo \etal\ 2014; Peacock \&
Zepf~2016; Peacock \etal\ 2017; Dage \etal\ 2019).  These studies have found
that the field LMXB XLF appears to have a steeper slope at $L \simgt 5 \times
10^{37}$~\lum\ compared to the GC XLF and shows no obvious galaxy-to-galaxy
variations among old elliptical galaxies, implying that the field LMXB
population is dominated by sources formed via the in-situ channel.
Furthermore, contrary to the findings of Zhang \etal\ (2012), Kim \& Fabbiano
(2010) and Lehmer \etal\ (2014) claimed an observed excess of luminous LMXBs in
{\it young} elliptical galaxies with $\simlt$5~Gyr stellar populations versus
{\it old} elliptical galaxies with $\simgt$8~Gyr.  These findings have been
supported by the observed increase in the average $L_{\rm X}$(LMXB)/$M_\star$
with increasing redshift among galaxy populations in deep \chandra\ surveys
(see, e.g., Lehmer \etal\ 2007; 2016; Aird \etal\ 2017), and are consistent
with population synthesis model predictions of the in situ LMXB XLF evolution
with increasing host stellar population age (see, e.g., Fragos \etal\ 2008,
2013a, 2013b).  

In this paper, we use the combined power of \chandra\ and \hst\ data to provide
new insight into the nature of the in-situ and GC formation channels, focusing
on the field LMXB population.  We study in detail a sample of \ngal\ elliptical
galaxies, using both archival and new data sets, with the aim of rigorously
testing whether there is evidence for GC seeding or a stellar-age dependence in
the field LMXB populations from XLFs.  This represents a factor of
three times larger study over any other published studies that analyze the GC
and field LMXB population XLFs separately (i.e., compared to the eight galaxies
studied by Peacock \& Zepf~2016 and Peacock \etal\ 2017).  In
Section~2 we describe our sample selection.  Section~3 provides details on the
various multiwavelength, \hst, and \chandra\ data analyses, and presents the
properties of the galaxies and their \xray\ point sources.  Section~4 details
our XLF fitting of the field, GC, and total LMXB populations, and culminates in
a global XLF model framework that self-consistently fits the XLFs of all
galaxies in our sample.  In Section~5, we discuss and interpret our results and
outline a way forward to establishing a universal physical parameterization of
XRB XLFs.  Full catalogs of the \chandra\ sources, \chandra\ images, and
additional supplementary data sets are provided publicly\footnote{{\ttfamily
https://lehmer.uark.edu/downloads/}} and archived in Zenodo [doi: {\ttfamily
10.5281/zenodo.3751108}].

Throughout this paper, we quote \xray\ fluxes and luminosities in the
\hbox{0.5--8~keV} bandpass that have been corrected for Galactic absorption,
{\it but not} intrinsic absorption.  Estimates of $M_\star$ and SFR presented
throughout this paper have been derived assuming a Kroupa~(2001) initial mass
function (IMF); when making comparisons with other studies, we have adjusted
all values to correspond to our adopted IMF.


\begin{deluxetable*}{lllccccccccc}
\tablewidth{1.0\columnwidth}
\tabletypesize{\footnotesize}
\tablecaption{Nearby Galaxy Sample and Properties}
\tablehead{
\multicolumn{1}{c}{\sc Galaxy}  & \colhead{} & \colhead{} & \colhead{} &\colhead{} & \colhead{} & \multicolumn{3}{c}{\sc Size Parameters} & \colhead{} & \colhead{} & \colhead{} \\
\vspace{-0.25in} \\
\multicolumn{1}{c}{\sc Name} & \multicolumn{1}{c}{\sc Alt.} &  \multicolumn{1}{c}{\sc Morph.} & \multicolumn{2}{c}{\sc Central Position} & \colhead{$D$} & \colhead{$a$} & \colhead{$b$} & \colhead{PA} & \colhead{$\log M_\star$} &  \colhead{} & \colhead{$\langle t_{\rm age} \rangle$}\\ 
\vspace{-0.25in} \\
\multicolumn{1}{c}{(NGC)} & \multicolumn{1}{c}{\sc Name} & \multicolumn{1}{c}{\sc Type} & \colhead{$\alpha_{\rm J2000}$} & \colhead{$\delta_{\rm J2000}$} & \colhead{(Mpc)} & \multicolumn{2}{c}{(arcmin)} & \colhead{(deg)} &  \colhead{($M_\odot$)}   & \colhead{$S_N$} &  \colhead{(Gyr)} \\ 
\vspace{-0.25in} \\
\multicolumn{1}{c}{(1)} & \multicolumn{1}{c}{(2)} & \multicolumn{1}{c}{(3)} & \colhead{(4)} & \colhead{(5)} & \colhead{(6)} & \colhead{(7)} & \colhead{(8)} & \colhead{(9)} & \colhead{(10)} & \colhead{(11)} & \colhead{(12)}
}
\startdata
      1023 &            &        SB0 &       02 40 24.0 & +39 03 47.7 &   11.43$\pm$1.00 &   3.02 &   1.15 &       82.0 &  10.62$\pm$0.01 &             1.71$\pm$0.10 &   8.76$\pm$0.20 \\
      1380 &            &       S0-a &     03 36 27.6 & $-$34 58 34.7 &   18.86$\pm$1.85 &   1.78 &   0.79 &        7.0 &  10.58$\pm$0.02 &             1.06$\pm$0.25 &   8.38$\pm$0.25 \\
      1387 &            &       E/S0 &     03 36 57.1 & $-$35 30 23.9 &   19.82$\pm$0.70 &   1.27 &   1.04 &      110.0 &  10.51$\pm$0.01 &             1.80$\pm$0.12 &   8.94$\pm$0.16 \\
      1399 &            &         E1 &     03 38 29.1 & $-$35 27 02.7 &   20.68$\pm$0.50 &   1.89 &   1.89 &      150.0 &  10.89$\pm$0.01 &             9.25$\pm$1.08 &   9.01$\pm$0.09 \\
      1404 &            &         E1 &     03 38 51.9 & $-$35 35 39.8 &   20.43$\pm$0.40 &   1.38 &   1.24 &      162.5 &  10.74$\pm$0.01 &             1.78$\pm$0.32 &   8.94$\pm$0.11 \\
\\
      3115 &            &         S0 &     10 05 14.0 & $-$07 43 06.9 &   10.00$\pm$0.50 &   2.74 &   1.07 &       45.0 &  10.59$\pm$0.01 &             1.84$\pm$0.27 &   8.90$\pm$0.11 \\
      3377 &            &         E5 &       10 47 42.4 & +13 59 08.3 &   11.04$\pm$0.25 &   1.41 &   0.82 &       48.0 &   9.84$\pm$0.02 &             2.00$\pm$0.16 &   6.20$\pm$0.45 \\
      3379 &       M105 &         E1 &       10 47 49.6 & +12 34 53.9 &   10.20$\pm$0.50 &   1.80 &   1.53 &       67.5 &  10.37$\pm$0.01 &             0.94$\pm$0.18 &   9.03$\pm$0.05 \\
      3384 &            &        SB0 &       10 48 16.9 & +12 37 45.5 &   10.80$\pm$0.77 &   2.07 &   1.06 &       50.5 &  10.08$\pm$0.06 &             0.76$\pm$0.19 &   4.54$\pm$1.07 \\
      3585 &            &         E7 &     11 13 17.1 & $-$26 45 18.0 &   21.20$\pm$1.73 &   1.85 &   1.17 &      104.5 &  10.90$\pm$0.01 &             0.57$\pm$0.19 &   8.88$\pm$0.16 \\
\\
      3923 &            &         E4 &     11 51 01.8 & $-$28 48 22.4 &   22.91$\pm$3.15 &   1.99 &   1.28 &       47.5 &  10.84$\pm$0.01 &             3.43$\pm$0.37 &   8.68$\pm$0.14 \\
      4278 &            &          E &       12 20 06.8 & +29 16 49.8 &   16.07$\pm$1.55 &   1.24 &   1.16 &       27.5 &  10.48$\pm$0.01 &             4.50$\pm$1.23 &   8.74$\pm$0.18 \\
      4365 &            &         E3 &       12 24 28.2 & +07 19 03.1 &   23.33$\pm$0.65 &   1.88 &   1.39 &       45.0 &  10.97$\pm$0.01 &             3.73$\pm$0.69 &   9.00$\pm$0.10 \\
      4374 &        M84 &         E1 &       12 25 03.8 & +12 53 13.1 &   18.51$\pm$0.61 &   1.92 &   1.76 &      123.0 &  10.92$\pm$0.01 &             4.89$\pm$1.37 &   8.53$\pm$0.18 \\
      4377 &            &         S0 &       12 25 12.3 & +14 45 43.9 &   17.67$\pm$0.59 &   0.60 &   0.52 &      170.0 &   9.84$\pm$0.01 &             1.19$\pm$0.52 &   8.66$\pm$0.26 \\
\\
      4382 &        M85 &         S0 &       12 25 24.1 & +18 11 26.9 &   17.88$\pm$0.56 &   2.46 &   1.65 &       12.5 &  10.88$\pm$0.02 &             1.40$\pm$0.23 &   7.91$\pm$0.27 \\
      4406 &        M86 &         E3 &       12 26 11.8 & +12 56 45.5 &   17.09$\pm$0.52 &   2.52 &   1.69 &      125.0 &  10.82$\pm$0.01 &             3.19$\pm$0.23 &   8.35$\pm$0.20 \\
      4472 &        M49 &         E2 &       12 26 11.8 & +12 56 45.5 &   17.03$\pm$0.21 &   2.99 &   2.42 &      162.5 &  11.07$\pm$0.01 &             5.21$\pm$0.60 &   9.09$\pm$0.04 \\
      4473 &            &         E5 &       12 29 48.9 & +13 25 45.6 &   15.25$\pm$0.51 &   1.56 &   0.84 &       95.0 &  10.34$\pm$0.02 &             1.78$\pm$0.46 &   8.44$\pm$0.21 \\
      4552 &        M89 &          E &       12 35 39.9 & +12 33 21.7 &   15.89$\pm$0.55 &   1.48 &   1.39 &      150.0 &  10.63$\pm$0.01 &             7.68$\pm$1.40 &   9.09$\pm$0.05 \\
\\
      4621 &        M59 &         E5 &       12 42 02.3 & +11 38 48.9 &   14.85$\pm$0.50 &   1.82 &   1.18 &      165.0 &  10.53$\pm$0.01 &             2.34$\pm$1.03 &   8.86$\pm$0.15 \\
      4649 &        M60 &         E2 &       12 43 40.0 & +11 33 09.4 &   17.09$\pm$0.61 &   2.44 &   1.98 &      107.5 &  11.09$\pm$0.01 &   4.35$\pm$0.54$^\dagger$ &   9.09$\pm$0.04 \\
      4697 &            &         E6 &     12 48 35.9 & $-$05 48 03.1 &   12.01$\pm$0.78 &   2.06 &   1.30 &       67.5 &  10.45$\pm$0.02 &             3.01$\pm$0.79 &   7.68$\pm$0.32 \\
      7457 &            &         S0 &       23 00 60.0 & +30 08 41.2 &   13.24$\pm$1.34 &   1.27 &   0.70 &      128.0 &   9.71$\pm$0.02 &             2.36$\pm$0.74 &   5.60$\pm$0.50 \\
\hline
{\bf Median} & \ldots & \ldots & \ldots & \ldots & 17.09 & 1.88 & 1.24 & \ldots & 10.62 & 2.34 & 8.76 \\ 
\enddata
$^\dagger$Value of $S_N$ has been corrected from H13 following the assumptions in Section~3.2.
\tablecomments{Col.(1): NGC number of galaxy. Col.(2): Alternative Messier designation, if applicable. Col.(3): Morphological type as reported in H13.  Col.(4) and (5): Right ascension and declination of the galactic center based on the 2 Micron All Sky Survey (2MASS) positions derived by Jarrett \etal\ (2003). Col.(6): Adopted distance and 1$\sigma$ error in units of Mpc.  For consistency with the H13 GC specific frequencies, we adopted the distances reported in H13 (see references within). Col.(7)--(9): $K_s$-band isophotal ellipse parameters, including, respectively, semi-major axis, $a$, semi-minor axis, $b$, and position angle east from north, PA.  The ellipses tract the 20~mag~arcsec$^{-2}$ surface brightness contour of each galaxy (derived by Jarrett \etal\ 2003) and are centered on the positions given in Col.(4) and (5).  Col.(10): Logarithm of the galactic stellar mass, $M_\star$, determined by our SED fitting.  These stellar masses are based on photometry from the areal regions defined in Col.(4)--(5) and Col.(7)--(8), excluding a central 3\arcsec\ circular region and any sky coverage that does not have \hst\ exposure (see Section 3.1 for details). The cumulative stellar mass of the sample is $\log (M_\star^{\rm tot}/M_\odot) = 12.1$. Col.(11): GC specific frequency, $S_N$, as reported by H13. Col.(12): Stellar-mass-weighted age of the population, based on the SED fitting techniques applied in Section~3.1.\\
}
\end{deluxetable*}

%
%
\begin{figure*}[t!]
\figurenum{1}
\centerline{
\includegraphics[width=19cm]{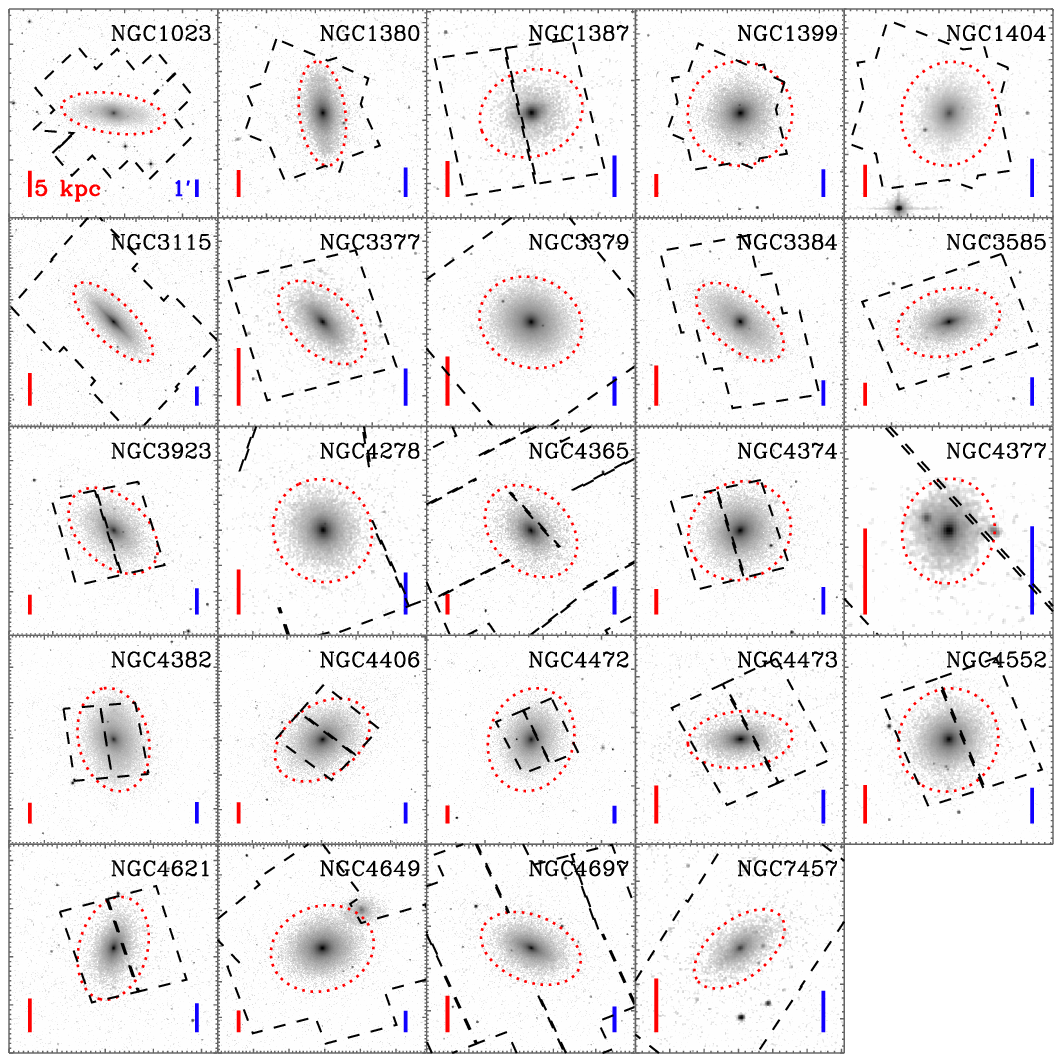}
\vspace{-0.3in}
}
\caption{
Log-scale 2MASS $K_s$-band images of the \ngal\ galaxies in our sample.  All
images have square dimensions with the length of each side being equal to two
times the $K_s$-band major axis corresponding to $\approx$20~mag~arcsec$^{-2}$
(as reported by Jarrett \etal\ 2003).  The galactic regions are highlighted
with red dotted ellipses, and the \hst\ ACS coverage of each galaxy has been shown
with dashed polygonal regions.  Note that some of the ACS regions are complex
(e.g., NGC~4278 and NGC~4365) due to chip gaps. For reference, vertical bars of
size 5~kpc and 1~arcmin are provided in the lower-left ({\it red}\/) and
lower-right ({\it blue\/}) of each panel, respectively.
}
\end{figure*}

\section{Sample Selection and Properties}

We began by selecting a sample of relatively nearby ($D \simlt 25$~Mpc)
early-type galaxies (from E to S0 morphologies across the Hubble sequence) that
had available deep \chandra\ ACIS ($\simgt$40~ks depth) data, as well as \hst\
ACS imaging over two bandpasses in the optical/near-IR (see Section~3.2
below).  These requirements allow us to identify \xray\
point-sources, isolate the faint optical counterparts, and effectively
classify these counterparts as GC or unrelated foreground/background
objects (e.g., Galactic stars and active galactic nuclei [AGN]).  We also
required that the galaxies have estimates of the GC richness, via
measurements of the GC specific frequency, $S_N$ (see Section 1).  

We chose to make use of the Harris \etal\ (2013; hereafter, H13) catalog of 422
galaxies with published measurements of GC population properties.  The H13
catalog consists of culled results, including values of $S_N$, from 112 papers
that had been published before 2012 December.  We note that the \hst\ data for
our sample is excellent for detecting and characterizing GCs and computing
$S_N$; however, the \hst\ footprints of these data are often constrained to
regions that do not encompass the full extents of the GC populations.
In this study, we are interested in characterizing GC-related LMXB
populations that are directly associated with GCs, as well as those ejected by
GCs that are observed in galactic fields.  For many galaxies in our sample, the
latter ``seeded'' LMXB populations are expected to have contributions from GCs
located well outside of the observational fields.  As such, we make use of
``local'' specific frequencies, $S_{N, {\rm loc}}$, which we calculate using
the \hst\ data presented here (see Section~3.2), when studying LMXB populations
directly associated with GCs.  We also make use of the ``global'' $S_N$ values derived
from H13 when studying seeded LMXB populations.

Using the criteria above, and rejecting galaxies that were very close to
edge-on (e.g., NGC~5866), had significant dust lanes (e.g., NGC~4526), or had
widely variable data coverage across the extents of the galaxies (e.g., M87 and
Cen~A), we identified \ngal\ elliptical galaxies from the H13 sample that were
suitable for our study.  These galaxies and their properties are tabulated in
Table~1.
In Figure~1, we show Two Micron All Sky Survey (2MASS) $K_s$-band
image cutouts of the sample.  Our sample spans the full morphological range of
our initial selection (E to S0).  The majority of the galaxies are in groups or
cluster environments, including three members from the M96 group, four Fornax
cluster galaxies, and eight Virgo cluster galaxies.  Our sample spans a
galactic stellar mass range of $\log M_\star =$~9.7--11.1, with a median $\log
M_\star^{\rm med}=$~10.6.  The GC specific frequency range is broad, spanning
$S_N =$~0.6--9.3, with a median value of $S_N^{\rm med} = $~2.0.  As such, this
sample is GC rich compared to similar mass late-type galaxies, which have a
median $S_N \approx 1$ (H13).

Given that our sample is selected from the complex combination of availability
of GC property measurements from the literature (as per H13) and the existence of
\chandra\ and \hst\ data, we do not regard this sample as representative of any
specific early-type galaxy population.  For instance, the selection bias of the
sample favors massive early-type galaxies with rich GC systems
(see, e.g., Brodie \& Strader~2006).  Despite the heterogeneous selection, our
approach here is to quantify how the XRB population XLFs in these galaxies are
correlated with host-galaxy properties and to assess how well these trends
describe all of the galaxies individually.  If such a ``global'' model is
successful for all galaxies, it is likely (though not guaranteed) to be
applicable to other galaxies with similar morphologies, mass ranges, and GC
$S_N$ ranges.  However, lower-mass early-type galaxies and galaxies with
different morphological types (e.g., late-type galaxies) can often have
different star-formation histories, GC $S_N$ values, and metallicities that can
have an effect on the XRB populations (see, e.g., Fragos \etal\ 2008, 2013a,
2013b; Basu-Zych \etal\ 2013, 2016; Brorby \etal\ 2016; Lehmer \etal\ 2014,
2019; Fornasini \etal\ 2019).

\section{Data Analysis}

To address the goal of quantifying how the {\it field}~LMXB XLF is influenced by stellar
ages and the injection of sources that originate in GCs, we require 
knowledge of (1) the star-formation histories (SFHs) of the galaxies, (2) the
GC source locations, and (3) the \xray\ source locations.  As such, we
calculate coarse SFHs using spectral energy distribution (SED) fitting
procedures applied to FUV--to--FIR data sets, directly identify GCs in our
galaxy sample using \hst\ imaging data, and identify \xray\ point sources using
\chandra\ data.  Our data analysis procedures and results are detailed below.


\begin{table*}
\begin{center}
\caption{Multiwavelength Coverage Used in SED Fitting}
\begin{tabular}{lcccccccc}
\hline\hline
&  \multicolumn{2}{c}{\galex} & \multicolumn{6}{c}{\swift} \\ 
\multicolumn{1}{c}{\sc Galaxy} &  \multicolumn{2}{c}{\rule{1.2in}{0.01in}} & \multicolumn{6}{c}{\rule{4.2in}{0.01in}} \\ 
\multicolumn{1}{c}{(NGC)} & FUV & NUV & UVW2 & UVM2 & UVW1 & $U$ & $B$ & $V$ \\ 
\hline
1023 &             $-0.25\pm0.06$ &             $ 0.36\pm0.06$ &                   $\ldots$ &                   $\ldots$ &                   $\ldots$ &                   $\ldots$ &                   $\ldots$ &                   $\ldots$ \\
1380 &             $-0.20\pm0.06$ &             $ 0.15\pm0.06$ &             $ 0.18\pm0.02$ &             $ 0.14\pm0.02$ &             $ 0.65\pm0.02$ &             $ 1.29\pm0.02$ &                   $\ldots$ &                   $\ldots$  \\
1387 &             $-0.28\pm0.06$ &             $ 0.10\pm0.06$ &                   $\ldots$ &                   $\ldots$ &                   $\ldots$ &                   $\ldots$ &                   $\ldots$ &                   $\ldots$  \\
1399 &             $ 0.33\pm0.06$ &             $ 0.49\pm0.06$ &             $ 0.57\pm0.02$ &             $ 0.51\pm0.02$ &             $ 0.90\pm0.02$ &             $ 1.50\pm0.02$ &             $ 2.05\pm0.02$ &             $ 2.39\pm0.02$  \\
1404 &             $-1.21\pm0.07$ &             $-0.23\pm0.06$ &             $ 0.69\pm0.02$ &             $ 0.64\pm0.02$ &             $ 1.22\pm0.02$ &             $ 1.79\pm0.02$ &             $ 2.36\pm0.02$ &             $ 2.69\pm0.02$  \\
\hline\hline
\end{tabular}
\end{center}
NOTE.---All columns, with the exception of the first column, provide the
logarithm of the flux, with 1$\sigma$ error, for each of the noted bandpass.
The fluxes are quoted in units of mJy and are appropriate for the regions
described in Section 3.1.  Only a portion of the table is shown here to
illustrate form and content.  The full table is available in machine-readable
form and provides flux measurements for all \ngal\ galaxies and 31 different
bandpasses.  
\end{table*}
%

%
\begin{figure*}
\figurenum{2}
\centerline{
\includegraphics[width=18cm]{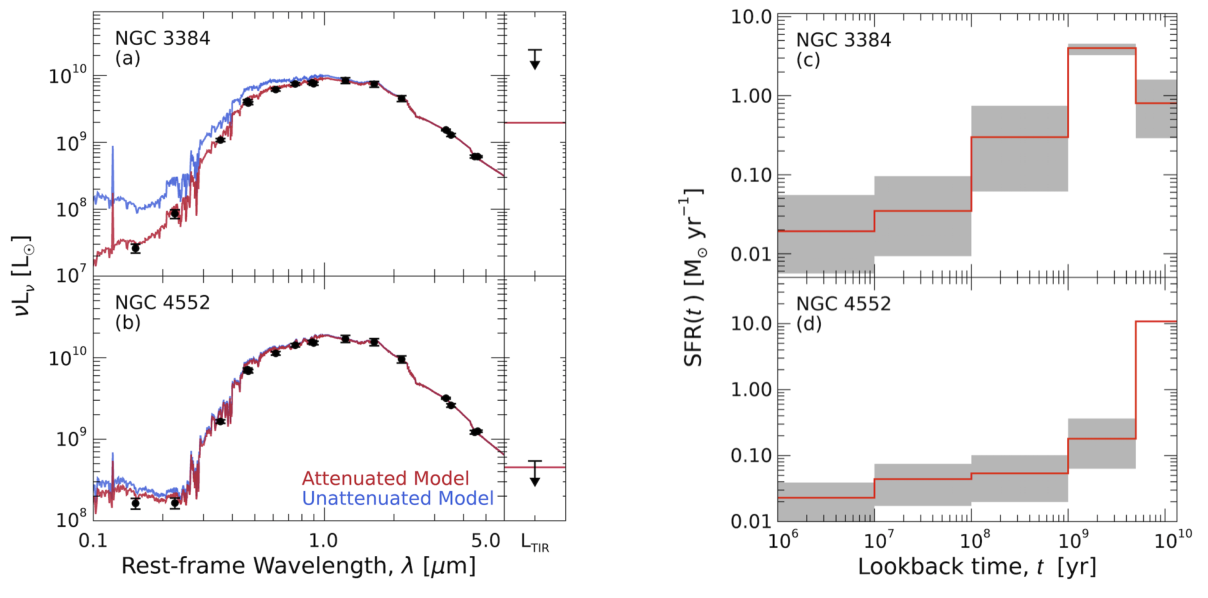}
}
\caption{ ({\it Left\/}) The observed UV--to--IR SED and the integrated total
infrared (TIR) luminosity using the 24$\mu$m-based calibration from Galametz \etal\ (2013)
are shown for NGC~3384 ($a$) and NGC~4552 ($b$), the galaxies with the
respectively youngest and oldest stellar-mass weighted ages in our sample. For
each galaxy, the best-fit model and unattenuated, intrinsic model are shown as
the red and blue curves, respectively. ({\it Right}) The resulting SFHs (median
values) and 16--84\% uncertainty ranges on the SFHs ({\it gray shaded
regions\/}) for NGC~3384 ($c$) and NGC~4552 ($d$).  The SFHs of all the
galaxies in our sample strongly favor large contributions from old stellar
populations, and have mass-weighted stellar ages ranging from
$\approx$4--9~Gyr, bracketed by the galaxies displayed here.  A full version of
this figure for all \ngal\ galaxies in our sample can be found in the
electronic version.}
\end{figure*}

\subsection{FUV to FIR Data Reduction and Star-Formation History Estimates}

The FUV--to--FIR SEDs for all \ngal\ galaxies were extracted using publicly
available data from \galex, \swift, \hst, SDSS, 2MASS, \wise, \spitzer, and
\herschel.  For a given galaxy, we limited our analyses to regions that
consisted of the intersection of the galactic extent, as estimated by an
ellipse approximating the 2MASS $K_{s}$-band $\approx$20~mag~arcsec$^{-2}$
isophotal contour (from Jarrett \etal\ 2003), and the \hst\ coverage of the
galaxy (see Section 3.2 below).  These regions (the galactic ellipses and \hst\
coverage areas) trace the bulk of the stellar mass of the galaxies, while
permitting us to directly identify GC and background-source counterparts.
Figure~1 highlights these areas for each galaxy and Table~1 provides their
sizes and orientations.  After visually inspecting \hst\ and \chandra\ images
of the galaxies, we chose to further exclude small, circular regions with
3~arcsec radii from the center of each galaxy to avoid complications from
potential AGN or extreme crowding of sources.  We note that all \ngal\ galaxies
harbor \xray\ detected sources within these nuclear regions, with NGC~1380,
NGC~1399, NGC~1404, NGC~3923, NGC~4278, NGC~4365, NGC~4374, NGC~4552, and
NGC~4649 containing sources in these regions with 0.5--8~keV luminosities in
the range of (1--20)~$\times 10^{39}$~\lum.  Such sources are not highly
luminous AGN, but are strong candidates for low-luminosity AGN.  
When constructing our SEDs, we extracted photometry from regions that were
within the ellipses that had \hst\ exposure, yet were outside the central
excluded core.  As such, these properties are not representative of the entire
galaxy, but in most cases are a significant fraction of the total stellar mass.
Hereafter, all quoted properties, with the exception of $S_N$, are derived from
these regions.

For each imaging data set from \galex\ FUV to \wise\ 4.6~$\mu$m, we masked full width at half maximum (FWHM)
circular regions at the locations of all foreground Galactic stars that were
within the galactic extents defined above and replaced the photometry with
local median backgrounds, following the procedure described in Section~2.2 of Eufrasio \etal\
(2017). 
We assumed that the contribution from foreground stars at $\simgt$5~$\mu$m is
negligible. Once foreground stars were removed and replaced, the total
photometry of the coverage region for each band was calculated by summing all
pixels within the region after the diffuse background emission was subtracted.
Uncertainties were determined from a combination of background and calibration
uncertainties using the methods described in Eufrasio \etal\ (2014). If a band
had incomplete coverage within the galactic extents, or if its total
photometry was less than 3$\sigma$ above the background level, that band was
excluded from our SED fitting analysis.  In Table~2, we summarize the bands that
satisfied the above criteria for each galaxy and were used in our
SED fitting.

To fit a given SED and estimate the corresponding SFH, we used the
{\ttfamily Lightning} SED fitting code (Eufrasio \etal\ 2017). {\ttfamily
Lightning} is a non-parametric SED fitting procedure, which fits the stellar
emission from the FUV to the NIR (through \wise\ 4.6$\mu$m), including
extinction that is restricted to be in energy balance with the dust emission in
the FIR (\wise\ 22$\mu$m to \herschel\ 250$\mu$m).  The stellar SED is based on
the {\ttfamily P\'EGASE} (Fioc \& Rocca-Volmerange~1997) population synthesis
models and a SFH model that consists of five discrete time steps, of constant
SFR, at \hbox{0--10~Myr}, 10--100~Myr, 0.1--1~Gyr, 1--5~Gyr, and 5--13.3~Gyr.
The stellar emission from these specific age bins provide comparable bolometric
contributions to the SED of a typical late-type galaxy SFH, and contain
discriminating features that can be discerned in broad-band SED fitting (see
Eufrasio \etal\ 2017 for details).  

The reprocessed dust emission is modeled as a single measurement of the
integrated 8--1000~$\mu$m total infrared luminosity, $L_{\rm TIR}$, which is
estimated based on the Galametz \etal\ (2013) scaling of 24~$\mu$m \spitzer\
(or 22$\mu$m from \wise\ when \spitzer\ was unavailable) to $L_{\rm TIR}$.
Since elliptical galaxies typically contain low levels of dust emission and
little obscuration, the 24~$\mu$m emission can have contributions from stellar
emission, as well as dust heated by stellar light from the central region of
the galaxy that is masked out (see above).  Furthermore, the Galametz \etal\
(2013) prescription is appropriate for dust heated by young stellar
populations, which is likely to overestimate $L_{\rm TIR}$ (e.g., Temi \etal\
2005).  As such, we treat our estimates of $L_{\rm TIR}$ as upper limits, with
high-luminosity Gaussian tails with 1$\sigma$ values set equal to the scatter-related
uncertainties provided in Table~2 of Galametz \etal\ (2013).

Figure~2 shows example UV--to--IR SED fit results (including SED models and
resulting SFHs from {\ttfamily Lightning}) for NGC~3384 and NGC~4552, which
have, respectively, the youngest and oldest mass-weighted stellar ages in our
sample: $4.54 \pm 1.07$~Gyr and $9.09 \pm 0.05$~Gyr, respectively (see Col.(12)
of Table~1).  We note that all galaxies, except for NGC~3384, have
mass-weighted stellar ages estimated to be in the narrow range of
$\simgt$5~Gyr, with a full-sample mean mass-weighted stellar age of $8.31 \pm
0.24$~Gyr (1$\sigma$ error on the mean).  All galaxies are fit well by our SED
models (fits for all \ngal\ galaxies are provided in
the electronic version in an expanded version of Figure~2).

The mass-weighted age range of our sample indicates that our galaxies are
expected to have XRB populations dominated by old LMXBs with little diversity
in host stellar-population age.  This lack of diversity is in contrast to the
$\approx$2--15~Gyr stellar age estimate range reported in the literature for
these same galaxies (e.g., Trager \etal\ 2000; Terlevich \& Forbes~2001; Thomas
\etal\ 2005; McDermid \etal\ 2006; S{\'a}nchez-Bl{\'a}zquez~2006).  However,
these other estimates are based primarily on absorption line strength
measurements from optical spectra that are appropriate for single or
light-weighted age stellar populations.  These ages are thus strongly sensitive
to metallicity and SFH variations (e.g., Rogers \etal\ 2010) and can differ in
value by as much as a factor of $\approx$4 between studies.  Furthermore,
mid-IR-based studies have shown that elliptical galaxies with young-age estimates
(e.g., $\simlt$5~Gyr) commonly overpredict the observed IR luminosities,
suggesting these galaxies are likely dominated by older ($\simgt$5~Gyr) stellar
populations (e.g., Temi \etal\ 2005).  We calculate from our SFH model SEDs
that $B$-band-luminosity-weighted ages are $\approx$0.3--3~Gyr younger than
mass-weighted ages.  More appropriate to LMXB studies are ``mass weighted''
stellar ages, which are difficult to derive from optical spectroscopy alone due
to low levels of optical emission from old stellar populations.  Our SED
fitting methods, by contrast, use information from UV, near-IR, and far-IR,
which allow for better decomposition of the SFHs of galaxies, with much less
sensitivity to metallicity variations, compared to optical spectroscopic line
indices.  We are therefore confident that our SED fitting results are
sufficiently robust for further interpretations throughout this paper.


\begin{table*}
\begin{center}
\caption{\hst\ Advanced Camera for Surveys (ACS) Observation Log}
\begin{tabular}{lcccccccc}
\hline\hline
  &   &  &  &  & \multicolumn{4}{c}{\sc Line Dither Pattern}\\ 
  &   &  & &   & \multicolumn{4}{c}{\rule{1.8in}{0.01in}}\\ 
  &   &  & \multicolumn{2}{c}{{\sc Exposure Time}}   &  & & \multicolumn{2}{c}{\sc Spacing} \\ 
%
%
  & {\sc Field}   & {\sc Obs. Start} &  \multicolumn{2}{c}{{\sc (s)}}  &   \multicolumn{2}{c}{$N_{\rm pts}$}  &  \multicolumn{2}{c}{(arcsec)}\\ 
\multicolumn{1}{c}{\sc Galaxy} & {\sc Number} & (UT)  &    \multicolumn{2}{c}{\rule{1.0in}{0.01in}}  &   \multicolumn{2}{c}{\rule{1.0in}{0.01in}} & \multicolumn{2}{c}{\rule{1.0in}{0.01in}}\\ 
\multicolumn{1}{c}{(1)}  & (2)  & (3) & (4) & (5) & (6) & (7) & (8) & (9)\\
\hline\hline
& & & F475W & F850LP & F475W & F850LP & F475W & F850LP \\
\hline
NGC~1023\ldots\ldots\dotfill 	&              1 &        2011-09-21 00:36 &	768 &	1308 &	2 &	3 &	0.145 &	0.145\\ 
&              2 &        2011-09-23 22:44 &	776 &	1316 &	2 &	2 &	0.145 &	0.145\\ 
&              3 &        2011-09-21 20:59 &	768 &	1308 &	2 &	3 &	0.145 &	0.145\\ 
&              6 &        2011-09-23 00:52 &	768 &	1308 &	2 &	3 &	0.145 &	0.145\\ 
&              7 &        2011-09-26 19:14 &	776 &	1316 &	2 &	2 &	0.145 &	0.145\\ 
&              8 &        2012-10-14 06:09 &	768 &	1308 &	2 &	3 &	0.145 &	0.145\\ 
\hline
& & & F475W & F850LP & F475W & F850LP & F475W & F850LP \\
\hline
NGC~1380\ldots\ldots\dotfill 	&              1 &        2004-09-06 23:54 &	760 &	1220 &	2 &	2 &	0.146 &	0.146\\ 
&              2 &        2006-08-03 23:53 &	680 &	\ldots &	2 &	\ldots &	2.8 &	\ldots\\ 
\hline
& & & F475W & F850LP & F475W & F850LP & F475W & F850LP \\
\hline
NGC~1387\ldots\ldots\dotfill 	&              1 &        2004-09-10 14:48 &	760 &	1220 &	2 &	2 &	0.146 &	0.146\\ 
\hline
& & & F475W & F850LP & F475W & F850LP & F475W & F850LP \\
\hline
NGC~1399\ldots\ldots\dotfill 	&              2 &        2004-09-11 07:59 &	760 &	1220 &	2 &	2 &	0.146 &	0.146\\ 
&              2 &        2006-08-02 23:54 &	680 &	\ldots &	2 &	\ldots &	2.8 &	\ldots\\ 
\hline
\hline
 \end{tabular}
 \end{center}
Note.---An abbreviated version of the table is displayed here to illustrate its
form and content.  Col.(1): Target name. Col.(2): Field number for \hst.
Col.(3): Observation start. Col(4)--(5): Exposure time for \hst\ filters
listed. Col.(6)--(7): Number of points in line dither pattern, $N_{\rm pts}$, for \hst\ filters
listed. Col.(8)--(9): Spacing in arcsec for \hst\ filters listed.
Instances with dots denote no data available. \\
(This table is available in its entirety in machine-readable form.)

 \end{table*}

\subsection{\hst\ Data Reduction}

By selection, our galaxies have \hst\ ACS coverage in both ``blue'' and ``red''
filters (defined below) and cover the bulk of the stellar mass within the
$K_s$-band ellipses of the galaxies.  Figure~1 shows the \hst\ footprints for
each of our galaxies and Table~3 provides an observational log of the
data sets used.  Half of our galaxies (i.e., 12) have \hst\ data covering the
full $K_s$-band ellipses.  The remaining twelve galaxies
only miss peripheral edges of the $K_s$ band
ellipses (see Fig.~1).  For 21 of the galaxies, we used the F475W ($g_{475}$)
and F850LP ($z_{850}$) bandpasses as our blue and red filters, respectively; however, when
this combination was not available, we utilized F475W and F814W ($z_{814}$) (NGC~1404 and NGC~4382) or F606W ($r_{606}$) and F814W (NGC~3923) filter pairs.

For each galaxy that had more than one \hst\ field of view, we created mosaicked
images.  These were constructed by first running the {\ttfamily Tweakreg} and
{\ttfamily Tweakback} tools (Fruchter \& Hook~2002), available in the
{\ttfamily Drizzlepac} version 2.1.14, STScI package.\footnote{For {\ttfamily
Drizzlepac} details, see
\url{http://www.stsci.edu/scientific-community/software/drizzlepac.html}.}
These tools first identify discrete sources that are common to all images
in a given
overlapping region, and then update the image headers
to align with one of the images (chosen as a reference), once an astrometric
solution is found.   Given the small overlaps between some image sets, we
implemented only small linear shifts in right ascension and declination to
align our images (typically only a few pixels, but up to 50~pixels in one
case).  After aligning all ACS fields of both filters for a given galaxy, we
then generated the mosaicked blue and red images by running {\ttfamily
astrodrizzle}.  The {\ttfamily astrodrizzle} procedure uses the aligned,
flat-field calibrated and charge-transfer efficiency (CTE) corrected images to
create a distortion-corrected mosaicked image with bad pixels and cosmic rays
removed. 

To construct \hst\ source catalogs, we ran
{\ttfamily SExtractor} (Bertin \& Arnouts 1996) on each image mosaic. 
We used a minimum of 10 above-threshold pixels for detection, with detection
and analysis thresholds set to 5$\sigma$. We used  
FWHM 2.5~pixels filtering Gaussians. Two apertures were used for photometry,
with radii of $r_1 = 0\farcs31$ (6.25~pixels) and $r_2 = 0\farcs63$
(12.5~pixels). The zeropoints used are from the ACS zeropoint calculator. Gains were
calculated using the exposure times and the CCDGAIN header keywords. The
background meshes were $8 \times 8$ pixels with a background filter size of
2.5~pixels FWHM.  We required that sources be present in both filters within a
tolerance of 0\farcs2 and have FWHM~$>$~1\farcs5 to
eliminate cosmic ray detections that were not rejected by {\ttfamily
astrodrizzle} (e.g., near image edges and gaps that have dithered exposures). 


\begin{table*}
\begin{center}
\caption{{\it HST} Source Classifications and $S_{N, {\rm loc}}$ Estimates}
\begin{tabular}{lccccccccc}
\hline\hline
 & & \multicolumn{4}{c}{\sc Optical}  & \multicolumn{4}{c}{\sc X-ray Detected} \\
\multicolumn{1}{c}{\sc Gal} & $M_g^{50}$ &\multicolumn{4}{c}{\rule{1.6in}{0.01in}}  & \multicolumn{4}{c}{\rule{1.3in}{0.01in}} \\
\multicolumn{1}{c}{(NGC)} & (mag) & $N_{\rm bkg}$ & $N_{\rm GC}$ & $M_{V, {\rm loc}}$ & $S_{N,{\rm loc}}$ & $N_{\rm X}$ & $N_{\rm bkg}$ & $N_{\rm GC}$ & $N_{\rm field}$  \\
\multicolumn{1}{c}{(1)} & (2) & (3) & (4) & (5) & (6) & (7) & (8) & (9) & (10) \\
\hline\hline
                     1023 &  $-$4.9 &  206 &  195 &        $-$20.2 &  1.65$\pm$0.12 &   66 &    3 &    7 &   56 \\
                     1380 &  $-$6.0 &   60 &  293 &        $-$20.2 &  3.03$\pm$0.18 &   35 &    0 &   15 &   20 \\
                     1387 &  $-$6.1 &  281 &  260 &        $-$19.8 &  4.23$\pm$0.26 &   13 &    1 &    8 &    4 \\
                     1399 &  $-$6.2 &  104 &  817 &        $-$20.8 &  5.09$\pm$0.18 &  146 &    8 &   75 &   63 \\
                     1404 &  $-$6.2 &  124 &  233 &        $-$20.5 &  1.94$\pm$0.13 &   62 &   12 &   14 &   36 \\
                     3115 &  $-$4.6 &  211 &  234 &        $-$20.2 &  2.10$\pm$0.14 &  131 &    9 &   32 &   90 \\
                     3377 &  $-$4.9 &   84 &  116 &        $-$18.8 &  3.60$\pm$0.33 &   14 &    0 &    7 &    7 \\
                     3379 &  $-$4.7 &  163 &   97 &        $-$19.6 &  1.48$\pm$0.15 &   86 &    8 &   11 &   67 \\
                     3384 &  $-$4.8 &   78 &   97 &        $-$19.4 &  1.69$\pm$0.17 &   22 &    1 &    1 &   20 \\
                     3585 &  $-$6.3 &   74 &  171 &        $-$21.0 &  0.95$\pm$0.07 &   60 &    2 &   13 &   45 \\
                     3923 &  \ldots &  225 &  519 &        $-$20.8 &  2.45$\pm$0.11 &   82 &    3 &   26 &   53 \\
                     4278 &  $-$5.7 &   61 &  346 &        $-$19.9 &  4.51$\pm$0.24 &  146 &    3 &   58 &   85 \\
                     4365 &  $-$6.5 &   75 &  634 &        $-$21.1 &  3.52$\pm$0.14 &  152 &    6 &   60 &   86 \\
                     4374 &  $-$6.0 &   98 &  411 &        $-$21.1 &  1.97$\pm$0.10 &   97 &    2 &   21 &   74 \\
                     4377 &  $-$5.9 &   46 &   54 &        $-$18.4 &  2.91$\pm$0.40 &    4 &    0 &    0 &    4 \\
                     4382 &  $-$6.0 &  176 &  514 &        $-$21.1 &  2.30$\pm$0.10 &   55 &    4 &   13 &   38 \\
                     4406 &  $-$5.8 &   62 &  324 &        $-$20.8 &  1.80$\pm$0.10 &   15 &    1 &    0 &   14 \\
                     4472 &  $-$5.8 &  112 &  617 &        $-$21.3 &  2.33$\pm$0.09 &  200 &    8 &   58 &  134 \\
                     4473 &  $-$5.5 &   72 &  176 &        $-$19.6 &  2.78$\pm$0.21 &   24 &    2 &    5 &   17 \\
                     4552 &  $-$5.6 &   64 &  311 &        $-$20.1 &  3.15$\pm$0.18 &  113 &    4 &   36 &   73 \\
                     4621 &  $-$5.5 &   60 &  242 &        $-$20.0 &  2.71$\pm$0.17 &   37 &    2 &    8 &   27 \\
                     4649 &  $-$5.8 &  519 & 1054 &        $-$21.3 &  3.78$\pm$0.12 &  286 &   22 &   95 &  169 \\
                     4697 &  $-$5.0 &   99 &  296 &        $-$20.1 &  2.96$\pm$0.17 &   83 &    3 &   32 &   48 \\
                     7457 &  $-$5.2 &   60 &  101 &        $-$18.6 &  4.05$\pm$0.40 &    8 &    0 &    0 &    8 \\
\hline
              {\bf Total} &  & 3114 & 8112 &     \ldots &     \ldots & 1937 &  104 &  595 & 1238 \\
\hline\hline
\end{tabular}
\end{center}
Note.---Breakdown of the \hst-based classifications for discrete optical sources detected within the footprints of the galaxies.  In Col.(2), we quote the effective 50\% completeness limit for the F475W band ($g_{475}$), appropriate for sources with GC-like light profiles.  In Col.(3) and (4), we include the total numbers of GC and background sources. Col.(5) and (6) provides the ``local' absolute $V$-band magnitudes of the host galaxy and GC specific frequencies, respectively, appropriate for the galactic footprints.  In Col.(7)--(10) we list the total number of \xray\ detected sources ($N_{\rm X}$), and the numbers of these sources classified as background sources, GCs, and field populations.
\end{table*}

We refined the absolute astrometry of our \hst\ data products and
catalogs by aligning them to either the Thirteenth Sloan Digital Sky Survey
(SDSS) Data Release (DR13; Albareti \etal\ 2017) frame or the United States
Naval Observatory catalog USNO-B (Monet \etal\ 2003) frame when SDSS DR13 data
were unavailable or inadequate (12 galaxies).  In this alignment procedure,
\hst\ products were matched to the reference catalogs using small shifts in
R.A. and Dec (ranging from 0\farcs15--0\farcs9).  Translational shifts were
then applied to the \hst\ images and catalogs to bring them into
alignment with the reference catalogs.  Based on the distributions of offsets
of source matches, we estimate the 1$\sigma$ uncertainties on the image
registrations to be in the range of 0\farcs04--0\farcs3 (median of
$\approx$0\farcs1).  

Using the shifted catalogs, we classified all individually-detected sources
that were present within the $K_s$-band ellipses defined in Table~1.  Sources
were classified as likely GCs if they had (1) colors in the range of \hbox{$0.6
\le g_{475}-z_{850} \le 1.6$}, \hbox{$0.5 \le g_{475}-z_{814} \le 1.3$}, or
\hbox{$0.3 \le r_{606}-z_{814} \le 0.8$}, depending on filter availability; (2)
absolute magnitudes (based on the distances to each galaxy) in the range of
$-12.5 \le M_z \le -6.5$; (3) extended light profiles in either of the blue or red
bandpasses, characterized as having {\ttfamily SExtractor} stellarity
parameters {\ttfamily CLASS\_STAR}~$\le$~0.9 or aperture magnitude differences
$m(r_1)-m(r_2) > 0.4$, where apertures consist of circles with radii $r_1$ and
$r_2$ (defined above); and (4) light profiles in both the blue and red bands
that were not too extended to be GCs, defined as $m(r_1)-m(r_2) \le 0.9$.  All
other \hst-detected sources were classified as unrelated background source
candidates (mainly background galaxies and some Galactic stars).  Visual
inspection of the GC and background sources classified using the above criteria
indicates that the misclassification rate is $\approx$1--2\%, and is unlikely
to have any important impact on our results.  In obvious cases where sources
were misclassified and are coincident with \xray\ detected sources, we manually
changed their classifications (see below).  However, all other sources were
classified using the above criteria.

In Table~4, we summarize the number of GCs and background sources classified
within the optical footprints of each galaxy (as defined in Table~1).  In
Appendix~A, we present simulations quantifying our completeness to detecting
GC-like sources and provide estimates of the ``local'' GC specific
frequencies, $S_{N, {\rm loc}}$, for the galaxies.  In Table~4, we summarize
our completeness findings and GC statistics (including $S_{N, {\rm loc}}$
values) for the galactic regions.
Our completeness limits span a range of $-6.3 \le M_g^{50} \le -4.7$, which is
always fainter than the peak of the GC luminosity functions at $M_g \approx
-7.1$~mag (e.g., Harris~2001; Kundu \& Whitmore~2001), allowing us to constrain
well the GC luminosity function and $S_{N, {\rm loc}}$.  

We find that the values of $S_{N, {\rm loc}}$ span 0.95--5.09 and are well
correlated with the global $S_N$ values.  As expected, the values of the local
$V$-band luminosities of our galaxy footprints are lower than the global values
reported by H13, with $M_{V,{\rm loc}} - M_{V, {\rm H13}} \approx$0.2--1.6~mag
(median of 1.1~mag).  Also, our estimated local numbers of GCs are smaller than
the global values provided by H13, with the exception of NGC~4649, in which we
estimate a $\approx$38\% larger number of GCs within our field of view.
Upon detailed inspection, we found that the value quoted in H13 for NGC~4649 was
taken directly from Jord{\'a}n \etal\ (2005), and is appropriate for the number
of GCs within the half-light radius and not the total number of GCs quoted by H13 for other galaxies.  As such,
the global $S_N$ value inferred for this source would be underestimated.  We
therefore estimated the global $S_N$ for NGC~4649 here by applying a correction
factor based on the average ratio of $S_N/S_{N, {\rm loc}} = 1.02 \pm 0.77$,
determined from the remaining 23 galaxies.  The resulting $S_N$ value and its
propagated uncertainty are quoted in Table~1 for NGC~4649, and are used for the
remainder of this study. 

In general, the relative galactic-light and GC-location profiles for our galaxies show
variations in the comparative values of $S_{N,{\rm loc}}$ and global $S_N$.
For the galactic regions used in this study, we find that $S_{N, {\rm loc}}$
tends to have somewhat larger (smaller) values compared to $S_N$, for $S_N
\simlt 3$ ($S_N \simgt 3$).  This trend appears to be driven primarily by the
differences between local and global numbers of GCs varying with $S_N$.  At
low-$S_N$, the global and local numbers of GCs are comparable, but as $S_N$
increases, the numbers of GCs are relatively small locally compared to the
global values.  Meanwhile, there are no strong trends in differences of
$M_{V,{\rm loc}}$ and $M_{V, {\rm H13}}$ with $S_N$.

%
%
\begin{figure*}[t!]
\figurenum{3}
\centerline{
\includegraphics[width=20cm]{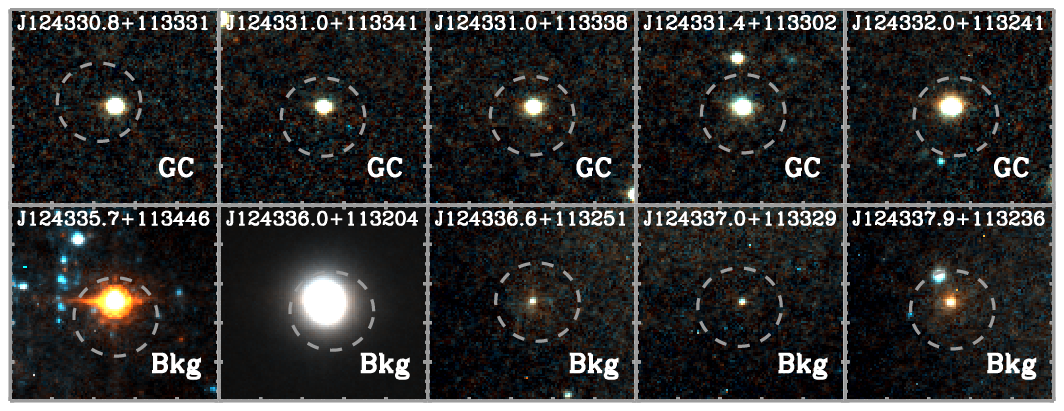}
}
\caption{
Sample ACS image cutouts for \xray\ detected GCs ({\it top panels\/}) and
background sources ({\it bottom panels\/}) in NGC~4469.  Our optical source
counterpart classification criteria are described in Section~3.2.  Each postage
stamp is centered on the optical source position and spans a $2\farcs5 \times
2\farcs5$ region.  For scaling purposes, we have included a circle with
1\arcsec\ radius centered on the location of the \xray\ source.
}
\end{figure*}

\subsection{ \chandra\ Data Reduction and \xray\ Catalog Production}

We made use of \chandra\ ACIS-S and ACIS-I data sets that had aim points within
5~arcmin of the central coordinates of the galaxy (given in Table~1).  The
observation logs for all galaxies are presented in Table~5.  In total 113
unique ObsIDs were used for the \ngal\ galaxies in our sample, representing
5.5~Ms of \chandra\ observation time.  The cumulative exposures ranged from
20--1127~ks, with the deepest observation reaching a minimum 50\% completeness
limit of $\approx$$10^{36}$~\lum\ ($f_{\rm 0.5-8~keV} \approx 10^{-16}$~\flux)
for NGC~3115 (Lin \etal\ 2015).

Our \chandra\ data reduction and cataloging procedures follow directly the
methods used in Sections~3.2 and 3.3 in Lehmer \etal\ (2019; hereafter, L19).
Briefly, for a given galaxy, we performed data reductions of all
ObsIDs (updated calibrations, flagged bad pixels, and removed flared
intervals), astrometrically aligned ObsIDs to the longest-exposure observation
(see Table~5), merged events lists, created images, and searched for point
sources using the 0.5--7~keV images for source detection purposes.  

Point source and background properties were extracted and computed using the
{\ttfamily ACIS Extract} ({\ttfamily AE}) v.~2016sep22 software package (Broos
\etal\ 2010, 2012), which calculates point-spread functions (PSFs)
from each ObsID, properly disentangles source event contributions from sources
with overlapping PSFs, and performs \xray\ spectral modeling of each source
individually using {\ttfamily xspec} v.~12.9.1 (Arnaud~1996).  As such, all
\xray\ point-source fluxes are based on basic spectral fits to data using an
absorbed power-law model with both a fixed component of Galactic absorption and
a free variable intrinsic absorption component ({\ttfamily TBABS $\times$ TBABS
$\times$ POW} in {\ttfamily xspec}).\footnote{The free parameters include the
intrinsic column density, $N_{\rm H, int}$, and photon index, $\Gamma$.  The
Galactic absorption column, $N_{\rm H, gal}$, for each source was fixed to the
value appropriate for the location of each galaxy, as derived by Dickey \&
Lockman~(1990).}  Throughout the remainder of this paper, we quote point-source
\xray\ luminosities, $L$, based on the Galactic column-density corrected
\hbox{0.5--8~keV} flux.  Following past studies, we do not attempt to correct
for intrinsic absorption of the sources themselves.


\begin{table*}
{\small
\begin{center}
\caption{\chandra\ Advanced CCD Imaging Spectrometer (ACIS) Observation Log}
\begin{tabular}{lcccccccc}
\hline\hline
& \multicolumn{2}{c}{\sc Aim Point} & {\sc Obs. Start} & {\sc Exposure}$^a$ & {\sc Flaring}$^b$ &  $\Delta \alpha$ & $\Delta \delta$  & {\sc Obs.} \\
\multicolumn{1}{c}{\sc Obs. ID} & $\alpha_{\rm J2000}$ & $\delta_{\rm J2000}$ & (UT) & (ks) & {\sc Intervals} & (arcsec) & (arcsec)  & {\sc Mode}$^c$ \\
\hline\hline
\multicolumn{9}{c}{{\bf NGC1023}} \\
\hline
4696  & 02 40 24.87 & +39 03 14.72 & 2004-02-27T18:26:26 & 10 & \ldots & $-$0.04 & $-$0.01 & V \\
8197  & 02 40 22.56 & +39 02 03.64 & 2007-12-12T11:56:14 & 48 & \ldots & $+$0.10 & $+$0.15 & V \\
8198$^d$ & 02 40 22.53 & +39 02 34.56 & 2006-12-17T19:15:53 & 50 & \ldots & \ldots & \ldots & V \\
8464  & 02 40 23.53 & +39 05 02.70 & 2007-06-25T17:54:05 & 48 & \ldots & $-$0.31 & $+$0.19 & V \\
8465  & 02 40 14.17 & +39 04 59.23 & 2007-10-15T09:09:05 & 45 & \ldots & $-$0.09 & $+$0.19 & V \\
Merged$^e$  &02 40 21.01 & +39 03 37.09 & & 201 & \ldots &  \ldots & \ldots & \ldots \\
\hline
\multicolumn{9}{c}{{\bf NGC1380}} \\
\hline
9526$^d$ & 03 36 25.01 & $-$34 59 43.63 & 2008-03-26T12:08:51 & 41 & 1, 0.5 & \ldots & \ldots & V \\
\hline
\multicolumn{9}{c}{{\bf NGC1387}} \\
\hline
4168$^d$ & 03 36 58.70 & $-$35 29 30.78 & 2003-05-20T22:56:28 & 46 & \ldots & \ldots & \ldots & V \\
\hline
\end{tabular}
\end{center}
Note.---The full version of this table contains entries for all \ngal\ galaxies and 113 ObsIDs, and is available in machine-readable form.  An abbreviated version of the table is displayed here to illustrate its form and content.  \\
$^a$ All observations were continuous. The times shown have been corrected for removed data that were affected by high background.\\
$^b$ Number of flaring intervals and their combined duration in ks.  These intervals were rejected from further analyses. \\
$^c$ The observing mode (F=Faint mode; V=Very Faint mode).\\
$^d$ Indicates Obs.~ID that all other observations are reprojected to for alignment purposes.  This Obs.~ID was chosen for reprojection as it had the longest initial exposure time, before flaring intervals were removed.\\
$^e$ Aim point represents exposure-time weighted value.
}
\end{table*}

To align the \chandra\ catalogs and data products, we matched the \chandra\
main catalogs of each galaxy to their corresponding astrometry-corrected \hst\
master optical catalogs using a matching radius of 1\farcs0.  In this exercise,
we limited our matching to \xray\ sources with more than 20 0.5--8~keV net
counts to ensure reasonable \chandra-derived positions.  Most galaxies had
respectably large numbers of matches ($\simgt$15 matches) and showed obvious
clusterings of points in $\delta$R.~A. and $\delta$dec diagrams, indicating
that a reliable astrometric registration could be obtained between \chandra\
and \hst.  For these galaxies, we applied additional simple median shifts in
R.A. and dec. (offsets ranged from 0\farcs08--0\farcs66 for
the galaxies) to the \chandra\ data products and catalogs to bring them into
alignment with the \hst\ and reference optical frames (by extension).  For
these galaxies, the final \hst\ and \chandra\ image and catalog registrations
have a 1$\sigma$ error of $\simlt$0\farcs25.  For the four galaxies (NGC~3384,
4125, 4377, and 4406) where the number of matches was too small ($<$3) to
reliably calculate cross-band offsets, we did not apply astrometric shifts to
the data.  For these galaxies, we estimate, based on the offsets of other
galaxies, that the cross-band registration error is $\simlt$0\farcs3.

After applying shifts to the \chandra\ catalogs and data products, we performed
a second round of matching with the \hst\ catalogs to identify reliable
counterparts to the \xray\ sources.  We ran simulations, in which we shifted
the \xray\ source locations by 5~arcsec in random directions and re-matched to
the \hst\ source catalogs, using a variety of source matching radii, to
determine the false-match rate.  From these simulations, we found that the
number of matches as a function of matching radius has a sharp peak around
0\farcs1 and declines rapidly with increasing radius.  We estimate that beyond
a matching radius of $\simgt$0\farcs5, the number of new matches (compared to
smaller matching radii) is equivalent to the expected number of false matches.
We therefore chose to utilize a matching radius of 0\farcs5 when identifying
reliable counterparts.  From the above analysis, the false-match rate is
calculated to be \hbox{4--6\%} for this adopted limit.  We note that this estimate is likely to be an overestimate, due to the inclusion of large numbers of sources that are truly associated with optical counterparts (see, e.g., Broos \etal\ 2011).
Figure~3 shows \hst\
cutout images for a random selection of \xray\ sources with GC and background
counterparts for the galaxy NGC~4649.  

In Appendix~B, we present the properties of all \nxsrc\ point sources detected in the 0.5--7~keV band
within the \chandra\ images, and include, when possible, \hst\ source
classifications.  Throughout the remainder of this paper, we focus our analyses
on the \nx\ sources within the galactic footprints defined above (i.e., within
the $K_s \approx 20$~mag~arcsec$^{-2}$ ellipses, in areas with \hst\ coverage,
and outside of the central removed regions).  In Table~4, we summarize for each
galaxy the number of these \chandra\ sources with \hst\ counterparts among the
three source categories defined above: i.e., background sources, GC, or field
LMXB candidate (when no counterpart is present).  In total, \nbkg, \ngc, and
\nfield\ sources are classified as background sources, GCs, and field LMXB
candidates, respectively.  

As we will describe below, our results rely on our GC LMXB designations being
highly complete, and not having a large number of field LMXBs that could
be associated with faint GCs below our optical detection thresholds.  In
Appendix~A, we address this in detail and show that our procedures are capable
of recovering the GC-LMXB designation for $\approx$96\% of the GC-LMXBs that
are among our \xray\ detected sources.  As such, our field LMXB population will
contain at most a negligible population of faint GCs that are simply
undetected.

\section{Results}

Our XLF fitting procedures followed the same techniques developed and presented
in Section~4.1 of L19; the salient details of this procedure are provided
below.  All XLF data are fit using a forward-fitting approach, in which detection
incompleteness, contributions from cosmic \xray\ background (CXB) sources
(hereafter, defined as unrelated Galactic stars and background AGN or normal
galaxies), and LMXB model components are folded into our models to fit {\it
observed} XLFs.  On occasion, we display completeness-corrected and
CXB-subtracted XLFs for illustrative purposes, but do not use such data in our
fitting.  Below, we describe the construction of the model components and
present our fitting results.

\subsection{Cosmic X-ray Background Modeling}

Many of the CXB sources can be directly classified using
our \hst\ data; however, our ability to accurately classify \xray\ detected
background objects depends on the \hst\ imaging depth.  In practice, there
will be a number of \xray\ detected CXB sources that have no \hst\
counterparts that we will classify here as LMXB candidates.  Even in
blank-field extragalactic \xray\ surveys with very deep extensive
multiwavelength follow-up (e.g., Nandra \etal\ 2015; Civano \etal\ 2016; Xue
\etal\ 2016; Luo \etal\ 2017; Kocevski \etal\ 2018), there are a number of
\xray\ sources with no multiwavelength counterparts.  The CXB
sources will be dominated by AGN that have optical fluxes that broadly
correlate with \xray\ flux, so the most likely sources to lack \hst\
counterpart identifications are those with the faintest \xray\ fluxes.  We assessed the level of
completeness by which we could reliably identify CXB source counterparts by
comparing the expected extragalactic number counts from blank-field surveys
with our background-object counts.  

In Figure~4, we show the number of background sources detected as a function of
0.5--8~keV flux, $S$, compared to the expected number from the extragalactic
number counts from Kim \etal\ (2007).  Note that the extragalactic number
counts curve has been corrected for \xray\ incompleteness of our data sets at
faint limits (see L19 for details).  

%
%
\begin{figure}
\figurenum{4}
\centerline{
\includegraphics[width=9cm]{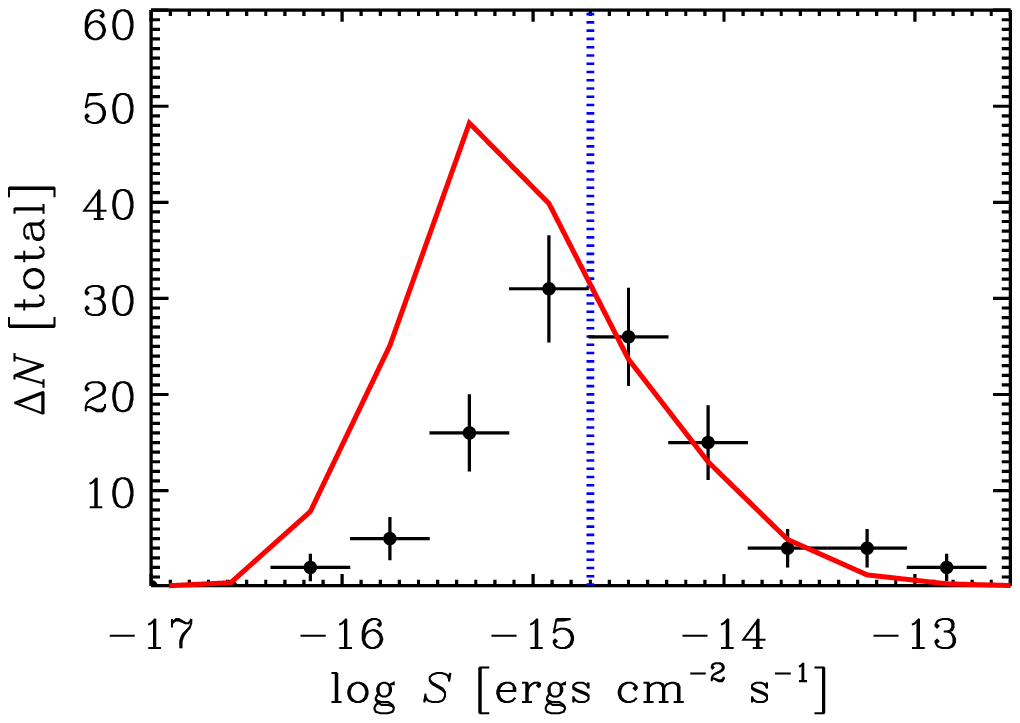}
}
\caption{
Observed number of sources $\Delta N$ per flux bin ($\Delta \log S = 0.42$~dex)
classified as CXB sources as a function of 0.5--8~keV flux $S$ for all galaxies
combined.  In total, \nbkg\ sources were classified as background sources.  The
predicted number of background objects for the full sample is shown as a solid
red curve (based on the Kim \etal\ 2007 extragalactic number counts).  This
prediction is based on the extragalactic number counts, and includes the
effects of incompleteness.  We find that our background counts are consistent
with the extragalactic counts above $S > 2 \times 10^{-15}$~\flux\ ({\it blue
dotted line\/}), but become highly incomplete ($\simlt$10--50\%) below this
flux level.  In our XLF fitting, we directly excluded CXB sources above this
flux level, and modeled the contributions from fainter sources using the Kim \etal\ (2007) results.
}
\end{figure}

%
%
\begin{figure*}[t!]
\figurenum{5}
\centerline{
\includegraphics[width=18cm]{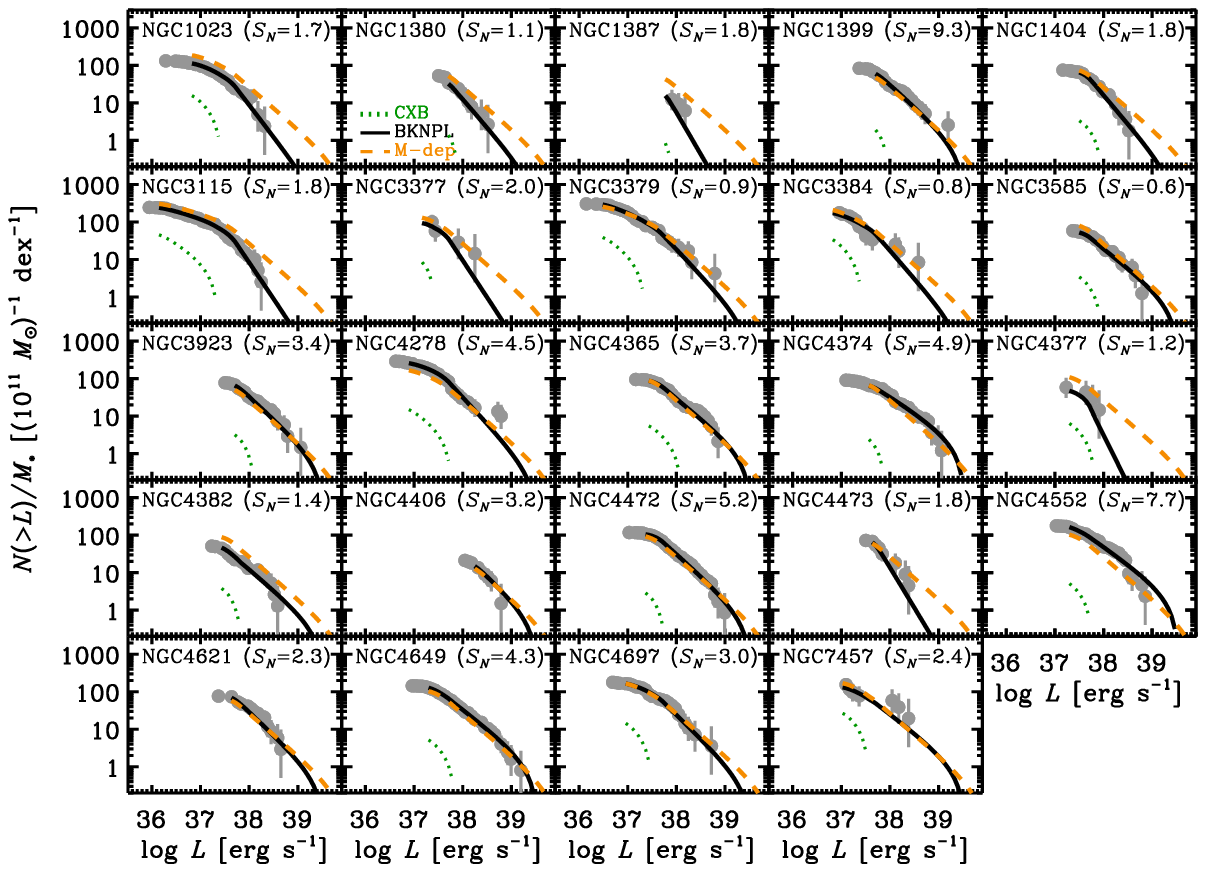}
}
\caption{
Observed {\it field} LMXB XLFs for all galaxies in our sample ({\it gray
circles with 1$\sigma$ error bars\/}).  Each panel provides the XLF of the
denoted galaxy and includes the GC specific frequency value, $S_N$, for
convenient reference.  These XLFs have been constructed by excluding all \xray\
sources designated as GCs, but include potential background sources.  They are
not corrected for incompleteness, explaining the perceptible turnovers at the
lowest luminosity values.  Model fits, which include contributions from the CXB
({\it green dotted curves\/}) and intrinsic point sources, are shown for the
broken power-law model ({\it black solid curves\/}) and the global stellar-mass
dependent model ({\it dashed orange curves\/}).  Displayed models (and CXB
contributions) include the effects of incompleteness, and are calculated down
to the 50\% completeness limit $L_{50}$.  As described in Section~4.1, directly
identified background sources with 0.5--8~keV fluxes $S \simgt 2 \times
10^{-15}$~\flux\ have been removed, so our CXB model only includes
contributions from sources with $S < 2 \times 10^{-15}$~\flux\ (or $L
=$~2.5--13~$\times 10^{37}$~\lum, depending on the galaxy).
}
\end{figure*}

We find that the observed CXB number counts for our sample match well the
expected number counts for $S \simgt 2 \times 10^{-15}$~\flux, suggesting that
our background source identification methods are reliable and highly complete
above this limit.  However, for sources below this limit, our observed source counts are
below the predictions by a significant margin ($\approx$10--50\% complete), indicating
there are some CXB sources in this regime that we are likely misclassifying as
field LMXB candidates.  Given these results, hereafter we chose to reject from
our field LMXB XLF analyses all CXB sources with $S \ge 2 \times
10^{-15}$~\flux\ (corresponding to $L \approx$~2.5--13~$\times 10^{37}$~\lum\
for our sample), but include all background sources fainter than this limit.
We account for these faint CXB sources when modeling the XLFs by implementing
the Kim \etal\ (2007) extragalactic number counts at $S < 2 \times
10^{-15}$~\flux.

%
%
\begin{figure}
\figurenum{6}
\centerline{
\includegraphics[width=9cm]{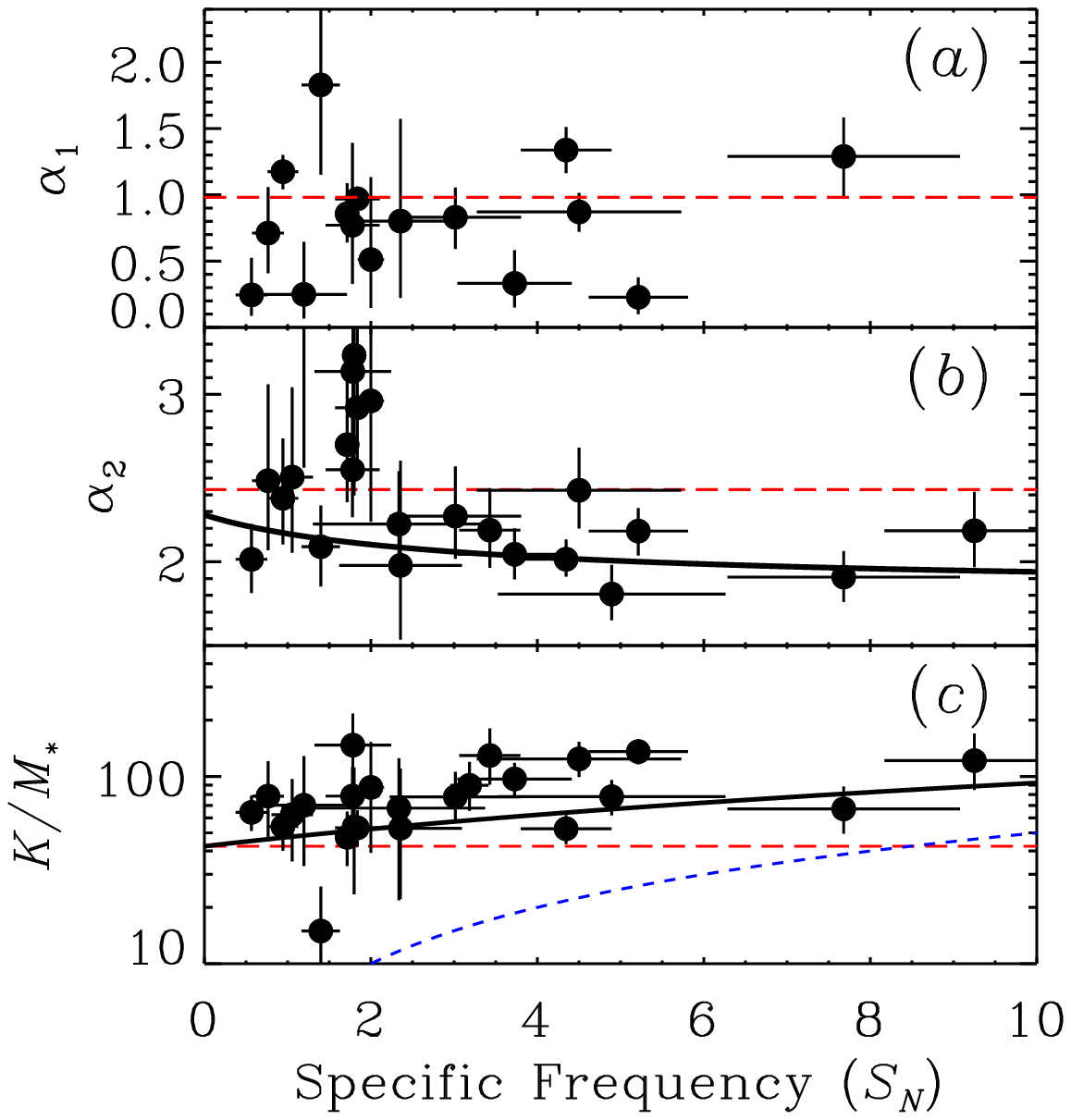}
}
\caption{
($a$)--($c$): Best-fitting broken power-law parameters for field LMXB
populations for each galaxy versus GC specific frequency. We include here the
($a$) low-luminosity XLF slope ($L < L_b$; $L_b = 3 \times 10^{37}$~\lum) for
galaxies with $L_{50} < L_b$, as well as ($b$) high-luminosity XLF slope ($L >
L_b$) for galaxies with $L_{50} < 10^{38}$~\lum, and ($c$) stellar-mass scaled
XLF normalization for all galaxies.
The predicted trends from our best-fit in-situ and GC-seeded model of field
LMXBs (discussed in Section 4.5) are shown for in-situ LMXBs ({\it red
long-dashed lines\/}), GC-seeded LMXBs ({\it blue short-dashed curves\/}), and
combined populations ({\it black solid cuves\/}).
}
\end{figure}

\subsection{Field LMXB X-ray Luminosity Functions By Galaxy}

We started by characterizing the field LMXB XLFs of each galaxy using basic
analytic models (i.e., power-law and broken power-law).  Since our focus here
is on the field LMXBs, we rejected all \xray\ sources coincident with GCs and
the subset of known CXB sources with $S > 2 \times 10^{-15}$~\flux (see
previous section).

In Figure~5, we show the observed stellar-mass-normalized field LMXB XLFs (in
cumulative form) for each of the \ngal\ galaxies in our sample.  Note that the
data displayed in Figure~5 have not been corrected for incompleteness and
therefore do not convey the intrinsic shapes of the XLFs.  Following L19, we
attempted to model the XLFs of each galaxy using both power-law and broken
power-law models:
\begin{equation}
\frac{dN}{dL} = K_{\rm PL} \left \{ \begin{array}{lr} 
L^{-\alpha}, & \;\;\;\;\;\;\;\;(L < L_c) \\
0, & (L \ge L_c) 
\end{array}
  \right.
\end{equation}

\begin{equation}
\frac{dN}{dL} = K_{\rm BKNPL}  \left \{ \begin{array}{lr} L^{-\alpha_1}  &
\;\;\;\;\;\;\;\;(L < L_b) \\ 
L_b^{\alpha_2-\alpha_1} L^{-\alpha_2},  & (L_b \le L < L_c) \\ 
0,  & (L \ge L_c) \\ 
\end{array}
  \right.
\end{equation}
where $K_{\rm PL}$ and $\alpha$ are the single power-law normalization and
slope, respectively, and $K_{\rm BKNPL}$, $\alpha_1$, $L_b$, and $\alpha_2$ are
the broken power-law normalization, low-luminosity slope, break luminosity, and
high-luminosity slope, respectively; both XLF models are truncated above,
$L_c$, the cut-off luminosity.  To make the normalization values more
intuitive, we take $L$, $L_b$, and $L_c$ to be in units of $10^{38}$~\lum\ when
using eqns.~(1) and (2).  For a given galaxy, we fit the data to determine all
constants, except for the break and cut-off luminosities, which we fixed at
$L_b = 3 \times 10^{37}$~\lum\ and $L_c = 3 \times 10^{39}$~\lum.  Also, when
the luminosity of the 50\% completeness limit,
$L_{50}$, was in the range of 1--2~$\times L_b$, the fit to
$\alpha_1$ was deemed unreliable, and was fixed to $\alpha_1 = 1.0$.  For one
galaxy, NGC~4406, $L_{50} > 10^{38}$~\lum, and we chose to fix $\alpha_1
= 1.0$ and $\alpha_2 = 2.1$ and fit only for the normalization.  The above
specific choices for fixed parameter values were motivated by global fits to
the full sample (see Section 4.3).

Following the procedures in L19, we modeled the observed XLF, $dN/dL({\rm
obs})$, using the intrinsic power-law and broken power-law model of the XRB
XLF, $dN/dL({\rm int})$, plus an estimated contribution from undetected
background sources, $dN/dL({\rm CXB})$, that were convolved with a
luminosity-dependent completeness function, $\xi(L)$ (see L19 for details on
the calculation of $\xi$):
\begin{equation}
dN/dL({\rm obs}) = \xi(L) [dN/dL({\rm int}) + dN/dL({\rm CXB})].
\end{equation}
For each galaxy, we constructed the observed $dN/dL({\rm obs})$ using
luminosity bins of constant $\delta \log L = 0.057$~dex that spanned the range
of \hbox{$L_{\rm min} = L_{50}$} (the 50\% completeness limit) to $L_{\rm max}
= 5 \times 10^{41}$~\lum.  We note that the size of these bins is chosen to be
comparable to distance-related uncertainties on the luminosities.  For most
galaxies, the majority of the bins contained zero sources, with other bins
containing small numbers of sources.  Therefore, when assessing maximum
likelihood we made use of a modified version of the $C$ statistic ({\ttfamily
cstat}; Cash~1979; Kaastra~2017):
\begin{equation}
C = 2 \sum_{i=1}^{n} M_i - N_i + N_i \ln(N_i/M_i),
\end{equation}
where the summation takes place over the $n$ bins of \xray\ luminosity, and
$N_i$ and $M_i$ are the observed and model counts in each bin.  We note that
when $N_i = 0$, $N_i \ln (N_i/M_i) = 0$, and when $M_i=0$ (e.g., beyond a
cut-off luminosity), the entire $i$th term in the summation is zero.


\begin{deluxetable*}{lccccccccccccc}
\tablewidth{0pt}
\tabletypesize{\scriptsize}
\tablecaption{Field X-ray Luminosity Function Fits By Galaxy}
\tablehead{
 \multicolumn{1}{c}{\sc Galaxy}  & \colhead{}  & \colhead{} & \colhead{}  & \multicolumn{4}{c}{\sc Single Power Law$^\dagger$} & \multicolumn{5}{c}{\sc Broken Power Law$^\ddagger$} & \colhead{}  \\
\vspace{-0.25in} \\
\multicolumn{1}{c}{\sc Name} &   \colhead{} & \colhead{$\log L_{50}$}  &  \colhead{$\log L_{90}$} & \multicolumn{4}{c}{\rule{1.6in}{0.01in}} & \multicolumn{5}{c}{\rule{2.3in}{0.01in}} &  \colhead{$\log L_{\rm X}$} \\
\vspace{-0.25in} \\
\multicolumn{1}{c}{\sc (NGC)} &  \colhead{$N_{\rm src}$} & \colhead{(\lum)} & \colhead{(\lum)}  & \colhead{$K_{\rm PL}$} & \colhead{$\alpha$} & \colhead{$C$} & \colhead{$P_{\rm Null}$} & \colhead{$K_{\rm BKNPL}$} & \colhead{$\alpha_1$} & \colhead{$\alpha_2$} & \colhead{$C$} & \colhead{$P_{\rm Null}$} & \colhead{(ergs~s$^{-1}$)} \\
\vspace{-0.25in} \\
\multicolumn{1}{c}{(1)} & \multicolumn{1}{c}{(2)} & \multicolumn{1}{c}{(3)} & \colhead{(4)} & \colhead{(5)} & \colhead{(6)} & \colhead{(7)} & \colhead{(8)} & \colhead{(9)} & \colhead{(10)} & \colhead{(11)} & \colhead{(12)} & \colhead{(13)}  & \colhead{(14)}  
}
\startdata
      1023 &  56 & 36.8 & 37.0 &    7.94$^{+2.19}_{-1.83}$ &             1.35$\pm$0.15 &   22 &      0.165 &      19.7$^{+7.3}_{-5.9}$ &    0.86$^{+0.23}_{-0.22}$ &    2.70$^{+0.40}_{-0.34}$ &   23 &      0.174 &              39.5$\pm$0.1 \\ 
      1380 &  20 & 37.7 & 37.8 &    7.47$^{+2.60}_{-2.15}$ &    1.88$^{+0.45}_{-0.49}$ &    6 &      0.005 &    23.6$^{+12.9}_{-10.3}$ &    0.63$^{+1.07}_{-0.43}$ &    2.51$^{+0.54}_{-0.45}$ &    9 &      0.034 &              39.7$\pm$0.2 \\ 
      1387 &   4 & 37.8 & 37.9 &    2.94$^{+1.81}_{-1.28}$ &    2.43$^{+0.66}_{-0.58}$ &   12 &      0.520 &     17.4$^{+18.6}_{-9.8}$ &               0.90$^\ast$ &    3.23$^{+0.96}_{-0.84}$ &   13 &      0.952 &              39.4$\pm$0.3 \\ 
      1399 &  66 & 37.7 & 38.0 &      29.2$^{+4.6}_{-4.2}$ &             2.18$\pm$0.16 &   22 &      0.034 &    94.4$^{+37.2}_{-28.6}$ &    0.40$^{+0.44}_{-0.25}$ &    2.19$^{+0.23}_{-0.22}$ &   22 &      0.171 &              40.3$\pm$0.1 \\ 
      1404 &  41 & 37.5 & 37.9 &      11.8$^{+2.5}_{-2.2}$ &             2.30$\pm$0.19 &   22 &      0.092 &    43.1$^{+18.7}_{-16.6}$ &    0.77$^{+0.62}_{-0.44}$ &    2.55$^{+0.32}_{-0.28}$ &   23 &      0.433 &              39.9$\pm$0.1 \\ 
\\
      3115 &  96 & 36.1 & 36.4 &    5.81$^{+1.12}_{-0.99}$ &             1.43$\pm$0.06 &   53 &      0.551 &      20.9$^{+5.1}_{-4.3}$ &             0.97$\pm$0.09 &    2.92$^{+0.50}_{-0.40}$ &   31 &      0.156 &              39.5$\pm$0.1 \\ 
      3377 &   7 & 37.2 & 37.3 &    1.09$^{+0.63}_{-0.45}$ &    1.90$^{+0.38}_{-0.35}$ &   21 &      0.973 &    6.08$^{+4.54}_{-3.36}$ &    0.51$^{+0.62}_{-0.36}$ &    2.96$^{+0.93}_{-0.72}$ &   22 &      0.250 &              39.0$\pm$0.2 \\ 
      3379 &  72 & 36.4 & 36.7 &    5.03$^{+1.08}_{-0.99}$ &             1.56$\pm$0.08 &   42 &      0.306 &      12.8$^{+4.1}_{-3.3}$ &             1.17$\pm$0.13 &    2.38$^{+0.36}_{-0.28}$ &   36 &      0.472 &              39.4$\pm$0.1 \\ 
      3384 &  21 & 36.8 & 36.9 &    2.13$^{+0.74}_{-0.61}$ &             1.59$\pm$0.17 &   33 &      0.729 &    9.34$^{+4.99}_{-3.98}$ &    0.71$^{+0.35}_{-0.30}$ &    2.48$^{+0.58}_{-0.42}$ &   33 &      0.759 &              39.2$\pm$0.2 \\ 
      3585 &  47 & 37.5 & 37.6 &      14.0$^{+2.5}_{-2.3}$ &    1.84$^{+0.16}_{-0.15}$ &   31 &      0.465 &    51.6$^{+11.7}_{-10.5}$ &    0.25$^{+0.28}_{-0.16}$ &    2.01$^{+0.21}_{-0.20}$ &   28 &      0.464 &              40.1$\pm$0.1 \\ 
\\
      3923 &  54 & 37.7 & 38.0 &      26.8$^{+4.5}_{-4.1}$ &             2.19$\pm$0.18 &   21 &      0.109 &    88.5$^{+34.8}_{-26.7}$ &    0.40$^{+0.37}_{-0.26}$ &    2.19$^{+0.25}_{-0.22}$ &   21 &      0.160 &              40.2$\pm$0.1 \\ 
      4278 &  87 & 36.8 & 37.3 &      10.0$^{+1.7}_{-1.6}$ &             1.65$\pm$0.08 &   54 &      0.494 &      37.2$^{+8.6}_{-7.4}$ &    0.87$^{+0.14}_{-0.15}$ &    2.43$^{+0.25}_{-0.23}$ &   39 &      0.935 &              39.8$\pm$0.1 \\ 
      4365 &  89 & 37.4 & 37.7 &      24.1$^{+3.1}_{-2.9}$ &    1.86$^{+0.11}_{-0.10}$ &   35 &      0.508 &    91.1$^{+20.0}_{-18.2}$ &    0.33$^{+0.25}_{-0.18}$ &             2.04$\pm$0.15 &   32 &      0.522 &              40.3$\pm$0.1 \\ 
      4374 &  75 & 37.5 & 37.9 &      21.3$^{+3.2}_{-3.1}$ &             1.74$\pm$0.13 &   33 &      0.353 &    64.6$^{+14.8}_{-13.2}$ &    0.24$^{+0.28}_{-0.15}$ &    1.81$^{+0.17}_{-0.16}$ &   30 &      0.372 &              40.3$\pm$0.1 \\ 
      4377 &   4 & 37.2 & 37.5 &    0.62$^{+0.52}_{-0.35}$ &    1.83$^{+0.57}_{-0.48}$ &   10 &      0.398 &    4.82$^{+4.01}_{-2.54}$ &    0.25$^{+0.40}_{-0.18}$ &    3.61$^{+0.88}_{-1.05}$ &    8 &      0.480 &      38.8$^{+0.2}_{-0.3}$ \\ 
\\
      4382 &  39 & 37.4 & 37.6 &    9.10$^{+2.04}_{-1.81}$ &             2.03$\pm$0.18 &   24 &      0.160 &      11.4$^{+8.3}_{-5.7}$ &    1.83$^{+0.70}_{-0.68}$ &    2.09$^{+0.25}_{-0.24}$ &   24 &      0.222 &      39.8$^{+0.4}_{-0.1}$ \\ 
      4406 &  14 & 38.2 & 38.4 &         82$^{+146}_{-49}$ &    3.30$^{+1.02}_{-0.77}$ &   17 &      0.991 &    59.1$^{+19.7}_{-15.9}$ &               0.90$^\ast$ &               2.20$^\ast$ &   16 &      0.386 &              40.1$\pm$0.1 \\ 
      4472 & 138 & 37.4 & 37.6 &      33.7$^{+3.3}_{-3.2}$ &             1.86$\pm$0.08 &   26 &      0.007 &   159.0$^{+25.8}_{-23.1}$ &    0.23$^{+0.15}_{-0.13}$ &    2.18$^{+0.14}_{-0.15}$ &   13 &      0.002 &      40.5$^{+0.0}_{-0.1}$ \\ 
      4473 &  17 & 37.6 & 37.8 &    6.33$^{+2.08}_{-1.72}$ &    2.70$^{+0.41}_{-0.36}$ &   15 &      0.351 &    32.5$^{+15.3}_{-14.1}$ &    0.59$^{+0.85}_{-0.38}$ &    3.14$^{+0.60}_{-0.56}$ &   15 &      0.711 &      39.7$^{+0.2}_{-0.1}$ \\ 
      4552 &  75 & 37.2 & 37.7 &      16.5$^{+2.4}_{-2.2}$ &             1.79$\pm$0.10 &   34 &      0.293 &      28.3$^{+9.0}_{-7.4}$ &    1.29$^{+0.29}_{-0.30}$ &             1.91$\pm$0.15 &   34 &      0.417 &              40.0$\pm$0.1 \\ 
\\
      4621 &  27 & 37.6 & 37.8 &      11.8$^{+2.7}_{-2.5}$ &    2.18$^{+0.26}_{-0.24}$ &   27 &      0.907 &    23.0$^{+19.5}_{-15.6}$ &    1.19$^{+1.67}_{-0.77}$ &    2.23$^{+0.32}_{-0.27}$ &   28 &      0.812 &      39.9$^{+0.8}_{-0.2}$ \\ 
      4649 & 180 & 37.2 & 37.6 &      36.4$^{+3.4}_{-3.2}$ &             1.86$\pm$0.07 &   31 &      0.109 &    65.1$^{+11.8}_{-11.1}$ &             1.34$\pm$0.17 &    2.02$^{+0.12}_{-0.10}$ &   29 &      0.111 &              40.3$\pm$0.0 \\ 
      4697 &  50 & 36.9 & 37.1 &    6.22$^{+1.39}_{-1.11}$ &    1.63$^{+0.10}_{-0.11}$ &   34 &      0.249 &      21.7$^{+7.8}_{-5.9}$ &    0.83$^{+0.22}_{-0.24}$ &    2.27$^{+0.30}_{-0.25}$ &   29 &      0.374 &              39.6$\pm$0.1 \\ 
      7457 &   8 & 37.0 & 37.2 &    1.06$^{+0.58}_{-0.43}$ &             1.59$\pm$0.32 &   19 &      0.396 &    2.72$^{+2.93}_{-1.57}$ &    0.80$^{+0.77}_{-0.58}$ &    1.98$^{+0.63}_{-0.44}$ &   21 &      0.938 &              38.9$\pm$0.3 \\ 
\enddata
\tablecomments{All fits include the effects of incompleteness and model contributions from the CXB, following Eqn.~(7).  A full description of our model fitting procedure is outlined in Section~4.2.  Col.(1): Galaxy NGC name, as reported in Table~1.  Col.(2): Total number of \xray\ sources detected within the galactic boundaries defined in Table~1.  Col.(3) and (4): Logarithm of the luminosities corresponding to the respective 50\% and 90\% completeness limits. Col.(5) and (6): Median and 1$\sigma$ uncertainty values of the single power-law normalization and slope, respectively -- our adopted ``best model'' consists of the median values.  Col.(7): C-statistic, $C$, associated with the best model. Col.(8): Null-hypothesis probability that the best model describes the data.  The null-hypothesis probability is calculated following the prescription in Kaastra~(2017). Col.(9)--(11):  Median and 1$\sigma$ uncertainty values of the broken power-law normalization and slope, respectively.  Col.(12) and (13): Respectively, C-statistic and null-hypothesis probability for the best broken power-law model. Col.(14): Integrated \xray\ luminosity, $L_{\rm X}$, for the broken power-law model. \\
$^\ast$Parameter was fixed due to shallow \chandra\ depth (see Section 4.2).\\
$^\dagger$Single power-law models are derived following Eqn.~(1) with a fixed cut-off luminosity of $L_c = 3 \times 10^{39}$~\lum.\\
$^\ddagger$Broken power-law models are derived following Eqn.~(2) with a fixed break luminosity of $L_b = 3 \times 10^{37}$~\lum\ and cut-off luminosity of $L_c = 3 \times 10^{39}$~\lum.\\
}
\end{deluxetable*}

%
%
\begin{figure*}
\figurenum{7}
\centerline{
\includegraphics[width=18cm]{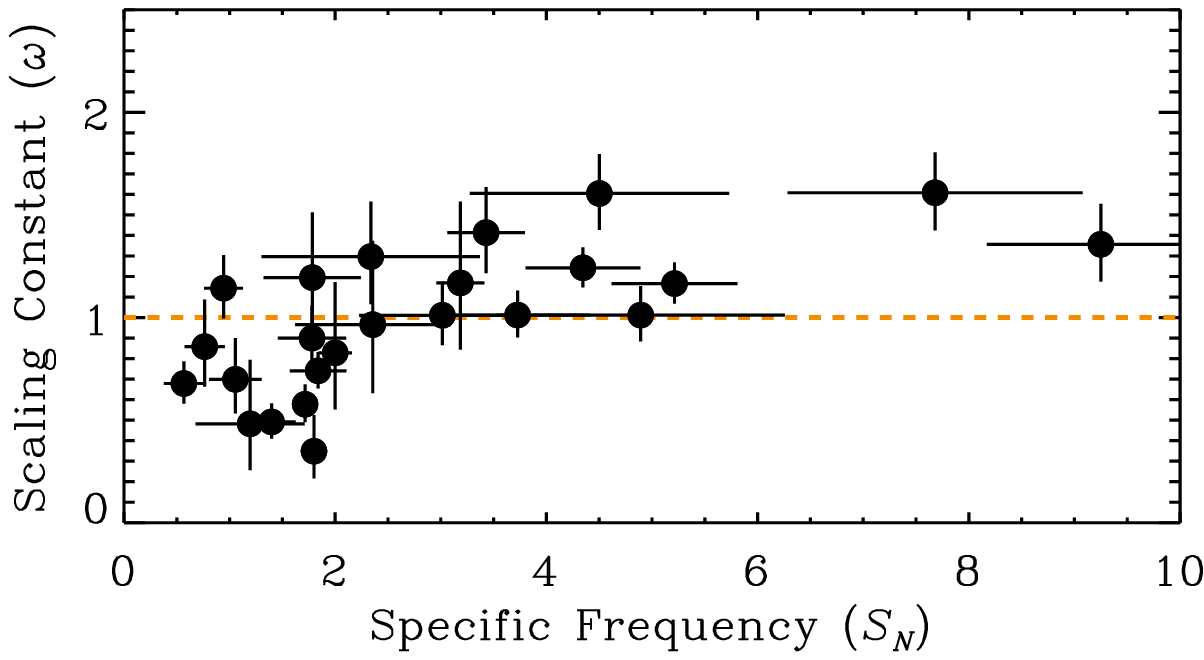}
}
\caption{
Scaling constant $\omega$, which multiplicatively scales
the stellar-mass dependent global XLF model to the XLF of each galaxy, versus
$S_N$ for all \ngal\ galaxies.  We find that $\omega$ is correlated with $S_N$ at the $>99.9\%$
confidence level, indicating field LMXBs are likely seeded by GCs (see
Section 4.2 discussion).
}
\end{figure*}

\subsection{Field LMXB XLF Dependence on Stellar Mass}

We calculated parameters, uncertainties, and uncertainty co-dependencies
following the Markov Chain Monte Carlo (MCMC) procedure outlined in Section~4.1
of L19.  In all fits, we adopted median values of the parameter distributions
as our quoted best-fit model parameters and the corresponding $C$ statistic.
We find that these values differ only slightly from those derived from the
global minimum value of the $C$ statistic.  We evaluated the goodness of fit
for our model based on the expected $C$-statistic value $C_{\rm exp}$ and its
variance $C_{\rm var}$, which were calculated following the procedures in
Kaastra~(2017).  The null hypothesis probability for the model was calculated
as:
\begin{equation}
P_{\rm null} = 1 - {\rm erf}\left( \sqrt{\frac{(C - C_{\rm exp})^2}{2 \; C_{\rm
var}}} \right).
\end{equation}

In Table~6, we tabulate the best-fit results for our power-law and broken
power-law models, including their goodness of fit evaluations.  In all cases,
the power-law and broken power-law fits result in $P_{\rm null} > 0.001$, with
only NGC~1380 and NGC~4472 showing some tension with the models (e.g., $P_{\rm
null} \simlt 0.02$).  Figure~5 displays the best-fit broken power-law model for
the observed XLFs of each galaxy ({\it black solid curves\/}), along with the
model contributions from CXB sources with $S < 2 \times 10^{-15}$~\flux\ ({\it
green dotted curves\/}), which in all relevant cases are much lower than the
LMXB contributions.

Figure~6 shows the parameters of the broken power-law versus GC $S_N$.  The
parameters include faint-end slope $\alpha_1$, bright-end slope $\alpha_2$, and
normalization per unit stellar mass $K_{\rm BKNPL}/M_\star$.  Displayed values
only include those that were determined via fitting, and exclude parameter
values that were fixed as a result of the \chandra\ data being too shallow.  We
find suggestive correlations between $\alpha_2$ and $K_{\rm BKNPL}/M_\star$
with $S_N$; based on Spearman's rank correlation tests, the correlation is
suggestive at the $\approx$97\% and $\approx$95\% confidence level.
This trend suggests that GCs may in fact provide some seeding to the field LMXB
population.  We explore this more in the sections below.

As discussed in Section~1, several past studies of LMXBs have focused on the
scaling of the LMXB XLF with stellar mass; however, very few studies have
attempted to isolate this relation for field LMXBs explicitly (however, see Peacock \etal\
2017).  Here, we determine the shape and scaling of the field LMXB XLF
appropriate for our sample as a whole, and subsequently revisit its application
to each galaxy in our sample to test for universality.  Our stellar-mass
dependent XLF can be quantified as:
\begin{equation}
\frac{dN_M}{dL} = M_\star \; K_M  \left \{
\begin{array}{lr} L^{-\alpha_1}  & \;\;\;\;\;\;\;\;(L < L_b) \\ 
L_b^{\alpha_2-\alpha_1} L^{-\alpha_2},  & (L_b \le L < L_c) \\ 
0,  & (L \ge L_c) \\ 
\end{array}
  \right.
\end{equation}
where $K_M$ is the normalization per stellar mass
(quoted in units of [$10^{11}$~\msol]$^{-1}$) at $L = 10^{38}$~\lum, and the remaining quantities
have the same meaning as they did in Equation~(2).  Here we are modeling all
data simultaneously, and we can thus allow all the parameters of the fit to be free and
directly determine their uncertainties.  Thus, we perform fitting for four
parameters: $K_M$, $\alpha_1$, $L_b$, and $\alpha_2$.  Due to the steep bright-end slope of the XLF, we are unable to constrain well $L_c$, and therefore fix its value at $L_c = 10^{40}$~\lum.  


\begin{deluxetable*}{lcccccccccccc}
\tablewidth{0pt}
\tabletypesize{\scriptsize}
\tablecaption{Summary of Field X-ray Luminosity Function Fits By Galaxy}
\tablehead{
\multicolumn{1}{c}{} & \multicolumn{8}{c}{} & \multicolumn{4}{c}{} \\
\multicolumn{1}{c}{} & \multicolumn{8}{c}{\sc Field LMXB XLF} & \multicolumn{4}{c}{} \\
\multicolumn{1}{c}{} & \multicolumn{8}{c}{\rule{3.2in}{0.01in}} & \multicolumn{4}{c}{} \\
\multicolumn{1}{c}{\sc Galaxy} &   \multicolumn{4}{c}{\sc Power Law} & \multicolumn{2}{c}{\sc $M_\star$-Dependent} & \multicolumn{2}{c}{\sc $M_\star$ and $S_N$ Dep.} & \multicolumn{2}{c}{\sc GC XLF} & \multicolumn{2}{c}{\sc Total XLF} \\
\multicolumn{1}{c}{\sc Name}  &  \multicolumn{2}{c}{\sc Single} &  \multicolumn{2}{c}{\sc Broken} & \multicolumn{2}{c}{\rule{0.8in}{0.01in}} & \multicolumn{2}{c}{\rule{0.8in}{0.01in}} & \multicolumn{2}{c}{\rule{0.8in}{0.01in}} & \multicolumn{2}{c}{\rule{0.8in}{0.01in}} \\
\multicolumn{1}{c}{\sc (NGC)} &  \colhead{$C$} & \colhead{$P_{\rm Null}^{\rm PL}$} & \colhead{$C$} & \colhead{$P_{\rm Null}^{\rm BKNPL}$} & \colhead{$C$} & \colhead{$P_{\rm Null}^{M}$} & \colhead{$C$} & \colhead{$P_{\rm Null}^{\rm field}$} & \colhead{$C$} & \colhead{$P_{\rm Null}^{\rm GC}$} & \colhead{$C$} & \colhead{$P_{\rm Null}^{\rm all}$}  \\
\multicolumn{1}{c}{(1)} & \multicolumn{1}{c}{(2)} & \multicolumn{1}{c}{(3)} & \colhead{(4)} & \colhead{(5)} & \colhead{(6)} & \colhead{(7)} & \colhead{(8)} & \colhead{(9)} & \colhead{(10)} & \colhead{(11)} & \colhead{(12)} & \colhead{(13)} 
}
\startdata
      1023 &   22 &      0.165 &   23 &      0.174 &   52 &      0.261 &   43 &      0.670 &         36 &      0.894 &   51 &      0.263 \\ 
      1380 &    6 &      0.005 &    9 &      0.034 &   17 &      0.096 &   15 &      0.182 &         32 &      0.132 &   36 &      0.101 \\ 
      1387 &   12 &      0.520 &   13 &      0.952 &   27 &      0.719 &   21 &      0.978 &         23 &      0.906 &   22 &      0.771 \\ 
      1399 &   22 &      0.034 &   22 &      0.171 &   41 &      0.435 &   43 &      0.281 &         56 &   $<$0.001 &   47 &      0.121 \\ 
      1404 &   22 &      0.092 &   23 &      0.433 &   30 &      0.516 &   28 &      0.702 &         24 &      0.639 &   28 &      0.570 \\ 
\\
      3115 &   53 &      0.551 &   31 &      0.156 &   47 &      0.497 &   40 &      0.305 &         43 &      0.411 &   49 &      0.703 \\ 
      3377 &   21 &      0.973 &   22 &      0.250 &   23 &      0.934 &   22 &      0.752 &         17 &      0.518 &   25 &      0.951 \\ 
      3379 &   42 &      0.306 &   36 &      0.472 &   47 &      0.845 &   52 &      0.207 &         33 &      0.788 &   56 &      0.163 \\ 
      3384 &   33 &      0.729 &   33 &      0.759 &   36 &      0.824 &   35 &      0.471 &         13 &      0.089 &   32 &      0.988 \\ 
      3585 &   31 &      0.465 &   28 &      0.464 &   45 &      0.271 &   39 &      0.224 &         27 &      0.761 &   37 &      0.597 \\ 
\\
      3923 &   21 &      0.109 &   21 &      0.160 &   28 &      0.625 &   27 &      0.748 &         36 &      0.160 &   32 &      0.880 \\ 
      4278 &   54 &      0.494 &   39 &      0.935 &   62 &      0.013 &   54 &      0.093 &         42 &      0.838 &   64 &      0.016 \\ 
      4365 &   35 &      0.508 &   32 &      0.522 &   41 &      0.803 &   38 &      0.774 &         36 &      0.846 &   37 &      0.927 \\ 
      4374 &   33 &      0.353 &   30 &      0.372 &   46 &      0.297 &   41 &      0.471 &         23 &      0.218 &   44 &      0.341 \\ 
      4377 &   10 &      0.398 &    8 &      0.480 &   15 &      0.214 &   13 &      0.291 &     \ldots &     \ldots &   14 &      0.206 \\ 
\\
      4382 &   24 &      0.160 &   24 &      0.222 &   58 &      0.009 &   41 &      0.256 &         40 &      0.236 &   47 &      0.108 \\ 
      4406 &   17 &      0.991 &   16 &      0.386 &   22 &      0.891 &   20 &      0.805 &     \ldots &     \ldots &   20 &      0.937 \\ 
      4472 &   26 &      0.007 &   13 &      0.002 &   29 &      0.097 &   24 &      0.064 &         28 &      0.399 &   43 &      0.962 \\ 
      4473 &   15 &      0.351 &   15 &      0.711 &   21 &      0.646 &   22 &      0.838 &         16 &      0.269 &   20 &      0.448 \\ 
      4552 &   34 &      0.293 &   34 &      0.417 &   52 &      0.048 &   35 &      0.744 &         39 &      0.461 &   34 &      0.613 \\ 
\\
      4621 &   27 &      0.907 &   28 &      0.812 &   36 &      0.165 &   36 &      0.063 &         31 &      0.220 &   46 &      0.005 \\ 
      4649 &   31 &      0.109 &   29 &      0.111 &   44 &      0.943 &   35 &      0.588 &         27 &      0.250 &   29 &      0.201 \\ 
      4697 &   34 &      0.249 &   29 &      0.374 &   36 &      0.765 &   35 &      0.855 &         44 &      0.181 &   43 &      0.679 \\ 
      7457 &   19 &      0.396 &   21 &      0.938 &   22 &      0.873 &   22 &      0.995 &     \ldots &     \ldots &   22 &      0.527 \\ 
\enddata
\tablecomments{Goodness of fit assessments for all galaxies for the field LMXB population (Col.(2)--(9); Sections 4.2, 4.3, and 4.5), the GC LMXBs (Col.(10)--(11); Section 4.4), and combined field-plus-GC LMXB model (Col.(12)--(13); Section 4.6).  Col.(1): Galaxy NGC name, as reported in Table~1.  Col.(2)--(5): $C$-statistic and null-hypothesis probability pairs for power-law and broken power-law models of the field LMXBs.  These columns are re-tabulations of Col.(7)--(8) and Col.(12)--(13) from Table~6.  Col.(6)--(7): $C$-statistic and null-hypothesis probability for the stellar-mass dependent model of the field LMXBs, which is based only on the $M_\star$ of the galaxy.  Col.(8)--(9): $C$-statistic and null-hypothesis probability for the stellar-mass and $S_N$ dependent model of the field LMXBs.  Col.(10)--(11):$C$-statistic and null-hypothesis probability for the GC LMXB population (Eqn.~(8)).  Col.(12)--(13): $C$-statistic and null-hypothesis probability for the global model, which includes contributions from both field LMXBs and GC LMXBs (see Section 4.6).\\
}
\end{deluxetable*}

When fitting for a global model, like the stellar-mass dependent model, we
determine best fit solutions and parameter uncertainties by minimizing the
cumulative $C$ statistic:
\begin{equation}
C = \sum_{i = 1}^{n_{\rm gal}} \left(2 \sum_{j=1}^{n_{\rm X}} M_{i,j} - N_{i,j}
+ N_{i,j} \ln(N_{i,j}/M_{i,j}) \right),
\end{equation}
where $C$ is now determined ``globally'' through the double summation over all
$n_{\rm gal} =$~\ngal\ galaxies ($i$th index) and $n_{\rm X}=100$ \xray\
luminosity bins, spanning \hbox{$\log L =$~35--41.7} ($j$th index).  Here, the
$C$ value for the $i$th galaxy is simply the contribution from the $i$th term
of Equation~(7) and can be compared with our individual fits from Section 4.2.
In total, \nfield\ \xray\ sources were used in the global model fit.

In Figure~5, we show the best-fit stellar-mass dependent global model applied
to each of the \ngal\ galaxies as orange dashed curves.  The assessed
galaxy-by-galaxy null-hypothesis probability, calculated using Equation~(5), is
tabulated in Table~7 (Col.~7) and the best-fit parameters for the global fit
are provided in Table~8.\footnote{We note that the use of the Kaastra~(2017)
tabulated values of $C_{\rm exp}$ and $C_{\rm var}$, as we use in Equation~(5),
do not incorporate uncertainties in the model terms for the global fits here
and in Section~4.5 (e.g., stellar mass and $S_N$ have uncertainties).
Appropriately incorporating such uncertainties into the estimates of $P_{\rm
null}$ requires computationally intensive simulations of the expected
distribution of $C$ for each best-fit model.  Unfortunately, due to the time
limitations, performing these simulations for all fits in this study is beyond
the scope of this paper.  However, in a few test cases, we find that the
incorporation of model-term uncertainties does not result in substantially
different estimates of $P_{\rm null}$ compared to those derived from
Equation~(5), since the distribution of $C$ values is dominated by Poisson
errors on the data alone.  Furthermore, we find that incorporating model-term
uncertainties tends to cause the goodness of the fits to yielded larger values
of $P_{\rm null}$ (e.g., due to larger values of $C_{\rm exp}$ and $C_{\rm
var}$).  We therefore regard our estimates of $P_{\rm null}$ to be lower limits
of more careful treatment, but generally good approximations on the goodness of
our fits.} While the stellar-mass dependent global model provides an acceptable
fit to the field LMXB data for the majority of the galaxies, there is some
tension (e.g., $P_{\rm null} \simlt 0.02$) in the fits to NGC~4278, NGC~4382,
and NGC~4552.  Despite these cases, the model is acceptable as globally
($P_{\rm null} = 0.136$).

Visual inspection of Figure~5 suggests that there are no obvious issues with
the parameterized shape of the the field LMXB XLF, but instead there is
noteworthy variation in the normalizations, with some galaxies having an
observed excess of sources (e.g., NGC~1399, NGC~4278, NGC~4472, and NGC~4552)
and others having a deficit of sources (e.g., NGC~1380, NGC~1387, NGC~3384, and
NGC~4382) compared to the stellar-mass dependent model prediction ({\it orange
dashed curves\/}).  When considering the GC specific frequencies of these
objects, the galaxies with apparent source excesses have high-$S_N$ and those
with apparent deficits have low-$S_N$.

To test the connection with $S_N$ further, we re-fit each galaxy XLF using a
model with fixed values from the best stellar-mass dependent model (i.e.,
$K_M$, $\alpha_1$, $L_b$, $\alpha_2$, and $L_c$ from Col.(3) in Table~8), but
multiplied by a scaling constant, $\omega$, that we fit for each galaxy.  By
definition, a galaxy XLF that follows the average behavior will have $\omega
\approx 1$, galaxies with excess (deficit) numbers of LMXBs will have $\omega >
1$ ($\omega < 1$).  In this fitting process, we followed the statistical
procedures above with only $\omega$ varying.  In all cases, statistically
acceptable fits were retrieved with this process, and in Figure~7, we show
the constant $\omega$ versus $S_N$.  A Spearman's ranking test indicates a
significant correlation between the $S_N$ and $\omega$ at the $\simgt$99.9\%
confidence level, providing a strong connection between the field LMXB
population and the GC population.  

Below, we consider a scenario in which the apparent shift in field LMXB XLF
shape from high-to-low $\alpha_2$ and increase in normalization per unit
stellar mass with increasing $S_N$ are due to increased contributions of a
``GC-seeded'' field LMXB population that scales with $S_N$. We
start, in Section~4.4, by modeling the GC LMXB XLF shape and normalization
scaling with $S_{N,{\rm loc}}$ and $M_\star$ for sources that are directly
coincident with GCs.  We then use the resulting direct-GC LMXB model shape as a
prior on the shape of the GC-seeded field LMXBs, the scaling of which we
determine in Section~4.5.  We note that a GC-seeded LMXB XLF need
not necessarily have the same shape as that of the direct-GC LMXB population;
however, to first order, we expect them to be similar. 

\subsection{Globular Cluster Population XLF}

Using our catalog of \ngc\ sources that were directly matched to GCs, we
generated GC LMXB XLFs for each galaxy.  In Figure~8, we show the co-added GC
LMXB XLF, in differential form (note this differs from the cumulative form
displayed in Fig.~5), for all galaxies combined.  Unlike Figure~5, we display
the completeness-corrected XLF here, since we use this representation to inform
the shape of our GC LMXB XLF model.  The shape of the observed GC LMXB XLF in
Figure~8 follows a smooth progression from a shallow-sloped power-law at $L
\simlt 10^{38}$~\lum\ to a steeply declining shape at higher luminosities.
Such behavior can be modeled as either a broken power-law or a power-law with a
high-$L$ exponential decline.  Given the apparent curvature of the XLF in
Figure~8, we chose to use the latter model.  We note that previous
investigations of GC LMXB XLFs in relatively nearby galaxies, e.g., Cen~A and
M31, have found similar shapes to those presented here, but with a further
flattening and potential decline in the GC LMXB XLF for $L \simlt
10^{37}$~\lum, just below the detection limits of our galaxies (e.g.,
Trudolyubov, \& Priedhorsky~2004; Voss \etal\ 2009).

Using the techniques discussed above, we fit the GC LMXB XLFs
of the full galaxy sample using the following model:
\begin{equation}
\frac{dN_{\rm GC}}{dL} = M_\star \; S_{N,{\rm loc}} \; K_{\rm GC} L^{-\gamma} \exp( -L/\lambda
),
\end{equation}
where $K_{\rm GC}$, $\gamma$, and $\lambda$ are unknown quantities to be determined
by data fitting.  As before, we utilized the global statistic in
Equation~(7) when determining our best-fit solution.

In Figure~8, the dotted purple curve shows our best-fit solution and residuals, and
Figure~9 provides probability distribution functions and co-variance contour
planes for the parameters $K_{\rm GC}$, $\gamma$, and $\lambda$.  The best-fit
model provides a good characterization of the broader shape and normalization
of the GC LMXB XLF for our sample.  We find a relatively shallow power-law
slope $\gamma \approx 1.1$ with a cut-off at $\lambda \approx 4 \times
10^{38}$~\lum, just above the Eddington limit of an $\approx$2--3~$M_\odot$
neutron star, a feature that has long been noted in LMXB XLFs (see, e.g., the
review by Fabbiano~2006).


\begin{longrotatetable}
\begin{table*}
{\footnotesize
\begin{center}
\caption{Best Fit Parameters for Global Fits}
\begin{tabular}{lccccccc}
\hline\hline
\multicolumn{1}{c}{} &  &  &  & \multicolumn{3}{c}{\sc $M_\star$-and-$S_N$-Dependent}  &  \\
\multicolumn{1}{c}{\sc Parameter} &  & {\sc $M_\star$-Dependent} & {\sc } & \multicolumn{3}{c}{\rule{3.5in}{0.01in}}  & {\sc Z12} \\
\multicolumn{1}{c}{\sc Name} & {\sc Units} & {\sc Field} & {\sc GC LMXBs} & {\sc Field (No Priors)} & {\sc Field (Priors)} & {\sc All LMXBs} & {\sc Value} \\
\multicolumn{1}{c}{(1)} & (2) & (3) & (4) & (5) & (6) & (7) & (8)\\
\hline\hline
LMXB Population  &  & Field & GC & Field & Field & Field + GC & Field + GC \\
$N_{\rm det}$  &  &1285$^\dagger$ & 595 & 1285$^\dagger$ & 1285$^\dagger$ & 1880$^\dagger$  & \\
\hline
\multicolumn{8}{c}{Field LMXB Component} \\
\hline
$K_{M}$ or $K_{\rm in\mbox{-}situ}$ & ($10^{11}$~\msol)$^{-1}$ & 60.9$^{+6.7}_{-6.9}$ & \ldots &42.4$^{+9.1}_{-7.9}$ & 42.7$^{+6.5}_{-6.0}$ &34.9$^{+9.5}_{-6.9}$ & $41.5 \pm 11.5$ \\
$\alpha_1$ &  & 1.00$^{+0.06}_{-0.06}$ & \ldots & 0.98$^{+0.09}_{-0.11}$ & 1.02$^{+0.07}_{-0.08}$ &1.07$^{+0.10}_{-0.12}$ & 1.02$^{+0.07}_{-0.08}$ \\
$L_b$ & $10^{38}$~\lum  & 0.49$^{+0.07}_{-0.04}$ & \ldots & 0.45$^{+0.07}_{-0.05}$ &0.45$^{+0.06}_{-0.04}$ &0.52$^{+0.17}_{-0.11}$ & $0.546^{+0.043}_{-0.037}$ \\
$\alpha_2$ &  &  2.12$^{+0.07}_{-0.06}$ & \ldots & 2.43$^{+0.18}_{-0.15}$ & 2.50$^{+0.18}_{-0.14}$ &2.27$^{+0.17}_{-0.13}$ & 2.06$^{+0.06}_{-0.05}$ \\
$L_{b,2}^\ddagger$ & $10^{38}$~\lum  & \ldots  & \ldots & \ldots & \ldots & \ldots & $5.99^{+0.95}_{-0.67}$ \\
$\alpha_3^\ddagger$ & \ldots & \ldots & \ldots & \ldots & \ldots & \ldots & $3.63^{+0.67}_{-0.49}$ \\
$\log Lc$ & $\log$ \lum\  & 40.0$^\ast$ & \ldots  &  \ldots &  \ldots & \ldots  & $40.04^{+0.18}_{-0.16}$\\
\hline
\multicolumn{8}{c}{GC-Related LMXB Component} \\
\hline
$K_{\rm GC}$ or $K_{\rm seed}$ & ($10^{11} M_\odot)^{-1}$~$S_N^{-1}$  & \ldots & 8.08$^{+0.42}_{-0.41}$ & 5.00$^{+0.67}_{-0.61}$ & 5.10$^{+0.53}_{-0.52}$ &12.63$^{+0.62}_{-0.59}$ &  \ldots \\
$\gamma$ &  & \ldots & 1.08$^{+0.04}_{-0.04}$ & 1.21$^{+0.11}_{-0.13}$ & 1.09$^{+0.04}_{-0.04}$ &1.12$^{+0.07}_{-0.08}$ & \ldots \\
$\log \lambda$ & $\log$ \lum\  & \ldots & 38.61$^{+0.05}_{-0.04}$ & 38.66$^{+0.10}_{-0.10}$ & 38.61$^{+0.04}_{-0.04}$ &38.50$^{+0.05}_{-0.05}$ & \ldots  \\
$C$ &  & 887 & 676 & 792 & 793 & 888 & \ldots \\
$C_{\rm exp}$ &  & 832 & 655 & 763 & 759 & 813 & \ldots \\
$C_{\rm var}$ &  & 1363 & 903 & 1273 & 1265 & 1403 & \ldots \\
$P_{\rm null}$ &  & 0.136 & 0.476 & 0.417 & 0.337 & 0.045 & \ldots \\
\hline
\multicolumn{8}{c}{Calculated Parameters} \\
\hline
$\log$($\alpha_{M}$ or $\alpha_{\rm in\mbox{-}situ}$) & $\log$ \lum~$M_\odot^{-1}$  & 29.17$^{+0.03}_{-0.03}$ &  \ldots & 28.76$^{+0.07}_{-0.07}$ & 28.75$^{+0.06}_{-0.06}$ & 28.86$^{+0.07}_{-0.08}$  & $29.2 \pm 0.1$ \\
$\log$($\kappa_{\rm GC}$ or $\kappa_{\rm seed}$) & $\log$ \lum~$M_\odot^{-1}$~$S_N^{-1}$  & \ldots & 28.55$^{+0.03}_{-0.03}$ & 28.38$^{+0.05}_{-0.06}$ & 28.38$^{+0.05}_{-0.05}$ & 28.67$^{+0.03}_{-0.04}$ &  \ldots \\
\hline
\end{tabular}
\end{center}
Note.---Col.(1) and (2): Parameter and units. Col.(3)--(7): Value of each parameter for the various global models applied throughout this paper. Col.(8): Comparison values of LMXB scaling relations from Zhang \etal\ (2012).\\
$^\dagger$Numbers include contributions from \nbkgf\ background sources with $S < 2 \times 10^{-15}$~\flux\ (see Section 4.1).\\
$^\ddagger$Parameter was used in Z12, but not in our study.\\
}
\end{table*}
\end{longrotatetable}

We find that our model provides a good overall characterization of the GC LMXB
XLFs for the sample ($P_{\rm null} = 0.476$; see Table~8).  On a
galaxy-by-galaxy basis, the model is a good fit ($P_{\rm null} \simgt 0.01$) to
the GC XLFs for 20 out of the 21 galaxies with \xray\ detected GCs (see
Table~7).  The three galaxies that did not have any GCs detected are consistent
with predictions, as a result of these galaxies having either low stellar mass
(NGC~4377 and NGC~7457) or shallow \chandra\ data (NGC~4406).  The one galaxy,
for which our GC-XLF model provides a poor characterization of the data,
NGC~1399, is the most GC-rich galaxy in our sample.

The failing of the GC LMXB XLF model in NGC~1399 is thus likely due to
unmodeled physical variations in the GC population.  For example, red,
metal-rich GC populations are observed to contain a larger fraction of bright
LMXBs than blue, metal-poor GCs (e.g., Kundu \etal\ 2007; Kim \etal\ 2013;
D'Abrusco \etal\ 2014; Mineo \etal\ 2014; Peacock \& Zepf~2016; Peacock \etal\
2017), and the fraction of metal-rich versus metal-poor GCs varies between
galaxies (e.g., Brodie \& Strader~2006).  For the case of NGC~1399, detailed
studies suggest that the red-to-blue ratio of GCs could be somewhat larger than
most galaxies in our sample (e.g., Paolillo \etal\ 2011; D'Ago \etal\ 2014).

In a forthcoming paper, we will assess in more
detail the properties of the LMXB populations in the GCs in our sample.  
Aside from the case of NGC~1399, the GC LMXB XLF model 
provides a good model to the GC LMXB data for the sample as a whole. 
We use parameters from our GC LMXB model in the next section to inform the shape of a GC-seeded
LMXB contribution to the field LMXB XLF.

%
%
\begin{figure}
\figurenum{8}
\centerline{
\includegraphics[width=9.3cm]{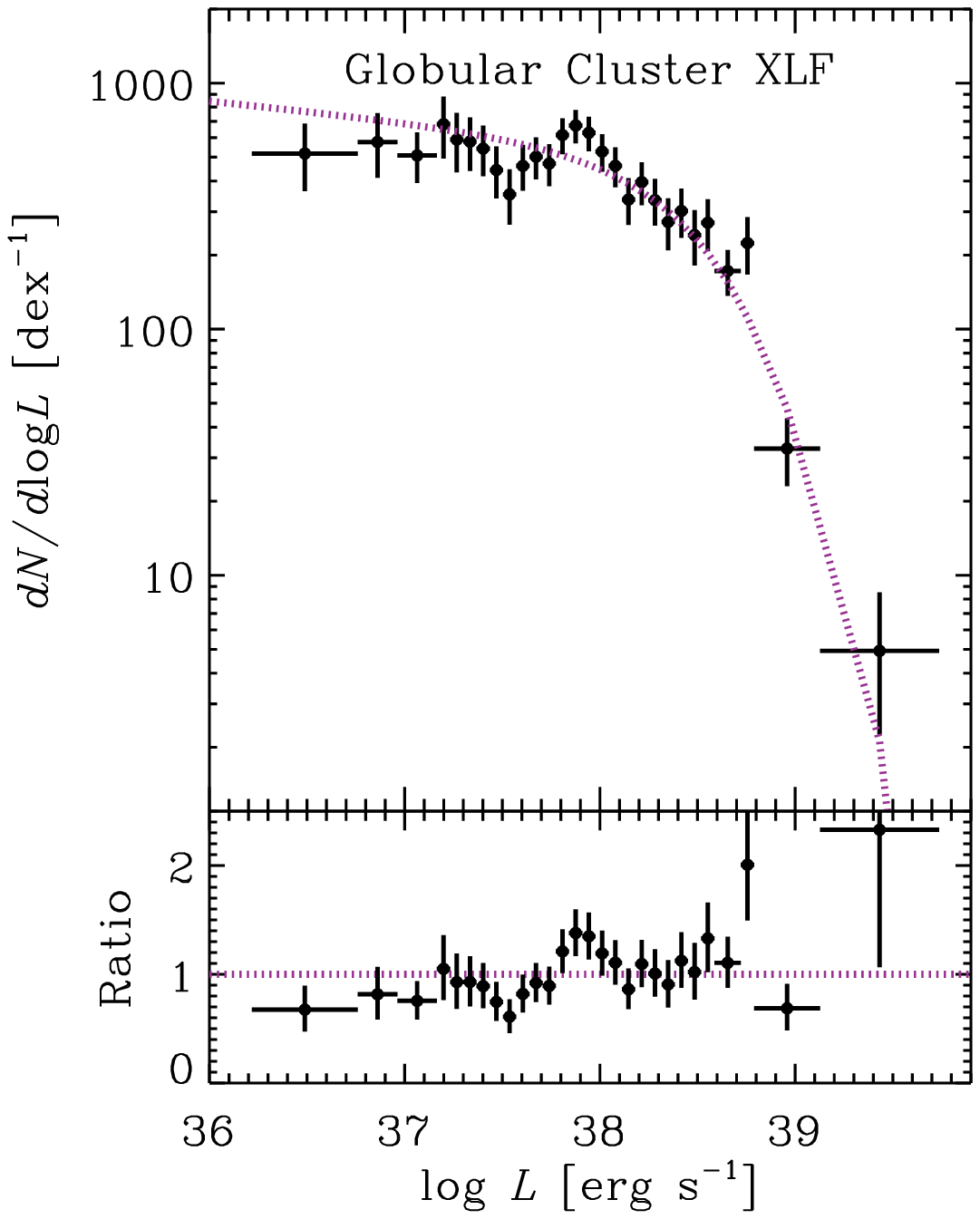}
}
\vspace{-0.3in}
\caption{
({\it Top\/}) Best-fit completeness-corrected XLF for \xray\ sources directly coincident with GCs
({\it filled circles} with 1$\sigma$ errors).  This sample includes \ngc\ such sources
collected from the \ngal\ elliptical galaxies in our sample.  Our best-fitting power-law with
exponential decay model is shown as a dotted purple curve.
({\it Bottom\/}) Ratio of data to model for our best-fit model.  The dotted purple
horizontal line at ratio~=~1 has been indicated for reference.
}
\end{figure}

%
%
\begin{figure}
\figurenum{9}
\centerline{
\includegraphics[width=9.3cm]{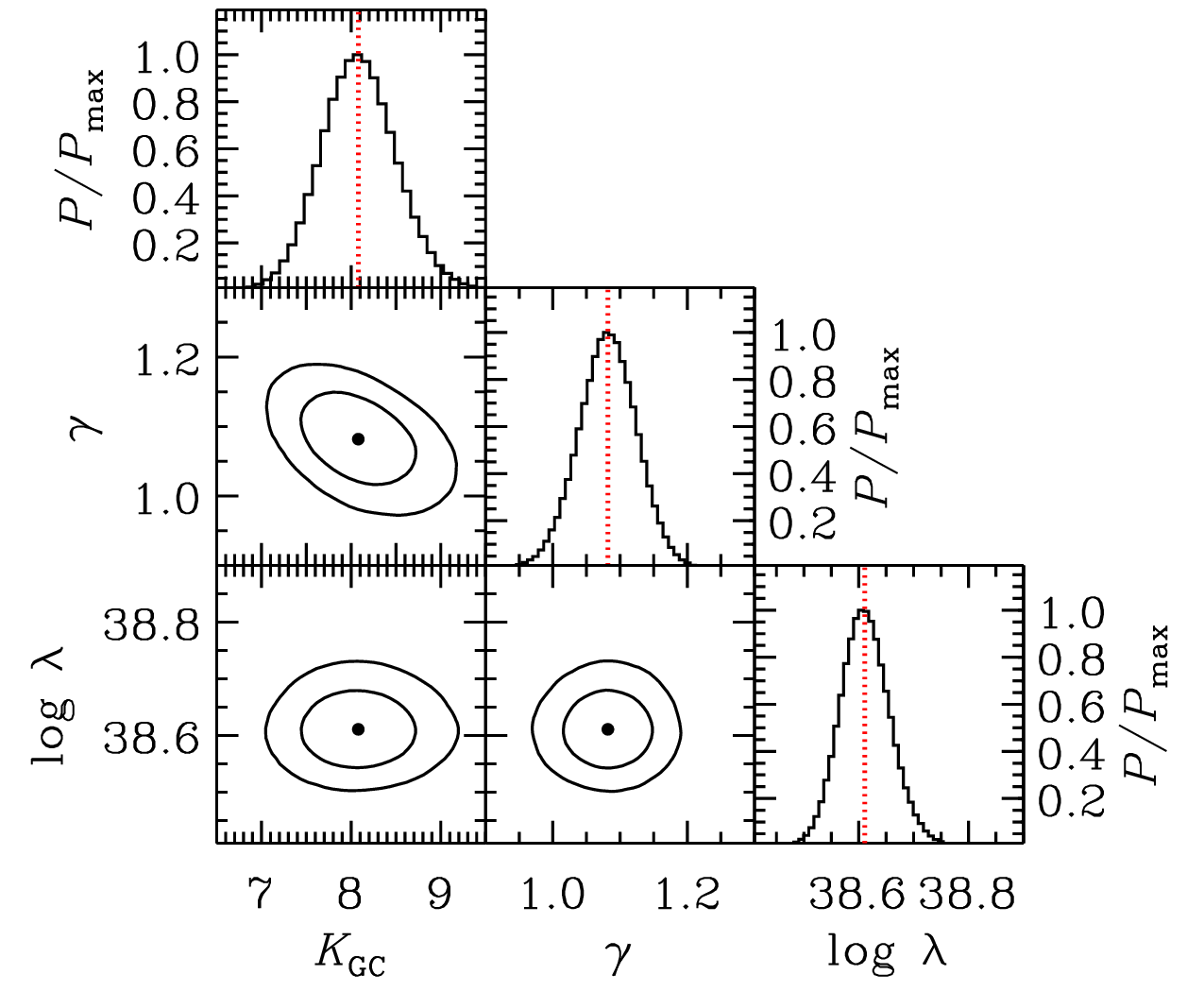}
}
\caption{
Probability distribution functions ($P$/$P_{\rm max}$) and confidence contours for
parameter pairs (showing 68\% and 95\% confidence contours drawn) for our
best-fit GC LMXB XLF model, which is based on \ngc\ sources in \ngal\
galaxies (see Section 4.4 for details).  The vertical red dotted lines and
solid black points indicate the median values of each parameter. 
}
\end{figure}

%
%
\begin{figure*}[h!]
\figurenum{10}
\centerline{
\includegraphics[width=18cm]{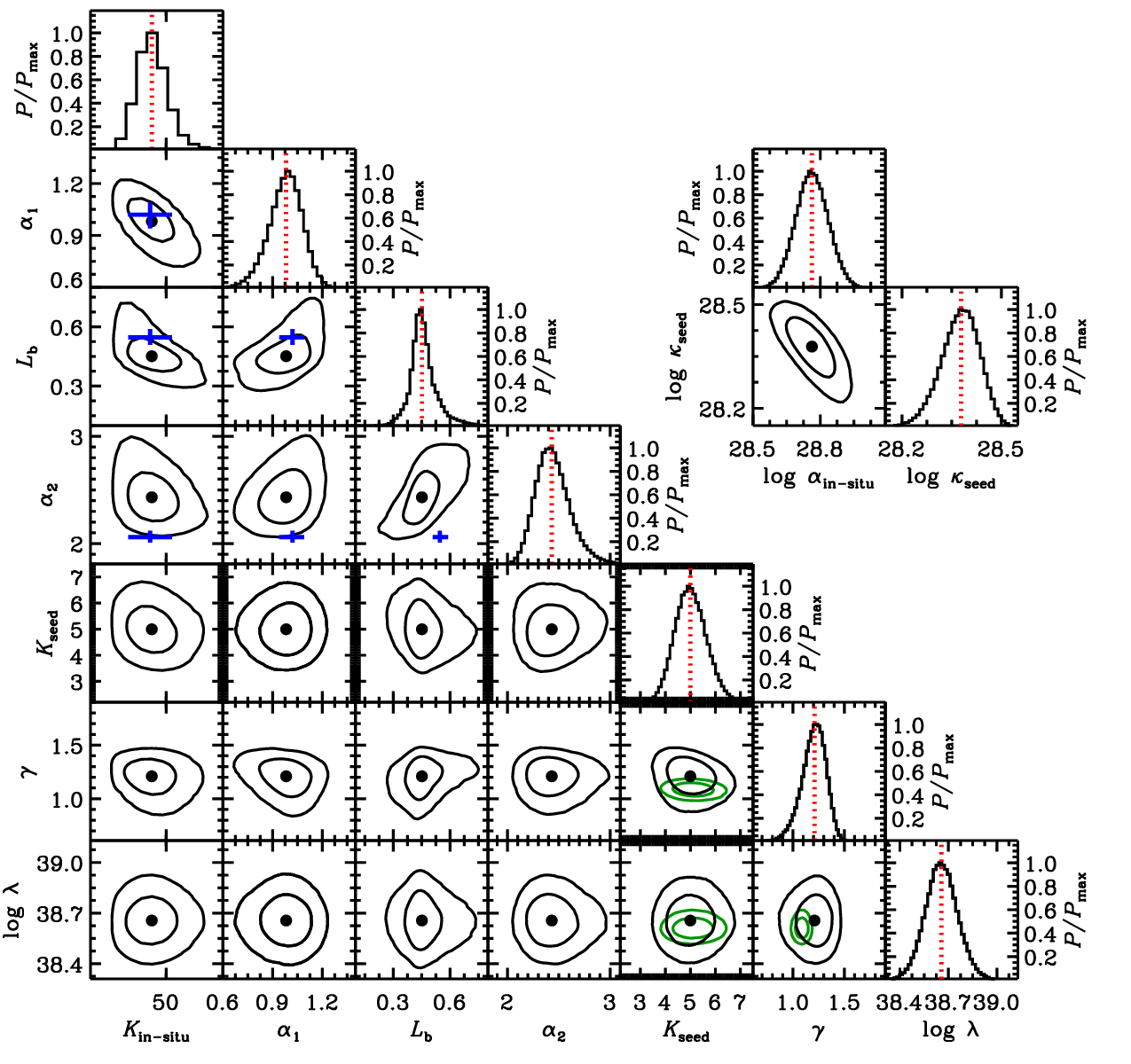}
}
\caption{
Probability distribution functions ($P$/$P_{\rm max}$) and confidence contours
for parameter pairs (showing 68\% and 95\% confidence contours) for our
best-fit field LMXB model, which is based on \nfield\ sources in \ngal\
galaxies and uses flat priors on all parameters (see Section 4.5 for details).
The parameters track the XLFs of LMXB populations that are presumed to form
both in situ ($K_{\rm in\mbox{-}situ}$, $\alpha_1$, $L_b$, and $\alpha_2$) and
those seeded from GCs ($K_{\rm seed}$, $\gamma$, and $\lambda$).  The
distribution functions for the integrated $L_{\rm X}$(in-situ)/$M_\star$
($\alpha_{\rm in\mbox{-}situ}$) and $L_{\rm X}$(seed)/$M_\star$/$S_N$
($\kappa_{\rm seed}$), implied by our model, are shown in the upper-right
panels.  Comparison values and 1$\sigma$ errors from Zhang \etal\ (2012) for
{\it all} LMXB populations within elliptical galaxies are indicated with blue
crosses in the in-situ parameterization.  We also show resulting
contours on $K_{\rm seed}$, $\gamma$, and $\lambda$ for the case where
direct-GC LMXB best-fit model priors are used on $\gamma$ and $\lambda$ ({\it
green contours\/}); these priors are informative, and the resulting values are
consistent with the case where flat priors are used.
}
\end{figure*}

%
%
\begin{figure*}
\figurenum{11}
\centerline{
\includegraphics[width=18cm]{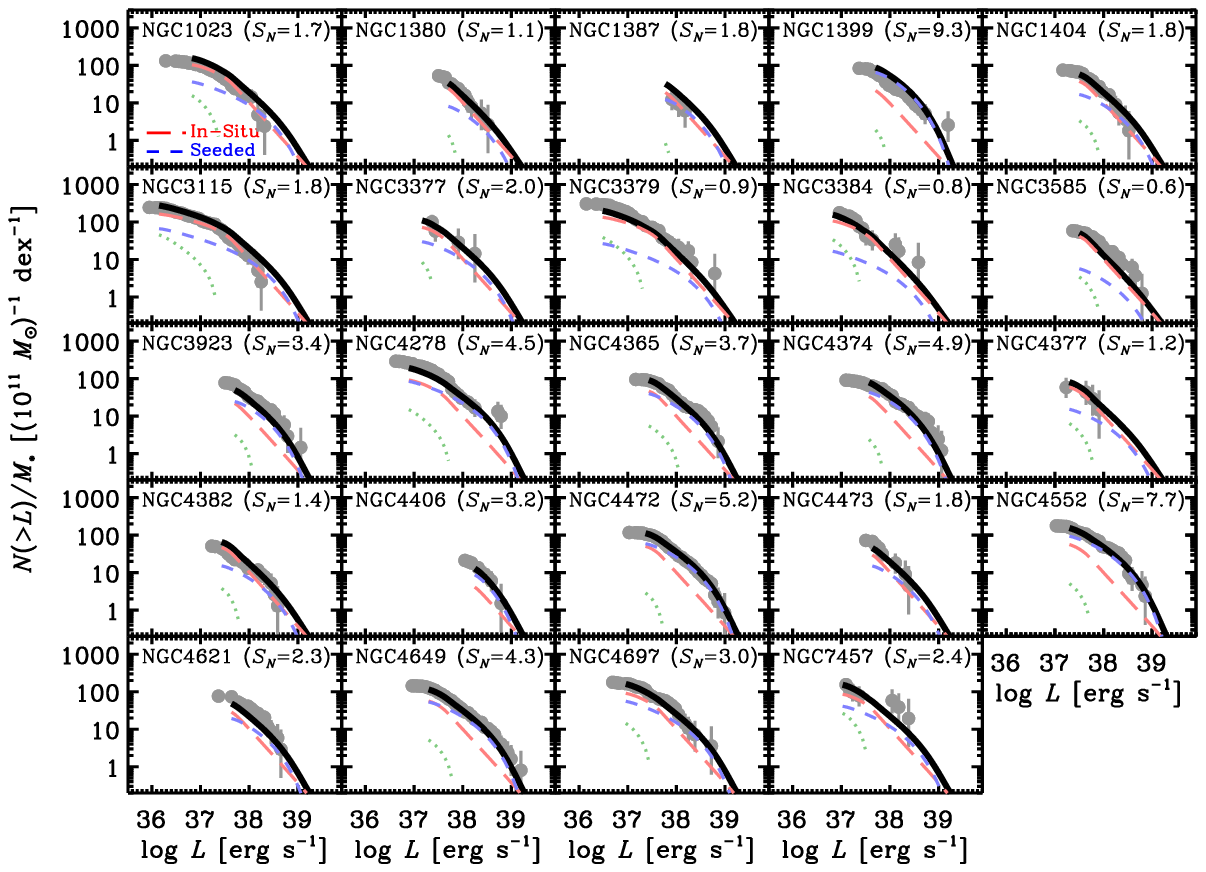}
}
\caption{
Same as Figure~5, but with the global model plotted.  This model includes contributions from
the CXB ({\it green dotted curves\/}), in-situ field LMXBs presumed to form within
the galactic field ({\it red long-dashed curves\/}), and GC-seeded field LMXBs ({\it
short-dashed blue curves\/}) that are expected to have originated in GCs.  The
model uses flat priors on all parameters in the fit (see Section 4.5 for
details).
}
\end{figure*}

\subsection{In-Situ and GC-Seeded Field LMXB XLF Model}

We chose to revisit our fitting of the field LMXB XLF data using a
two-component model consisting of an LMXB population that forms in situ and has
an XLF that scales with stellar mass, plus a GC-seeded LMXB population with XLF
normalization that scales with stellar mass and global $S_N$.  The observed
field LMXB XLF for a given galaxy is thus modeled following:
\begin{equation}
\frac{dN_{\rm field}}{dL} = \xi(L) \left[\frac{dN_{\rm in\mbox{-}situ}}{dL} +
\frac{dN_{\rm seeded}}{dL} + {\rm CXB} \right],
\end{equation}
where $dN/dL$(in-situ) and $dN/dL$(seeded) follow the functional forms provided
in Eqns.~(6) and (8), respectively, with $S_N$ being used here instead of $S_{N,{\rm loc}}$ in Eqn.~(8).

In total, we fit for 7 parameters, including four parameters related to the
in-situ component ($K_{\rm in\mbox{-}situ}$, $\alpha_1$, $L_b$, and $\alpha_2$; via Eqn.~(6)) and
three for the GC-seeded component ($K_{\rm seeded}$, $\gamma$, and $\lambda$; via Eqn.~(8)).
Following the fitting procedures discussed above (i.e., calculating the $C$
statistic via eqn~(7) and using an MCMC technique to determine uncertainties)
we determined the best-fit solution and parameter uncertainties for our model.
We chose to fit the data for two scenarios: one in which all parameters varied
freely without informative priors (i.e., flat priors), as well as a scenario in which
informative priors were implemented on $\gamma$ and $\lambda$, based on the GC LMXB fit
PDFs determined in Section~4.4. 

In Table~8 (Col.~(5) and (6)), we list the best-fit parameter values,
uncertainties, and statistics for our model, including the cases with and
without informative priors on $\gamma$ and $\lambda$.  We graphically show the
parameter PDFs and their correlations in Figure~10 (for the case with
informative priors).  From this representation, it is clear that all seven
parameters are well constrained by our data, and we find this to be the case
whether or not informative priors are implemented.  The model provides an
improvement in fit quality over the stellar-mass dependent model presented in
Section~4.3 ($P_{\rm null} = 0.337$ and 0.417 with and without informative
priors, respectively).  Furthermore, we find that $K_{\rm GC}$ is greater than
zero at the $>$99.999\% confidence level whether or not informative priors are
implemented, providing further strong evidence that this component is required.

The above analysis confirms that the field LMXB population has a non-negligible
contribution from sources that are correlated with the GC $S_N$, strongly
indicating that GCs seed the field LMXB population.  Further support for this
scenario is seen in the good agreement between the shape of the GC-seeded and
GC LMXB XLFs.  Specifically, when informative priors are not implemented, the
best-fit values for $\gamma$ and $\lambda$ are well constrained for the seeded
population, and the values of these parameters are in good agreement with those
from the direct GC population (Col.~(3) of Table~8), consistent with a
connection between the populations.  However, we find that our GC LMXB priors
on $\gamma$ and $\lambda$ are informative on the GC-seeded field LMXB
population, and when implemented, result in tighter constraints on all
parameters.

%
%
\begin{figure*}[t!]
\figurenum{12}
\centerline{
\includegraphics[width=18cm]{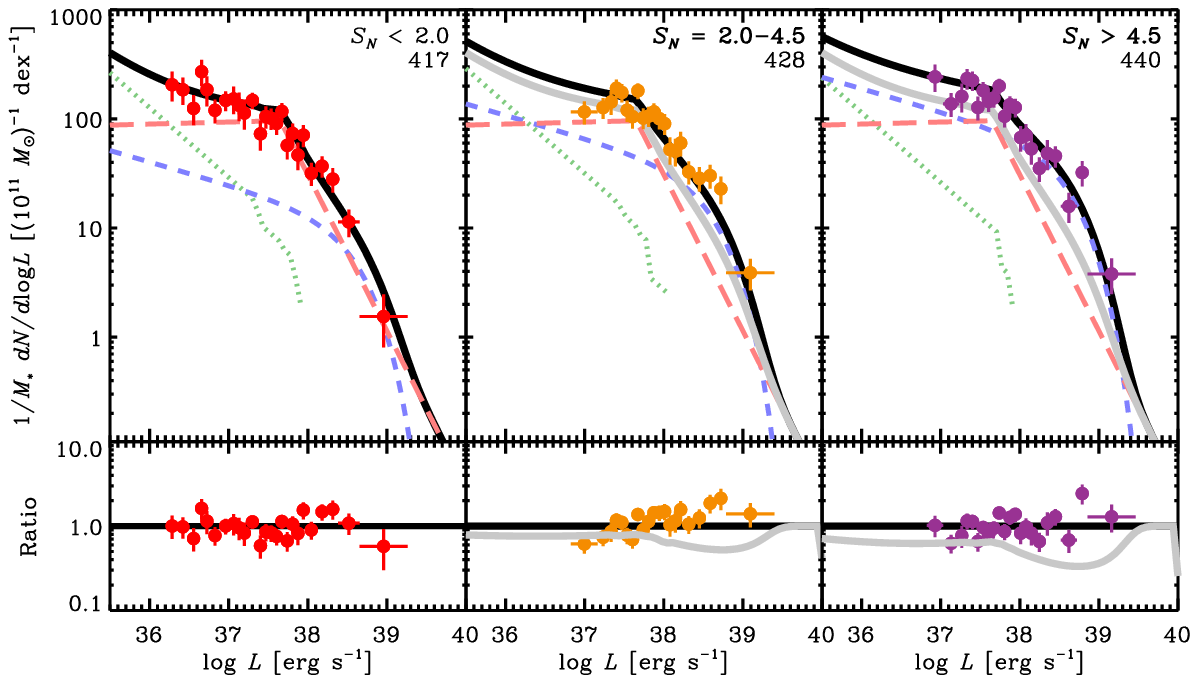}
}
\caption{
Completeness-corrected and stellar-mass-normalized field LMXB XLFs for
subsamples of galaxies in bins of global $S_N$.  All \nfield\ field LMXBs and \nbkgf\
background sources with $S < 2 \times 10^{-15}$~\flux\ are represented here.
For each panel, the $S_N$ range and number of \xray\ detected sources are
indicated in the upper right-hand corners.  Our best-fit XLF model
(based on flat priors of all parameters) is shown as a
solid black curve, and contributions from in-situ LMXBs ({\it red long-dashed
curves\/}), GC-seeded LMXBs ({\it blue short-dashed curves\/}), and CXB sources
({\it green dotted curves\/}) are indicated.  Residuals (ratio of
data-to-model) are provided in the bottom panels.  For ease of comparison
between panels, we repeated the $S_N < 2$ best-fit curve ({\it gray curve\/})
in subsequent panels corresponding to higher $S_N$ values.  As $S_N$ increases,
the influence of GC seeding becomes more prominent, leading to more field LMXBs
per unit stellar mass and a shallower-sloped XLF at $L \simgt 3~\times
10^{37}$~\lum.
}
\end{figure*}

%
%
\begin{figure*}[t!]
\figurenum{13}
\centerline{
\includegraphics[width=18cm]{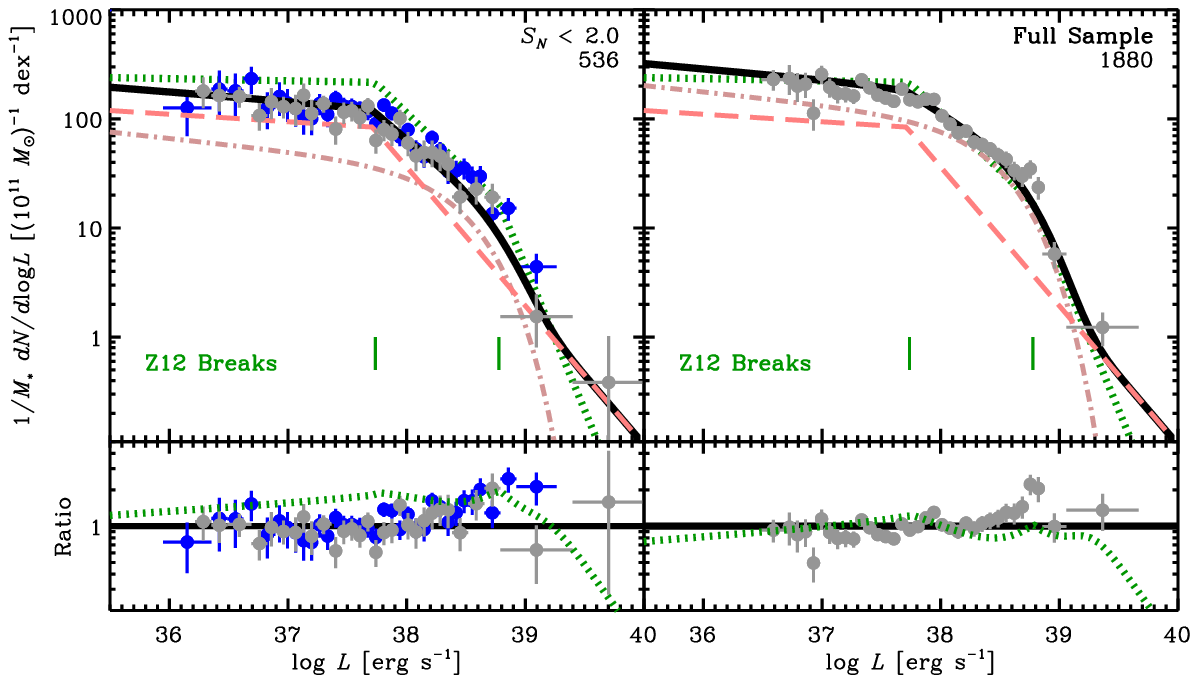}
}
\caption{
Completeness-corrected and stellar-mass-normalized {\it total} LMXB XLFs for
galaxies with $S_N < 2$ ({\it left\/}) and all sources ({\it right}), with
residuals ({\it bottom panels\/}).  In each panel, we provide our best-fit
overall model ({\it black}), and the in-situ field ({\it long-dashed faded red\/}) and
GC-related (combined field and direct GC; {\it dot-dashed faded brown\/}) model
contributions.  In the left panel, we also show the LMXB XLF from L19 ({\it
blue data points\/}), which is measured from low SFR/$M_\star$ regions in
primarily late-type galaxies that have $S_N < 2$.  We also show the best-fit
model from Zhang \etal\ (2012) as dotted green curves and indicate the
locations of their two power-law break luminosities.  
}
\end{figure*}

To better assess the quality of our model, we evaluated the fit quality it
provides to the field LMXB XLF of each galaxy.  In Figure~11, we display the
stellar-mass normalized observed XLF along with the best-fit
$M_\star$-and-$S_N$-dependent model (based on flat priors) and its in-situ and
GC-seeded model components shown separately.  In Table~7, we list the
statistical fit quality for each galaxy for the case of flat
priors.  In all cases, the individual field LMXB XLF is well described by this
model, with $P_{\rm Null} \ge 0.069$ for all galaxies (see Col.(8) and (9) in
Table~7).

With the exception of galaxies with $S_N \simlt 2$, our model suggests that the
field LMXB XLFs of our galaxy sample has significant, and often dominant,
contributions from seeded GCs at $L \simgt 10^{38}$~\lum.  At lower
luminosities, $L \simlt 10^{38}$~\lum, the in situ LMXB population is generally
dominant for most galaxies with $S_N \simlt 4$.  

Figure~12 shows the {\it completeness corrected}, stellar-mass-normalized
field LMXB XLFs (in $dN/d\log L$ differential form) for combined subsamples of
galaxies divided into bins of $S_N$.  This view demonstrates that as $S_N$
increases, the field LMXB XLF increases in normalization and transitions from a
broken power-law with a single obvious break to a shallower slope between $L
\approx$~(3--100)~$\times 10^{37}$~\lum.  For the highest $S_N$ bin ($S_N >
4.5$), the field LMXB XLF appears to take on a three-sloped power-law, with
breaks near $5 \times 10^{37}$~\lum\ and $5 \times 10^{38}$~\lum.  The apparent
break locations are consistent with those that have been reported in the
literature (see, e.g., Gilfanov~2004; Zhang \etal\ 2012), based on global fits
to LMXB XLFs that include both field and GC sources combined. For example, the
Zhang et al. (2012) break locations are at $L \approx 3 \times 10^{37}$~\lum\
and $5 \times 10^{38}$~\lum.
From our analysis, the low and high $L$ breaks can be attributed to the in-situ
and GC-seeded populations, respectively; however, it is unclear from our data
whether the GC seeded LMXB population also has a low-$L$ break ($L \approx 3 \times
10^{37}$~\lum), although some studies suggest this may be the case (e.g., Voss
\etal\ 2009).  We discuss the physical origins of this break in the Discussion
section (Section~5) below.

%
%
\begin{figure*}
\figurenum{14}
\centerline{
\includegraphics[width=9cm]{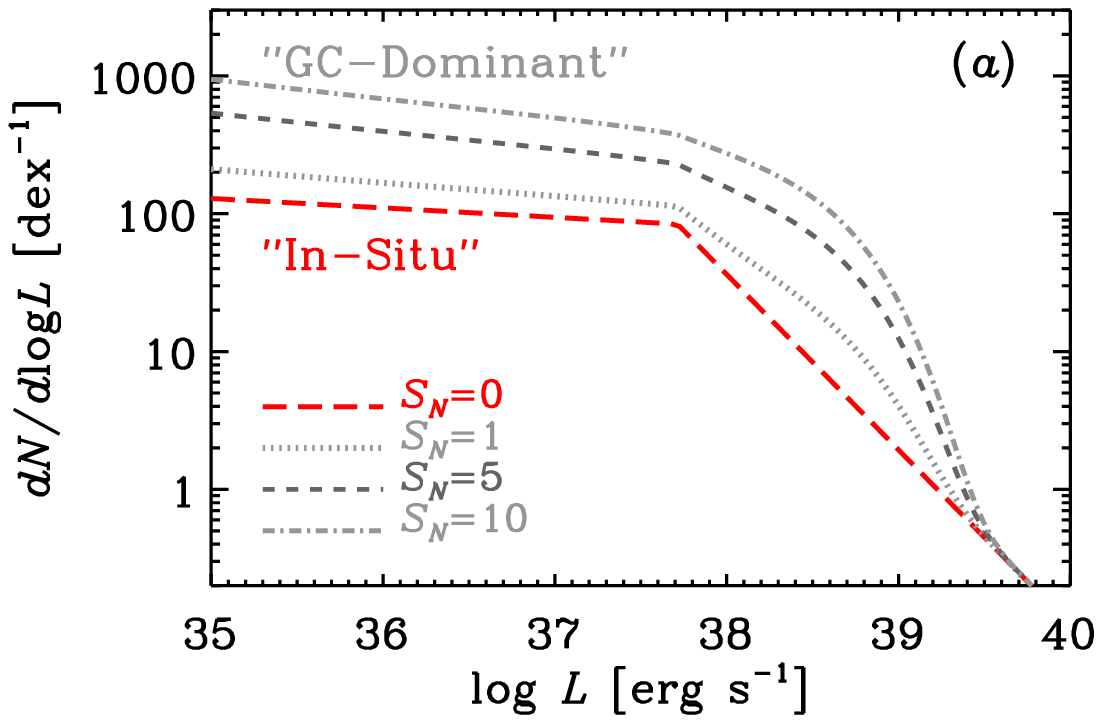}
\hfill
\includegraphics[width=9cm]{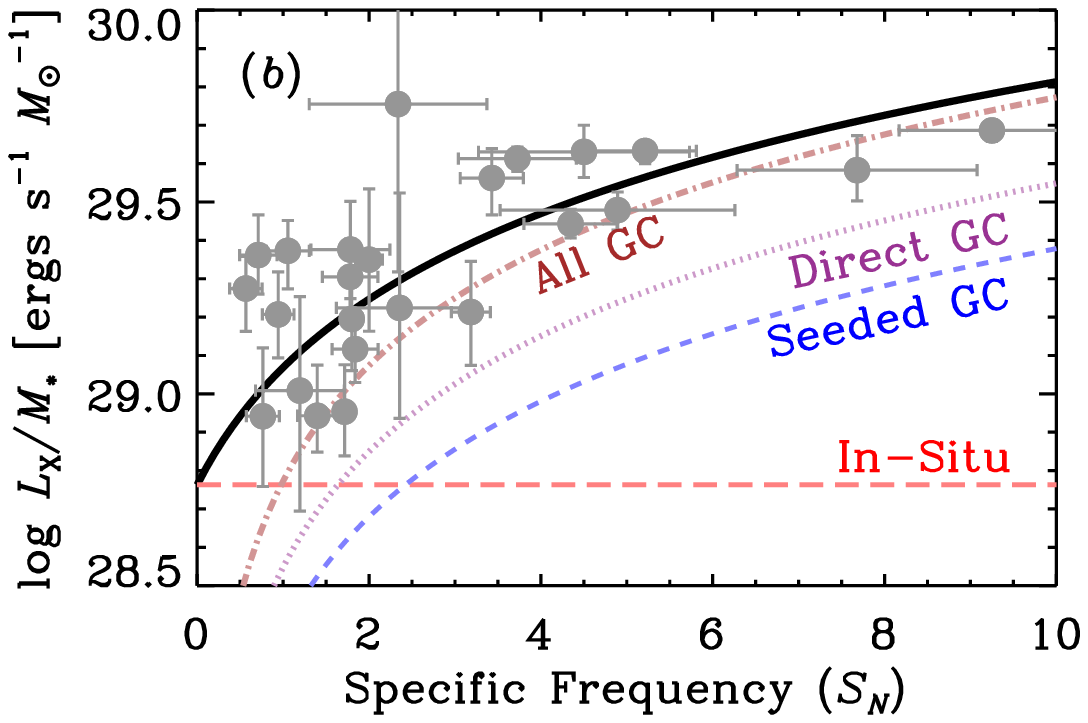}
}
\caption{
($a$) LMXB XLF models (based on Col.~(4) and (5) from
Table~8) for various values of GC $S_N$ (see annotations).  The
XLF models are all appropriate for a galaxy mass of $M_\star = 10^{11}$~\msol.
The model transitions from a single-break broken power-law at $S_N=0$ to a more
complex shape and higher normalization at larger $S_N$ due to the added
contribution from GC LMXBs.  
($b$) Integrated LMXB $L_{\rm X}/M_\star$ versus $S_N$ for our LMXB model ({\it
black solid curve\/}), with contributions from in situ ({\it red
long-dashed\/}), GC-seeded field ({\it blue short-dashed\/}), and direct GC
({\it purple dotted\/}) LMXBs shown.  The overall LMXB population originating
in GCs is shown as a brown dot-dashed curve.  Values of $L_{\rm X}/M_\star$,
and 1$\sigma$ uncertainties, for each of the \ngal\ galaxies in our sample are
shown as gray points with error bars.  These values were determined by
integrating best-fit broken power-law models to the total XLFs of each galaxy,
as per the techniques described in Section~4.2.
}
\end{figure*}

\subsection{Putting it All Together: A Global LMXB XLF Model for Elliptical Galaxies}

The above analyses of the field and GC LMXB XLFs indicate that we can
successfully model the LMXB XLF of a given galaxy as consisting of both in-situ
and GC seeded field LMXBs, as well as direct-counterpart GC LMXBs, with the
seeded and direct-counterpart GC LMXB XLFs having similar shapes.
When combining the field and GC LMXB data sets and model statistics
(i.e., combining Col.~(5) and (4) in Table~8), we find $P_{\rm
null} = 0.280$, suggesting a very good overall characterization of both field
and direct-counterpart GC LMXB populations. Given the success of
this framework, as well as the fact that our galaxy sample does not have
substantial diversity in stellar-mass-weighted age (see Section~3.1), we do not
attempt to model how the in situ field LMXB population evolves with age.
However, in Section~5, we contextualize the constraints placed on the field
LMXB populations studied here and in previous investigations, as well as the
constraints on the age-dependence of LMXB populations (e.g., Fragos \etal\
2013a, 2013b; Lehmer \etal\ 2016; Aird \etal\ 2017).

Given the similarities between the seeded and direct-counterpart GC
LMXB XLF solutions, we attempted to fit the entire data set (i.e., both field
and GC LMXB populations taken together) using a single model for the GC population.  Using the
priors on the direct-counterpart GC LMXB and field LMXB models determined in
Sections~4.4 and 4.5, respectively (i.e., the models summarized in Table~8,
Col.(4) and (5), respectively), we fit the {\it total} LMXB XLFs (including
both direct-counterpart GC LMXBs and field LMXBs) to test whether our
cumulative model fits are acceptable for all galaxies and cumulatively for the
whole sample.  In practice, we made use of Equation~(8) when modeling GC LMXBs
XLF components, using normalizations that consist of $K_{\rm GC}$ and $K_{\rm
seeded}$, which scale with $S_{N, {\rm loc}}$ and $S_N$, respectively.  To
track the relative scalings in our MCMC procedure fit to all sources, we drew
from previous MCMC chains originating from our fits to direct-counterpart and
seeded GC LMXB populations to implement priors on each of the respective
normalizations, and we quote a single normalization that includes the sum of
the priors (i.e., $K_{\rm seed}+K_{\rm GC}$).  We implemented flat priors for
all other parameters in the fits.

The resulting fit parameters are listed in Col.(7) of Table~8.  The fit quality
is acceptable ($P_{\rm null} \approx 0.045$), albeit less
favorable than the case where we utilize separate direct-counterpart and seeded
GC XLF solutions (i.e., the combined models from Col.(4) and (5) of Table~8).

In Figure~13, we show the completeness-corrected and CXB subtracted XLFs for a
subsample of galaxies with $S_N < 2$ and the full galaxy sample.  The $S_N < 2$
subsample is shown for comparison with late-type galaxy samples, which
primarily fall into this $S_N$ regime (see H13).  We show the L19 LMXB XLF data
({\it blue points\/}), which were constrained from subgalactic regions with
SFR/$M_\star$~$\simlt 10^{-10}$~yr$^{-1}$.  We find a factor of $\approx$1.5--2
elevated residuals above $10^{38}$~\lum\ for the late-type galaxy LMXBs.
Excesses of LMXBs in this luminosity regime were also noted by L19 in regions
of late-type galaxies with relatively active SF activity ($-10.5 <
\log$~SFR/$M_\star < -10$) compared to lower SF activity ($\log$~SFR/$M_\star <
-10.5$).  This could potentially implicate an age effect, in which younger LMXB
populations have an excess of $10^{38}$~\lum\ sources.  Quantifying this effect
is beyond the scope of the current paper.

In both panels of Figure~13, we compare our best-fit
$M_\star$-and-$S_N$-dependent XLF model ({\it black curves\/}) with the
stellar-mass-dependent model from Zhang \etal\ (2012; Z12; {\it green dotted
curves\/}), which is based on a three-sloped broken power-law model with two
break locations (indicated in Figure~13).  Consistent with what we found for
the field LMXB XLF, our model reproduces the two-break nature of the LMXB XLF,
with the break locations consistent with those seen by Z12.  We find that the
Z12 model itself overpredicts the observed total LMXB XLFs for $S_N < 2$
galaxies by a factor of $\approx$2 for the $L \simlt 10^{39}$~\lum\ population.
Comparison between the Z12 model and our full-sample LMXB XLF shows very good
agreement.  This can be reconciled by the fact that the Z12 sample has
substantial overlap with our own sample, which is dominated by galaxies with
high-$S_N$ and GC LMXBs.  Thus, a single $M_\star$-dependent model (i.e., the
Z12 model) would not be applicable for low-$S_N$ galaxies, which are dominated
by field LMXBs at $L \simlt 10^{38}$~\lum.  Given that the late-type galaxy
LMXB XLF is in good agreement with the $S_N < 2$ LMXB XLF at $L \simlt
10^{38}$~\lum, this result further implies that $M_\star$-dependent LMXB XLFs
derived for massive elliptical galaxies overpredict the numbers of LMXBs in
this luminosity range for late-type galaxies.  However, at $L \simgt 3 \times
10^{38}$~\lum, the late-type galaxy LMXBs may exceed such predictions (see
bottom-left panel of Fig.~13).

\section{Discussion}

We find that, after excluding background \xray\ sources and LMXBs coincident
with GCs, the remaining field LMXB populations in elliptical galaxies show both
signatures of LMXBs that originated in GCs (via GC seeding), as well as a
non-negligible population of LMXBs that formed in situ through secular binary
evolution.  We construct a framework describing how the LMXB XLF in elliptical
galaxies varies with GC $S_N$, that provides a statistically acceptable model
to all \ngal\ elliptical galaxies in our sample.

Figure~14 illustrates our final model of LMXB populations in elliptical
galaxies, which is derived from the model parameters in Col.(4) and
(5) from Table~8.  In Figure~14$a$, we show the
stellar-mass-normalized LMXB XLF model at various $S_N$ values, illustrating
the variation of the LMXB XLF going from a broken power-law XLF for an
in-situ-dominated population at $S_N \simlt 2$ to a more numerous population of
GC-related LMXBs with a more complex XLF shape at higher $S_N$.

Figure~14$b$ tracks the stellar-mass-normalized integrated luminosity, $L_{\rm
X}/M_\star$, of LMXB populations as a function of $S_N$.  For our model, the
expectation value of $L_{\rm X}/M_\star$ can be calculated following:
\begin{equation}
L_{\rm X}/M_\star = \int_{L_{\rm lo}}^{L_c} \left( \frac{dN_{\rm
in\mbox{-}situ}}{dL} + \frac{d N_{\rm seed}}{dL} + \frac{dN_{\rm GC}}{dL}
\right) \frac{L}{M_\star} \; dL. \\
\end{equation} 
Integrating from $L_{\rm lo} = 10^{36}$~\lum\ to the cut-off luminosity at $L_c
= 10^{40}$~\lum\ gives:
\begin{eqnarray}
L_{\rm X}/M_\star = \alpha_{\rm in\mbox{-}situ} + \kappa_{\rm GC+seed} S_N,
\nonumber \\
\log (\alpha_{\rm in\mbox{-}situ}/[{{\rm erg~s^{-1}~M_\odot^{-1}}}]) = 28.76 \pm 0.07, \nonumber \\
\log (\kappa_{\rm GC+seed}/[{{\rm erg~s^{-1}~M_\odot^{-1}}}]) = 28.77^{+0.06}_{-0.07}. 
\end{eqnarray}
Here, we define $\alpha_{\rm in\mbox{-}situ} \equiv L_{\rm X, in~situ}/M_\star$
and $\kappa_{\rm GC+seed} \equiv (L_{\rm X, GC} + L_{\rm X,
seed})/M_\star/S_N$, as integrated luminosity scaling relations for in-situ and
combined direct-GC plus GC-seeded populations, respectively.  In Figure~14$b$,
we show our best-fit model $L_{\rm X}/M_\star$ versus $S_N$ as a solid black
line, and highlight the contributions from in-situ, GC-seeded, and GC
populations separately.  For comparison, we have overlaid the estimated $L_{\rm
X}/M_\star$ values for each of the \ngal\ galaxies, based on integrating
best-fitting broken power-law models, appropriate for the full LMXB population
of each galaxy (following the methods described in Section~4.3).

In terms of integrated luminosity, we find that galaxies above (below) $S_N
\approx 1.5$ are dominated by GC-produced (in-situ) LMXBs.  As discussed in
Section~4.6 above, typical late-type galaxies are observed to have $S_N \approx
1$, with the majority having $S_N < 2$, suggesting that although GC-produced
LMXBs dominate the elliptical galaxies in our sample, they are not expected to
dominate the integrated LMXB luminosities of more typical late-type galaxies in
the nearby universe.  Furthermore, as presented in $\S$4.6, there is evidence
that the late-type galaxy LMXB XLF contains an excess of luminous LMXBs ($L
\simgt 10^{38}$~\lum) compared with the $S_N < 2$ elliptical galaxy LMXB
population (see Fig.~13$a$), potentially due to an underlying age-dependence in
the LMXB population.

%
%
\begin{figure}
\figurenum{15}
\centerline{
\includegraphics[width=9.3cm]{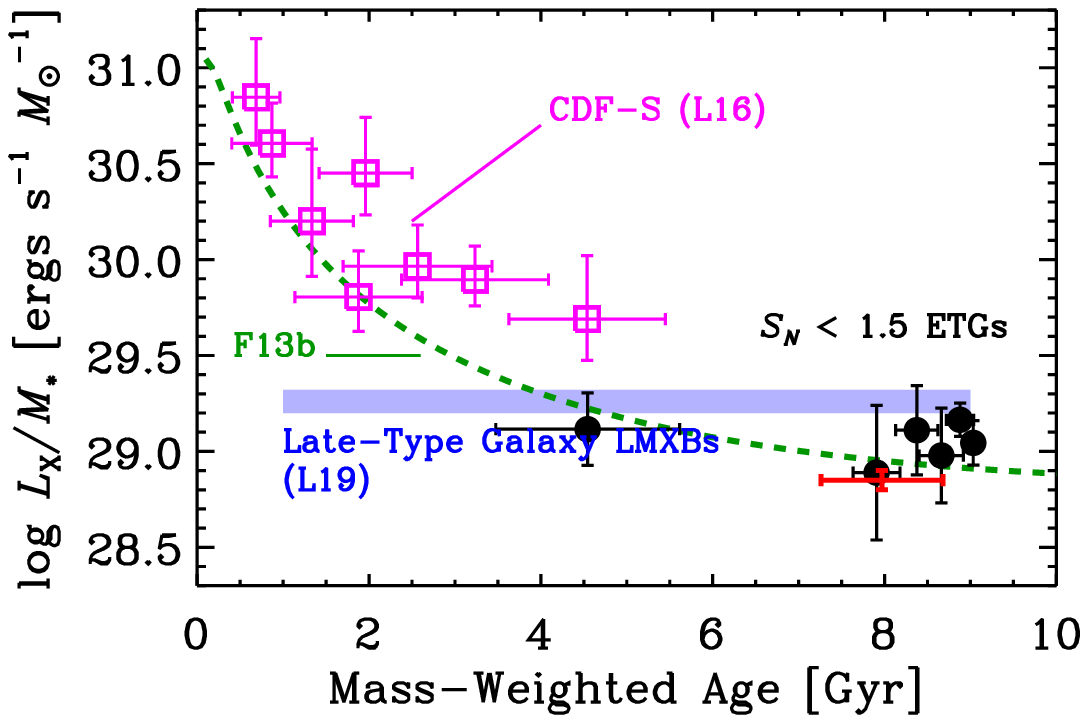}
}
\caption{
Integrated field LMXB XRB luminosity per unit stellar mass, $L_{\rm
X}/M_\star$, versus stellar-mass weighted age for elliptical galaxies with $S_N
< 1.5$ ({\it black circles} with 1$\sigma$ uncertainties).  These galaxies are
expected to have field LMXB populations dominated by the in-situ formation
channel, for which we expect $L_{\rm X}/M_\star$ to be stellar-age dependent.
The red error bars indicate 1$\sigma$ standard deviations on the mass-weighted
stellar ages and the uncertainty on the in-situ $L_{\rm X}/M_\star$ value
derived for the full sample.  The blue band shows the mean $L_{\rm X}/M_\star$
value for LMXBs derived for late-type galaxies from Lehmer \etal\ (2019), and
the magenta open squares (with 1$\sigma$ uncertainties) show estimates from the
Lehmer \etal\ (2016) stacking analyses from the 6~Ms CDF-S (see text for
details).  The green dashed curve shows the predicted XRB population synthesis
model trend from Fragos \etal\ (2013b), based on the evolving mass-weighted
stellar age of the Universe.   The collection of constraints thus far are
basically consistent with the trends predicted by the Fragos \etal\ (2013b)
population synthesis framework, however, significant uncertainties remain.
}
\end{figure}

In addition to the differences between late-type and elliptical galaxy LMXB
XLFs, there is evidence that the average integrated LMXB luminosity scaling
with stellar mass ($L_{\rm X}/M_\star$) increases with redshift between $z
\approx$~0--2 (Lehmer \etal\ 2007, 2016; Aird \etal\ 2017), signifying younger
LMXB populations (at high redshift) contain more luminous in-situ LMXBs.  Using
the mass-weighted stellar ages for our galaxies, based on the SFHs presented in
Section~3.1, we searched for any residual trends between the field-LMXB $L_{\rm
X}/M_\star$ and age.  To avoid contamination from GC LMXBs, we limited our
sample to elliptical galaxies with $S_N < 1.5$, which includes seven of the \ngal\
galaxies in our sample (NGC~1380, 3379, 3384, 3585, 4377, 4382, and 4697).
These galaxies span a mass-weighted stellar age range of 
$\approx$4--9~Gyr and have comparable values of field-LMXB $L_{\rm
X}/M_\star$ ($\log L_{\rm X}/M_\star \approx 29$), which we show in Figure~15.

In Figure~15, we have overlaid estimated values of $L_{\rm X}$(LMXB)/$M_\star$
versus mass-weighted stellar age for stacked constraints from Lehmer \etal\
(2016; L16), which are based on results from a 6~Ms exposure of the \chandra\
Deep Field-South (CDF-S; Luo \etal\ 2017).  These are mean values of $L_{\rm
X}$(LMXB)/$M_\star$ at different redshifts, in which the redshift has been
converted into a mass-weighted stellar age for the population.  These
mass-weighted stellar ages were calculated by first extracting synthesized
galaxy catalogs from the {\ttfamily Millenium~II} cosmological simulation from
Guo \etal\ (2011) that had the same SFR and $M_\star$ selection ranges as those
adopted by Lehmer \etal\ (2016).  These galaxy catalogs contain estimates of
the mass-weighted stellar ages for each galaxy.  The mass-weighted stellar age
of the entire galaxy population (catalog) is then estimated, and a standard
deviation of the population is calculated to estimate the uncertainty.  Given
this highly model-dependent procedure, we provide these points only for
guidance and note that their true uncertainties are likely to be much larger
than those shown.

The L16 CDF-S constraints indicate that galaxies with mass-weighted stellar
ages in the range of $\approx$\hbox{0.5--5}~Gyr, have $\log (L_{\rm X}$[LMXB]/$M_\star)
=$~\hbox{29.5--31}, generally well above the $L_{\rm X}/M_\star$ values for the
elliptical galaxies in our sample (including those with the highest $S_N$).
Since we do not expect that LMXBs that originate in GCs would have XLFs or
$L_{\rm X}$[GC-LMXB]/$M_\star$ values that depend on stellar age, it is likely
that these high redshift LMXB populations would be dominated by in-situ LMXBs.
In Figure~15, we show $L_{\rm X}/M_\star$ versus mass-weighted stellar age from
the XRB population synthesis model of Fragos \etal\ (2013b; F13b, {\it green
dashed curve\/}), which is based entirely on the in situ formation channel.
Qualitatively, the constraints for $S_N < 1.5$ elliptical galaxies studied here, the
late-type galaxies presented in Lehmer \etal\ (2019), and the CDF-S stacked
data follow the F13b trend, providing support for a rapidly evolving in situ
LMXB population that dominates for most galaxies, except for the most GC-rich
elliptical galaxies.  

Unfortunately, we are unable to find quantitative evidence for stellar-age
dependence in the in situ formation rates of the elliptical galaxies studied
here, due to a lack of diverse stellar ages in our sample.  In future work, we
can mitigate this limitation by using SFH information, similar to that
presented in Section~2.3, combined with \chandra\ constraints on XRB
populations for a combined late-type and elliptical galaxy sample.  The XRB
XLFs for members of this combined galaxy sample would contain non-negligible
contributions from XRBs associated with stellar populations of all ages.  In
Lehmer \etal\ (2017), we presented preliminary work for the single case of the
XRB population within M51, and were able to show that a stellar-age
parameterized XLF model suggested that the XRB population integrated $L_{\rm
X}/M_\star$ declines by 2.5--3 orders of magnitude from 10~Myr to 10~Gyr,
similar to that shown in Figure~15.  With the combined late-type and elliptical
galaxy sample, we can constrain the evolution of the XRB XLF as a function of
age without an explicit parameterization with age.

\acknowledgements We thank the anonymous referee for their
thoughtful and helpful comments on the paper.  These comments led to notable
improvements in the quality and presentation of this work.  We
gratefully acknowledge support from the National Aeronautics and Space
Administration (NASA) Astrophysics Data Analysis Program (ADAP) grant
NNX16AG06G (B.D.L., A.F., K.D., R.T.E.), \chandra\ X-ray Center grant
GO7-18077X (B.D.L. and A.F.), and Space Telescope Science Institute grant
HST-GO-14852.001-A (B.D.L. and A.F.).  W.N.B. thanks NASA ADAP grant
80NSSC18K0878 and the V.M. Willaman Endowment.  G.R.S. acknowledges support
from NSERC Discovery Grant RGPIN-2016-06569.

We made use of the NASA/IPAC Extragalactic Database (NED), which is
operated by the Jet Propulsion Laboratory, California Institute of Technology,
under contract with the National Aeronautics and Space Administration.

Our work includes observations made with the NASA {\it Galaxy Evolution
Explorer} (\galex). \galex\ is operated for NASA by the California Institute of
Technology under NASA contract NAS5-98034. This publication makes use of data
products from the Two Micron All Sky Survey (2MASS), which is a joint project
of the University of Massachusetts and the Infrared Processing and Analysis
Center/California Institute of Technology, funded by NASA and the National
Science Foundation (NSF). This work is based on observations made with the {\it
Spitzer Space Telescope}, obtained from the NASA/IPAC Infrared Science Archive,
both of which are operated by the Jet Propulsion Laboratory, California
Institute of Technology under a contract with NASA.
We acknowledge the use of public data from the {\it
Swift} data archive.

{\it Facilities:} {\it Chandra}, {\it Herschel}, {\it Hubble}, {\it GALEX}, Sloan, {\it Spitzer}, {\it Swift}, 2MASS

\software{DrizzlePac (v2.1.14; Fruchter \& Hook~2002), SExtractor (v2.19.5; Bertin\&Arnouts~1996); ACIS Extract (v2016sep22; Broos et al. 2010, 2012), MARX (v5.3.2; Davis et al. 2012), CIAO (v4.8; Fruscione~et al. 2006), xspec (v12.9.1; Arnaud~1996)}

%

%

\appendix

%
%
\begin{figure*}[t!]
\figurenum{A1}
\centerline{
\includegraphics[width=6cm]{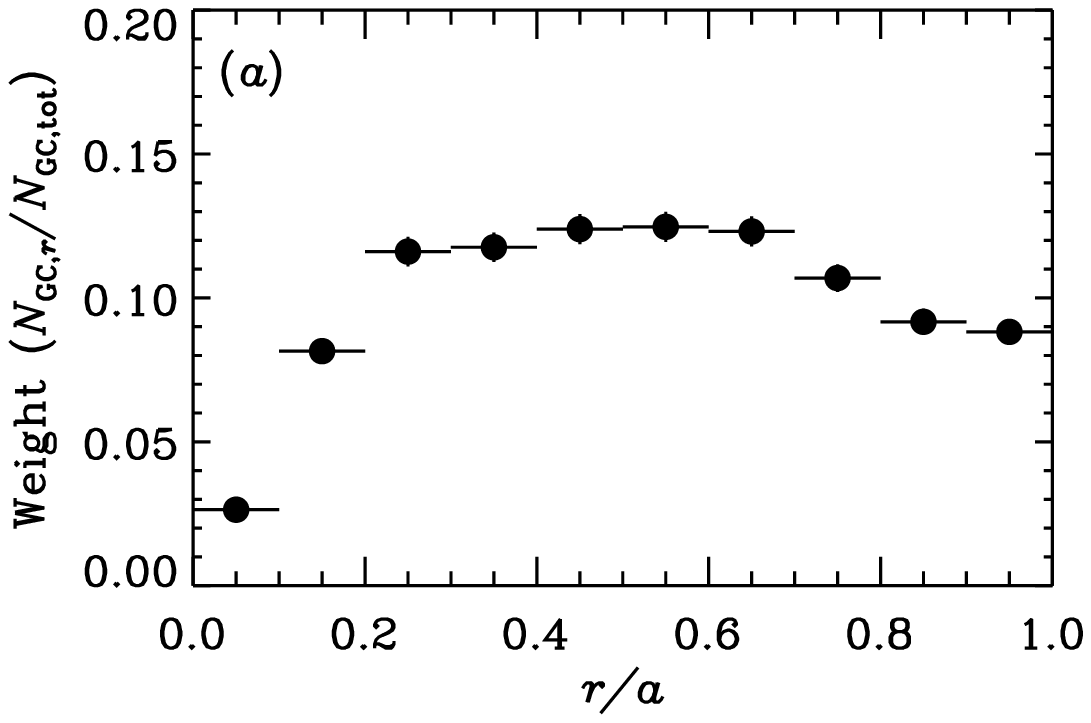}
\hfill
\includegraphics[width=6cm]{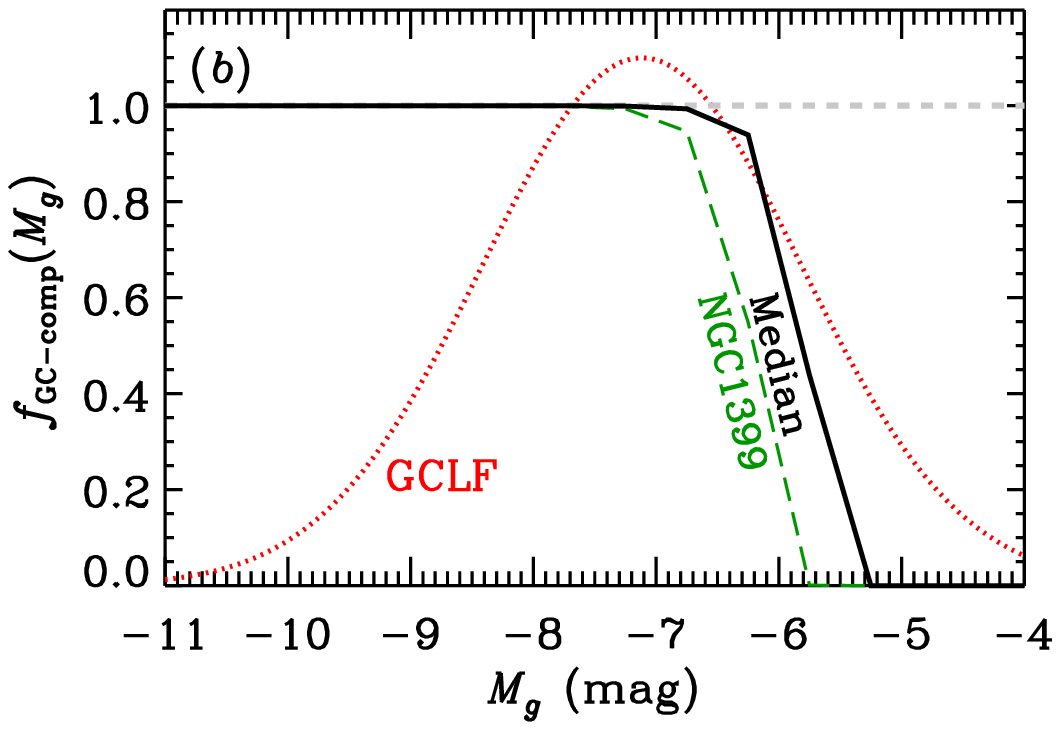}
\hfill
\includegraphics[width=6cm]{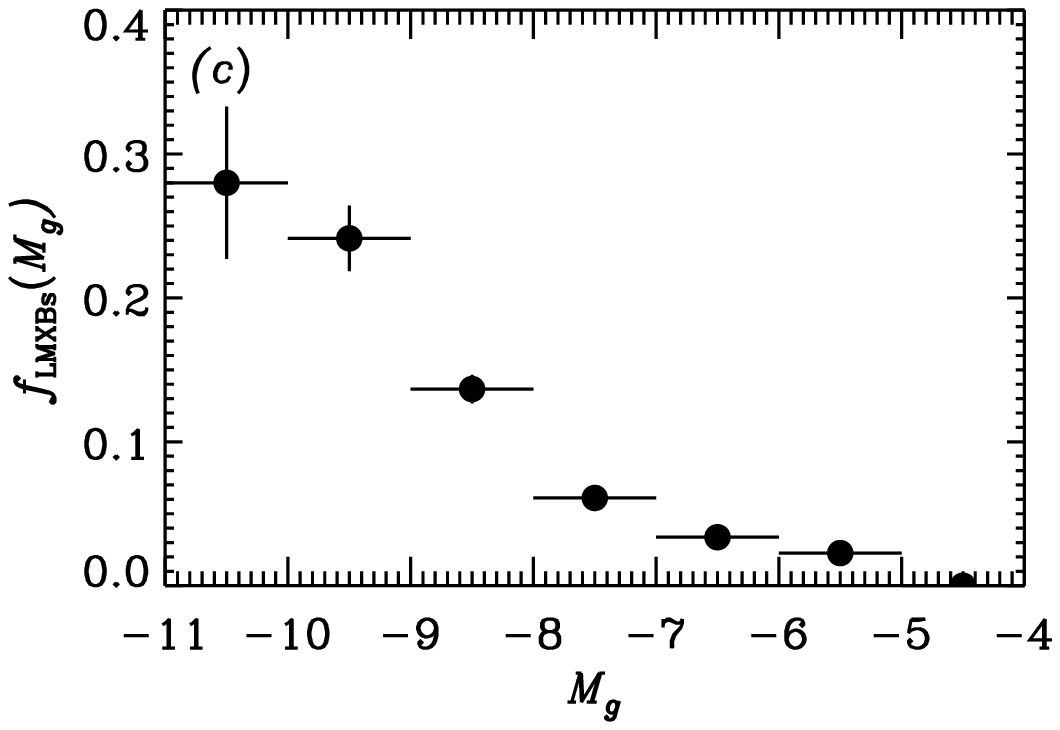}
}
\caption{
($a$) Fraction of bright GCs ($M_g < -7$) as a function of galactocentric
semi-major axis, $r/a$, for all \ngal\ galaxies in our sample.  Each annulus is
elliptical in shape with position angle and axis ratio matching that of the
$K_s$-band ellipse defined in Table~1.  This distribution is used to weight GC
completeness functions, which vary across the galaxies due to background
variations.
($b$) Weighted GC completeness functions for the example case of NGC~1399 ({\it
green dashed curve\/}) and the median of our sample ({\it black solid
curve\/}).  For comparison, the shape of the GC optical luminosity function
from Kundu \& Whitmore~(2001) has been overlaid ({\it red dotted curve\/}).
($c$) Fraction of GCs with \chandra-detected point sources as a function of
absolute magnitude.  The fraction of \xray\ detected sources drops rapidly with
decreasing GC optical luminosity suggesting that most of the GCs below the
\hst\ detection thresholds will not host LMXBs.
}
\end{figure*}

\section{Globular Cluster Completeness and Local Specific Frequency}

Quantifying the relative contributions of LMXBs associated with direct-GC
counterparts and those seeded in the field requires that we obtain estimates of
the local specific frequency, $S_{N, {\rm loc}}$, as well as the completeness
of our \hst\ data to detecting GCs that host LMXBs.  We define the GC-LMXB
completeness, $f_{\rm GC-LMXB}^{\rm recover}$ as the fraction of \xray-detected
sources for which we could identify a direct GC counterpart, if one were
present.  For a given galaxy, $f_{\rm GC-LMXB}^{\rm recover}$ will depend on
\hst\ exposure depths, the variation of the background light throughout the
galaxy (primarily due to the galaxy light profile), the distribution of GCs
throughout the galaxy, the intrinsic optical luminosity function of the GC
population, and finally, the intrinsic fraction of \xray\ sources that are
associated with GCs as a function of optical luminosity.

We measured each of the above completeness factors using a series of
techniques.  We began by using simulations to measure the magnitude
and background dependent completeness function for GCs in the galaxy NGC~1399,
which contains the largest range of background levels for our galaxy sample and
a rich GC population.  Due to computation time limitations, our strategy was to
determine a magnitude and background dependent parameterization for the
completeness function for NGC~1399, and then apply that parameterization to
other galaxies in our sample to assess completeness variations.

First, working with \hst\ images for NGC~1399 in units of counts, we added 120,000
fake GCs, with absolute magnitudes in the range of $-10 \le M_g \le -5$.  The
surface brightness distributions of our sources followed empirical King light profiles with
parameters chosen to mimic the light profiles of detected GCs.  The sources
were implanted at random positions across the extent of the galaxies, covering
all background levels.  Following the procedures outlined in Section 3.2, we
searched the fake images for source detections and calculated the GC recovery
fraction as a function of background counts and apparent magnitude.  

From these simulated recovery fractions, we generated an absolute-magnitude and
background-count dependent parameterization that followed the observed
behavior.  To test whether our parameterization was robust, we repeated the
above simulation process for NGC~1404 to generate recovery fraction diagrams in
bins of background counts.  We found that our NGC~1399-based parameterization,
given simply local background counts in NGC~1404, correctly reproduced the
simulated completeness functions very well with only a $\approx$0.5~mag
deviation appearing at the highest background levels.  To account for this
deviation at high background levels, we based our final recovery fraction
parametrization on the simulations for both NGC~1399 and NGC~1404, and we take
this 0.5~mag deviation to be a conservative limit on the robustness of our
parameterization.

For the remaining galaxies, we calculated completeness levels using the
measured background levels within each of the galaxies and our parameterization
for completeness as a function of background.  the mean background levels in
several concentric annuli from the galactic centers to the edges of the
$K_s$-band ellipses.  The annuli were elliptical in size and followed the axis
ratios and position angles provided in Table~1.  At each annulus, our
completeness parameterization provides a magnitude-dependent completeness
function.  To assess the overall GC detection completeness for a given galaxy,
we required knowledge of the spatial distribution of GCs throughout the galaxy,
since the total population completeness will depend on the relative weightings
on the local completeness functions.  Using all GCs in our total elliptical
galaxy sample with $M_g< -7$~mag (bright GCs that we are $\approx$100\%
complete to), we compiled the galaxy-sample-total distribution of GCs with
respect to galactocentric offset (in units of fractional distance to the
nearest edge of the galaxy), $w_{\rm GC}(r) \equiv N_{{\rm GC},r}/N_{\rm GC,tot}$,
where $N_{{\rm GC},r}$ is the number of GCs within annuli of galactocentric
semi-major axes $r/a$ and $N_{\rm GC,tot}$ is the total number of GCs within the whole
sample.  Figure~A1($a$) shows $w_{\rm GC}(r)$ measured from the whole sample.
Examination of the distributions of GCs within individual galaxies show
consistency with this distribution, when such distributions can be measured
reliably.

Using the galactocentric offset distribution of GCs, $w_{\rm GC}(r)$,  as
statistical weights, we derived a weighted completeness function for each
galaxy following:
\begin{equation}
f_{\rm GC-comp}(M_g) = \sum_i^{n_{\rm r}} w_{\rm GC}(r_i) f_{\rm GC-comp}(M_g,
r_i),
\end{equation}
where the summation takes place over $n_{\rm r} = 10$ annuli. Here, $r_i$,
$w_{\rm GC}(r_i)$, and $f_{\rm GC-comp}(M_g, r_i)$ are the radius (in units of the
semimajor axis), fraction of GCs located in the annulus, and completeness
function, respectively, appropriate for the $i$th annulus.  

Figure~A1(b) shows the weighted magnitude dependent completeness function for
NGC~1399 and the median galaxy in our sample.  In these cases, the \hst\ data
are complete to a 50\% limiting F475W absolute magnitudes of $M_g \approx -6.2$
and $-5.8$~mag for NGC~1399 and the median, respectively.  For context, in
Figure~A1(b), we overlay the shape of the best-fit Gaussian GC luminosity
function from Kundu \& Whitmore~(2001),
\begin{equation}
\frac{dN_{\rm GC}}{dM_g} = A_{\rm GC} \exp\left[ - \frac{ (M_g - M_0)^2 }{2
\sigma^2} \right],
\end{equation}
where $A_{\rm GC}$ has been normalized arbitrarily, $M_0 = -7.1$~mag (converted
from $M_V = -7.4$~mag), and $\sigma = 1.3$~mag.  Based on these diagrams, it is
clear that our \hst\ data will not be fully complete to all GCs present within
the galaxy.

We note, however, that the
fraction of GCs that host \xray\ detected sources is known to decline with
decreasing GC luminosity.  Using the known GC populations within all galaxies,
we calculated the fraction of GCs coincident with \xray\ detected objects as a
function of GC absolute magnitude, $f_{\rm LMXBs}(M_g)$.  In Figure~A1($c$), we
show our empirical version of $f_{\rm LMXBs}(M_g)$, which indeed shows a
precipitous decline with decreasing GC optical luminosity.

Given the above ingredients, we can finally calculate the fraction of GCs that
host LMXBs that would be identified as such using the following equation:
\begin{equation}
f_{\rm GC-LMXB}^{\rm rec} \approx \frac{ \sum_{j=0}^{n_{\rm M}} (dN_{\rm
GC}/dM_g)_j f_{\rm GC-comp}(M_g)_j f_{\rm LMXBs}(M_g)_j}{\sum_{j=0}^{n_{\rm M}}
(dN_{\rm GC}/dM_g)_j f_{\rm LMXBs}(M_{g})_j} ,
\end{equation}
where the summations take place over $n_M = 100$ bins of absolute magnitude
from $-11 \simlt M_g \simlt -3$ using linearly-interpolated values from our
$f_{\rm GC-comp}$ and $f_{\rm LMXBs}$.  Since the completeness functions vary
somewhat from galaxy-to-galaxy, primarily due to distance variations, the value
of $f_{\rm GC-LMXB}^{\rm rec}$ varies too.  We find a full range of $f_{\rm
GC-LMXB}^{\rm rec} =$~89.0--99.6\%, with a median completeness of 96.1\%.  This
implies that our GC LMXB classifications are highly complete, and we are very
unlikely to be misclassifying a substantial fraction of LMXBs coincident with
GCs as field sources.

The GC completeness information obtained above allows us to
estimate $S_{N, {\rm loc}}$ for our sample.  For each galaxy, we first obtained
an estimate of the total number of GCs that were present locally within the
\hst\ observational fields, $N_{\rm GC, loc}$, using Equation~(A2) and our
completeness functions (Equation~A1).  Specifically, we determined the value of
$A_{\rm GC}$ from Equation~(2) using the following equation:
$$A_{\rm GC} = \frac{N_{\rm GC, obs}}{\int \frac{dN_{\rm GC}}{dM_g} f_{\rm
GC-comp}(M_g) dM_g} = \frac{N_{\rm GC, obs}}{\int f_{\rm GC-comp}(M_g)
\exp\left[ - \frac{ (M_g - M_0)^2 }{2 \sigma^2} \right] dM_g},$$
where $N_{\rm GC,obs}$ is the number of GCs observed in the \hst\ field of
view.  Given $A_{\rm GC}$, we estimated $N_{\rm GC, loc}$ as
$$N_{\rm GC, loc} = \int \frac{dN_{\rm GC}}{dM_g}dM_g.$$
Finally, values of $S_{N, {\rm loc}}$ were calculated following
\begin{equation}
S_{N, {\rm loc}} = N_{\rm GC, loc} 10^{0.4(M_{V,{\rm loc}} + 15)},
\end{equation}
where estimates of local absolute $V$-band magnitude, $M_{V, {\rm loc}}$, were
obtained from our SED fitting procedure (see Section~3.1).

\section{X-ray Point Source Catalog}


\begin{table*}
\renewcommand\thetable{B1}
{\footnotesize
\begin{center}
\caption{X-ray point-source catalog and properties}
\begin{tabular}{llcccccccccc}
\hline\hline
\multicolumn{1}{c}{\sc Gal.} &  &  $\alpha_{\rm J2000}$ & $\delta_{\rm J2000}$ & $\theta$ & $N_{\rm FB}$ & $N_{\rm H}$  & & $\log F_{\rm FB}$ & $\log L_{\rm FB}$  & {\sc Loc.} & {\sc Opt.}\\
 \multicolumn{1}{c}{(NGC)} & \multicolumn{1}{c}{\sc ID} &  (deg) & (deg) & (arcmin) & (counts) & ($10^{22}$~cm$^{-2}$) & $\Gamma$ & (cgs) & (\lum)  & Flag & {\sc Class} \\
 \multicolumn{1}{c}{(1)} & \multicolumn{1}{c}{(2)} & (3) & (4) & (5) & (6)--(7) & (8)--(9) & (10)--(11)  & (12) & (13) & (14) & (15) \\
\hline\hline
1023 & 1 & 02 40 04.07 & +39 03 53.89 & 3.9 & 13.7$\pm$4.5 & 0.056 & 1.7 &  $-$15.2 & 37.0& 3 & U \\
  & 2 & 02 40 04.83 & +39 04 25.65 & 3.8 & 24.5$\pm$5.7 & 0.056 & 1.7 &  $-$15.0 & 37.2& 3 & U \\
  & 3 & 02 40 04.94 & +39 00 40.53 & 4.8 & 16.9$\pm$5.0 & 0.056 & 1.7 &  $-$14.9 & 37.3& 3 & U \\
  & 4 & 02 40 05.13 & +39 05 06.94 & 3.9 & 110.1$\pm$12.6 & 0.823$\pm$0.539 & 1.90$\pm$0.64 &  $-$14.2 & 38.0& 3 & U \\
  & 5 & 02 40 05.81 & +39 04 32.56 & 3.6 & 30.4$\pm$7.5 & 1.527$\pm$1.276 & $<$2.57 &  $-$14.8 & 37.4& 3 & U \\
\\
  & 6 & 02 40 07.52 & +39 03 43.77 & 3.2 & 73.2$\pm$10.4 & 0.291$\pm$0.340 & 2.37$\pm$0.87 &  $-$14.6 & 37.6& 3 & U \\
  & 7 & 02 40 07.76 & +39 03 14.00 & 3.2 & 118.1$\pm$12.8 & $<$0.056 & 1.92$\pm$0.30 &  $-$14.3 & 37.9& 3 & U \\
  & 8 & 02 40 08.37 & +39 01 59.18 & 3.5 & 76.2$\pm$10.6 & 0.504$\pm$0.493 & 2.25$\pm$0.94 &  $-$14.5 & 37.7& 3 & U \\
  & 9 & 02 40 08.64 & +39 04 42.73 & 3.1 & 14.2$\pm$4.5 & 0.056 & 1.7 &  $-$15.2 & 37.0& 3 & U \\
  & 10 & 02 40 10.43 & +39 04 09.94 & 2.7 & 15.5$\pm$4.4 & 0.056 & 1.7 &  $-$15.2 & 37.0& 3 & U \\
\\
  & 11 & 02 40 10.45 & +39 02 25.88 & 3.0 & 66.9$\pm$8.5 & 0.056 & 1.7 &  $-$14.6 & 37.6& 3 & U \\
  & 12 & 02 40 11.35 & +39 05 29.79 & 3.0 & 48.5$\pm$8.8 & $<$0.056 & 2.14$\pm$0.51 &  $-$14.8 & 37.4& 3 & U \\
  & 13 & 02 40 11.37 & +39 06 09.86 & 3.4 & 25.6$\pm$5.7 & 0.056 & 1.7 &  $-$15.0 & 37.2& 3 & U \\
  & 14 & 02 40 12.56 & +39 06 49.39 & 3.8 & 23.7$\pm$5.9 & 0.056 & 1.7 &  $-$15.0 & 37.2& 3 & U \\
  & 15 & 02 40 13.04 & +39 00 51.77 & 3.6 & 3735.1$\pm$65.6 & $<$0.056 & $<$3.30 &  $-$12.9 & 39.3& 3 & U \\
\\
  & 16 & 02 40 13.53 & +39 01 33.09 & 3.0 & 181.2$\pm$15.5 & $<$0.056 & 1.62$\pm$0.23 &  $-$14.1 & 38.1& 3 & U \\
  & 17 & 02 40 13.70 & +39 04 04.12 & 2.0 & 93.6$\pm$11.6 & 7.333$\pm$1.447 & $<$3.30 &  $-$14.1 & 38.1& 1 & F \\
  & 18 & 02 40 14.37 & +39 02 50.95 & 2.1 & 8.9$\pm$3.5 & 0.056 & 1.7 &  $-$15.5 & 36.7& 1 & F \\
  & 19 & 02 40 15.56 & +39 00 15.03 & 3.9 & 24.2$\pm$5.8 & 0.056 & 1.7 &  $-$14.9 & 37.3& 3 & U \\
  & 20 & 02 40 15.90 & +39 07 23.95 & 3.9 & 23.3$\pm$6.0 & 0.056 & 1.7 &  $-$15.0 & 37.2& 3 & U \\
\\
\hline
\end{tabular}
\end{center}
Note.---The full version of this table contains 4206 sources.  An abbreviated version of the table is displayed here to illustrate its form and content.  A description of the columns is provided in the text of Appendix~B.\\
}
\end{table*}
In Table~B1, we provide the \xray\ point source catalogs, based on the analyses
presented in $\S\S$3.2 and 3.3.  The columns include the following: Col.(1):
Name of the host galaxy. Col.(2): point-source identification number within the
galaxy. Col.(3) and (4): Right ascension and declination of the point source.
Col.(5): Offset of the point source with respect to the average aim point of
the \chandra\ observations. Col.(6) and (7): 0.5--7~keV net counts (i.e.,
background subtracted) and 1$\sigma$ errors. Col.(8)--(9) and (10)--(11):
Best-fit column density $N_{\rm H}$ and photon index $\Gamma$, respectively,
along with their respective 1$\sigma$ errors, based on spectral fits to an
absorbed power-law model ({\ttfamily TBABS$_{\rm Gal} \times$ TBABS $\times$
POW} in {\ttfamily xspec}).  For sources with small numbers of counts ($<$20
net counts), we adopted only Galactic absorption appropriate for each galaxy
and a photon index of $\Gamma = 1.7$.  Col.(12) and (13): the respective
0.5--8~keV flux and luminosity of the source. Col.(14): Flag indicating the
location of the source within the galaxy.  Flag=1 indicates the source is
within the $K_s$-band footprint adopted in Table~1, and outside a central
region of avoidance, if applicable.  All XLF calculations are based on Flag=1
sources.  Flag=2 indicates that the source is located in the central region of
avoidance due to either the presence of an AGN or very high levels of source
confusion.  Flag=3 indicates that the source is outside the
20~mag~arcsec$^{-2}$ $K_s$-band ellipse of the galaxy. Col.(15): \hst\
classification of the source, whenever the source is within the footprint of
the galaxy, as defined in Table~1.  Classifications include field LMXBs
(``F''), direct-GC LMXBs (``G''), background sources (``B''), and unclassified sources (``U'').

\end{document}